%% file: paper_2024_arxiv_update.tex
\documentclass[11pt]{article}

\usepackage{cite}

\usepackage{amssymb, amsmath}
\usepackage{graphicx}
\usepackage{color}
\usepackage[dvipsnames]{xcolor}
\usepackage{transparent}
\usepackage{svg}
\usepackage{import}
\usepackage{subcaption}	
\usepackage{placeins}

\usepackage{floatrow}
\usepackage{float}
\floatstyle{plaintop}
\restylefloat{table}

\usepackage{tabularx}
\usepackage{booktabs}

\def\calF{{\cal F}}
\def\calG{{\cal G}}
\def\Rea{\mbox{Re}}
\def\Imm{\mbox{Im}}

\def\w{\textit{w}}	
\def\imag{{\rm i}}

\newcommand{\prti}[2]{\displaystyle{\partial #1\over\partial #2}}

\def\eq{eq.~\hspace*{-.2em}}	
\def\eqs{eqs.~\hspace*{-.2em}}

\def\fig{fig.~}	
\def\figs{figs.~}	
\def\tab{tab.~}	
\def\sec{sec.~}	
\def\secs{secs.~}

\def\App{App.~}	
\newcommand{\Refr}{Ref.~}
\newcommand{\Refrs}{Refs.~}

\newcommand{\eg}{e.\,g.\ }
\newcommand{\ie}{i.\,e.\ }
\newcommand{\etal}{et al.\ }

\def\rhs{right-hand side}	

\newcommand{\SVGmathsize}{\fontsize{7}{2}\selectfont}
\newcommand{\SVGlabelsize}{\fontsize{9}{2}\selectfont}
\newcommand{\SVGunitsize}{\fontsize{7}{2}\selectfont}

\newcommand{\svglabelsize}{\fontsize{9}{2}\selectfont}
\newcommand{\svgunitsize}{\fontsize{7}{2}\selectfont}
\newcommand{\svglegendsize}{\fontsize{7}{2}\selectfont}

\begin{document}

\begin{center}
{\Large
{\bf Modulational instability of Geodesic-Acoustic-Mode packets}
}

\renewcommand*{\thefootnote}{\fnsymbol{footnote}}

\vskip 15pt
{D.~Korger$^{1,2,\ast}$, E.~Poli$^{1}$, A.~Biancalani$^3$, A.~Bottino$^1$, O.~Maj$^1$, J.~N.~Sama$^4$}

\vskip 10pt
\noindent
$^1$ {\it Max Planck Institute for Plasma Physik, D-85748 Garching, 
	Germany}\\
$^2$ {\it Ulm University, D-89081 Ulm, Germany}\\
$^3$ {\it L\'{e}onard de Vinci P\^{o}le Universitaire, Research Center, F-92916 Paris, La D\'{e}fense, France} \\
$^4$ {\it Université de Lorraine, CNRS, IJL, F-54000 Nancy, France}

\vskip 10pt
\noindent
$^\ast$ Corresponding author: david.korger@ipp.mpg.de

\end{center}

\small{\noindent{\bf Abstract}:
Isolated, undamped geodesic-acoustic-mode (GAM) packets have been demonstrated to obey a (focusing) nonlinear Schr\"{o}dinger equation (NLSE) [E. Poli, Phys. Plasmas 2021]. This equation predicts susceptibility of GAM packets to the modulational instability (MI). The necessary conditions for this instability are analyzed analytically and numerically using the NLSE model. The predictions of the NLSE are compared to gyrokinetic simulations performed with the global particle-in-cell code ORB5, where GAM packets are created from initial perturbations of the axisymmetric radial electric field $E_r$. An instability of the GAM packets with respect to modulations is observed both in cases in which an 
initial perturbation is imposed and when the instability develops spontaneously. However, significant differences in the dynamics of the small scales are discerned between the NLSE and gyrokinetic simulations. 
These discrepancies are mainly due to the radial dependence of the strength of the nonlinear term, which we do not retain in the solution of the NLSE, and to the damping of higher spectral components.
The damping of the high-$k_r$ components, which develop as a consequence of the nonlinearity, can be understood in terms of Landau damping. The influence of the ion Larmor radius $\rho_i$ as well as the perturbation wavevector $k_\text{pert}$ on this effect is studied. 
For the parameters considered here the aforementioned damping mechanism hinders the MI process significantly from developing to its full extent and is strong enough to stabilize some of the (according to the undamped NLSE model) unstable wavevectors.  
}


\section{Introduction}
\label{sec:intro}
The Geodesic Acoustic Mode (GAM) is a plasma oscillation observed in fusion reactors with toroidal geometry, such as tokamaks or stellarators. It develops when the $E\!\times\!B$ drift velocity of the zonal flows (ZFs) varies so strongly along the poloidal coordinate that the corresponding $m = \pm 1, n = 0$ flow divergence (where $m$ and $n$ are the poloidal and toroidal mode numbers, respectively) cannot be compensated by parallel flows \cite{Conway21}. As a result, a characteristic $m = \pm 1, n = 0$ ``up-down antisymmetric'' pressure mode emerges, which leads to an oscillation of the ZFs, \ie the GAM \cite{Conway21,Qiu18,ScottNJP05}. GAMs are thus recognized to be the non-stationary branch of the zonal flows \cite{Conway21,DiamondZF} and their associated electric potential is (to the leading order) an $m = n = 0$ structure. The name GAM stems from the geodesic magnetic field line curvature, which is responsible for plasma compressibility and thus a necessary condition for the emergence of the characteristic pressure mode. 

The interaction between GAMs and turbulence is fairly complex. Nonlinear self-interactions of drift-wave (DW) turbulence are one of the main mechanisms for generating the perturbations of the electric potential, which are the origin of ZFs and GAMs \cite{Itoh06PoP}. Meanwhile, similarly to the ZFs, GAMs are understood to suppress DW turbulence and regulate cross-field turbulence, thus enhancing energy confinement \cite{Conway21}. Still, their direct effect on turbulence is not clear at the moment \cite{SmolyakovPPR16}, as GAMs are known to deplete the energy available to ZFs and transfer part of the energy of the system back to turbulence \cite{ScottPPCF03,ScottPLA03}. This complex contribution to the turbulence dynamics makes GAMs highly interesting in current fusion research.

It was recently shown in \Refr\cite{PoP21} that in the regime of moderate nonlinearity the dynamics of undamped isolated GAM packets is well described by a (cubic) nonlinear Schr\"{o}dinger equation (NLSE). The NLSE is a standard model \cite{Spatschek} for describing nonlinear dispersive oscillations with a (linear) dispersion relation of the form 
\begin{equation}
    \omega(k_r) = a + b k_r^2,
\end{equation}
which in the limit $k_r^2 \rho_i^2 \ll 1$ (where $\rho_i$ denotes the ion Larmor radius and $k_r$ is the radial wavevector associated with the GAM radial electric field) approximates to the standard gyrokinetic GAM dispersion relation \cite{Smolyakov08}, as will be discussed in \sec\ref{sec:nlse-introdtion}. Adopting an NLSE as a model equation for GAM packets is furthermore supported by the general result that plasma eigenmodes (more exactly, their radial envelope), arising in toroidal systems as a consequence of various types of instabilities, can be described by a nonlinear Schr\"{o}dinger equation with integro-differential coefficients \cite{ZoncaPoP14I,ZoncaPoP14II,Zonca15,ChenZoncaRMP}. The NLSE model has been applied and studied extensively in the contexts of deep water waves, light traveling through optical fibres, Bose-Einstein condensates and others \cite{Agrawal,kuznetsov86,kengne21}. Some predictions of the NLSE are well-known, like the emergence of solitons, nonlinear wave breaking, the nonlinear phase shift and susceptibility to the modulational instability (MI). While some of these phenomena have already been observed in gyrokinetic simulations of GAMs in \Refr\cite{PoP21}, thus confirming the NLSE as a valid description of the dynamics, in this report the focus is set on the MI, which to the best of the authors' knowledge has not yet been studied in the context of GAMs. 

The MI is usually analyzed for wave envelopes which consist of a (nearly) constant phase front that is modulated by a sinusoidal perturbation with a wavelength $\lambda_\text{pert} = 2\pi/k_\text{pert}$. One finds that this perturbation is unstable under the conditions that the NLSE is self-focusing (which for GAMs is the case when $\tau_e = T_e/T_i \lesssim 5.45$ \cite{SmolyakovPPR16}) and the wavevector $k_\text{pert}$ of the modulation of the envelope is within a certain range, which will be discussed in further detail in \sec\ref{sec:mi}. Unstable perturbations will grow exponentially at the expense of the constant envelope component until the sinusoidal modulation dominates the shape of the oscillation and saturates.

In the field of plasma physics the MI has been observed in numerous waves and oscillations (\eg \Refrs\cite{Schamel75,murtaza82}). Most notable for the context of this paper are DWs, which as explained before are one of the driving mechanisms of GAMs and ZFs. The MI of DWs has been shown to spontaneously excite ZFs \cite{Chen00} or increase their amplitude \cite{Itoh06}. Since in this study the focus is set on \textit{isolated} GAMs, where the effects of the generation mechanisms such as DWs are excluded, it is stressed here that in the present analysis the MI stems only from the self-interaction of GAMs and is not directly connected to DW MI.

After analyzing the conditions for MI for the case of GAMs, the analytic predictions of the NLSE are first confirmed by numerical simulations of the NLSE and then validated against gyrokinetic simulations obtained from the global particle-in-cell code ORB5 \cite{Bottino15,Lanti20}. Details about the numerical tools are given in \sec\ref{sec:numerics}. The results, which are presented in \sec\ref{sec:results}, demonstrate that the MI does in fact appear in gyrokinetic GAM simulations, however, significant differences between the gyrokinetic and NLSE simulations are observed that can in part be explained by the (currently not included) radial dependency of the nonlinear strength $\alpha_\text{NL} = \alpha_\text{NL}(r)$ in the NLSE. Moreover, the radial spectra of the simulations indicate that a damping term should be included in the NLSE model in this context. The need to consider damping in the current context may at first seem surprising, since simulations are performed using a high safety factor ($q_s = 15$) and an initial $k_r$ spectrum for which, according to theoretic predictions by \Refrs\cite{SugamaJPP06,SugamaErr,Qiu09}, it is expected that damping of GAMs is weak to negligible. However, the nonlinear evolution of the packet leads naturally to the generation of shorter and shorter wavelengths which are more efficiently damped. This effect is the nonlinear analogue of the enhanced Landau damping discussed in \Refrs\cite{PalermoEPL,Biancalani16}, where the shorter wavelengths were generated by the linear dynamics in the presence of gradients. 

In this paper, the theoretical predictions for the MI growth rate are modified with the inclusion of the damping rate derived in \Refr\cite{Qiu09}, which is appropriate in the in the underlying high safety factor $q_s$ and high spectral wavevectors (high $q_s, k_r$) regime and is furthermore consistent with the approximation of adiabatic electrons employed in the GK simulations. Good agreement between theory and gyrokinetic simulations is found if the damping rate of  \Refr\cite{Qiu09} is multiplied by a factor of approx. 2.5 (see \secs \ref{sec:results-kpert} and \ref{sec:results-rhoi}). This is not unexpected, as a similar discrepancy has been already reported in previous benchmarks \cite{Biancalani17}. The following \sec\ref{sec:results-damped-NLSE} includes these findings in the NLSE model and compares damped NLSE simulations with the gyrokinetic results, which are observed to show good agreement.

Sections \ref{sec:results-self-focusing} and \ref{sec:results-breather} illustrate the self-focusing of a GAM with unperturbed Gaussian initial condition, which is a phenomenon closely related to the MI, and a long GAM simulation depicting breather behaviour (see \eg\Refrs\cite{Dudley09,Akhmediev86}) of the GAM MI.

\section{Nonlinear Schr\"{o}dinger equation model}
\label{sec:nlse}

\subsection{Introduction}\label{sec:nlse-introdtion}
The nonlinear Schr\"{o}dinger equation (NLSE) model \cite{PoP21} describes the dynamics of an isolated GAM packet via the complex wavefunction $\psi(r,t)$, where its real part
\begin{equation}
\Rea \left[ \psi(r,t) \right] \equiv E_r(r,t),
\end{equation}
represents the axisymmetric component of the GAM radial electric field $E_r(r,t)$. The function $\psi$ obeys the cubic nonlinear Schr\"{o}dinger equation given by
\begin{equation}
\imag\frac{\partial \psi}{\partial t}=\calF\psi-\frac{\partial}{\partial r}\left(\frac{\calG}{2}\frac{\partial\psi}{\partial r}\right)-\alpha_\text{NL} |\psi|^2\psi, 
\label{eq:nlse}
\end{equation}
where the first two terms on the right hand side characterize the linear GAM dispersion, while the last term introduces the contribution from nonlinear self-interactions. Their respective strengths are determined by the parameters $\calF$, $\calG$ and $\alpha_\text{NL}$, which in most sections of this study will be assumed to be real values and independent of the radial coordinate $r$, which is equivalent to assuming no damping and a uniform plasma background, respectively. Damping in the NLSE will be considered later in this paper in \secs\ref{sec:results-damping} and \ref{sec:results-damped-NLSE}, and thus is discussed separately there.

The values of the parameters $\calF$ and $\calG$ are obtained from the analytical gyrokinetic result for the linear GAM dispersion relation, which according to \Refr\cite{SmolyakovPPR16} is given by
\begin{align}
	\omega(k_r) &= \omega_0\sqrt{1 + \frac{1}{2} k_r^2 \rho_i^2 D(\tau_e)}, \label{eq:Dispersion-Relation}
\end{align}
and holds only when $k_r^2 \rho_i^2 \ll 1$. Here, $\omega_0$ is the dispersionless GAM frequency given by \eq\eqref{eq:omega0} below, $k_r$ is the radial wavevector of the GAM spectrum, $\rho_i$ is the ion Larmor radius and $D(\tau_e)$ is a coefficient characterizing the strength of the dispersive corrections, which depends on the electron-to-ion temperature ratio $\tau_e = T_e/T_i$
\begin{align}
D(\tau_e) &= \frac{3}{4} - \frac{\frac{13}{4} + 3\tau_e + \tau_e^2}{\frac{7}{4} + \tau_e} + \frac{\frac{747}{32} + \frac{481}{32}\tau_e + \frac{35}{8}\tau_e^2 + \frac{1}{2}\tau_e^3}{\left(\frac{7}{4} + \tau_e\right)^2}. \label{eq:Analytic-D}
\end{align}
For $\omega_0$ the expression derived in \Refr\cite{SugamaJPP06} is utilized and damping is neglected
\begin{align} \label{eq:omega0}
\omega_0^2 &= \left[ 1 + \frac{2(23 + 16\tau_e + 4\tau_e^2)}{q_s^2(7 +4\tau_e)^2}\right]\left(\frac{7}{4} + \tau_e\right) \frac{v^2_{Ti}}{R_0^2},
\end{align}
where $v_{Ti} = \sqrt{2T_i/m_i}$ is the thermal ion velocity, $R_0$ the major tokamak radius and $q_s$ the safety factor. In order to obtain expressions for the parameters $\calF$ and $\calG$, the square root in the dispersion relation, \eq\eqref{eq:Dispersion-Relation}, is expanded using the approximation $k_r^2 \rho_i^2 \ll 1$ that was already assumed to hold during the derivation of the dispersion relation 
\begin{equation}
    \omega \approx \omega_0 + \frac{1}{4} k_r^2 \rho_i^2 \omega_0 D =: \calF + \frac{1}{2}\calG k_r^2,\\
\end{equation}
\begin{equation}
    \calF := \omega_0,\\
\end{equation}
\begin{equation}
    \calG := \frac{1}{2}\omega_0 \rho_i^2 D. \label{eq:Definition-G}
\end{equation}

Here, a definition is denoted by ``$:=$''. For the strength $\alpha_\text{NL}$ of the nonlinear self-interaction term there currently exists no analytical expression. As a consequence in this study the values for $\alpha_\text{NL}$ are obtained through comparisons to gyrokinetic GAM simulations, which were generated with the particle-in-cell code ORB5 \cite{Bottino15, Lanti20}. The results of these comparisons are presented in \App\ref{app:alpha_NL} and show that the parameter is positive, $\alpha_\text{NL} > 0$, increases with $\tau_e$ and depends approximately inversely on the ion Larmor radius, $\alpha_\text{NL} \propto 1/\rho_i$. Furthermore, $\alpha_\text{NL}$ increases when approaching the center of the plasma crosssection $r=0$. No significant impact of the safety factor $q_s$ on $\alpha_\text{NL}$ was found.

\subsection{Dynamics} \label{sec:nlse-dynamics}
This section gives a short introduction to the terms in the NLSE, \eq\eqref{eq:nlse}, that will be relevant for the modulational instability. The first term on the \rhs, $\calF \psi$, is responsible for the coherent oscillation of the GAM at the dispersionless frequency $\calF = \omega_0$. The corresponding dynamics can be split off of the wavefunction $\psi$ through the following transformation
\begin{equation} \label{eq:NLSE-Transform}
    \psi(r,t) = \hat{\psi}(r,t) e^{-\imag \calF t},
\end{equation}
where $\hat{\psi}$ is the envelope of the GAM packet. This ansatz reduces the NLSE to the usually reported form
\begin{equation}
	\imag\prti{\hat{\psi}}{t}=-\frac{\calG}{2}\prti{^2\hat{\psi}}{r^2}-\alpha_\text{NL} |\hat{\psi}|^2\hat{\psi}.
	\label{eq:NLSE-Envelope}
\end{equation}

\begin{figure}[b!]
	\centering
	\def\svgwidth{.4\textwidth}
	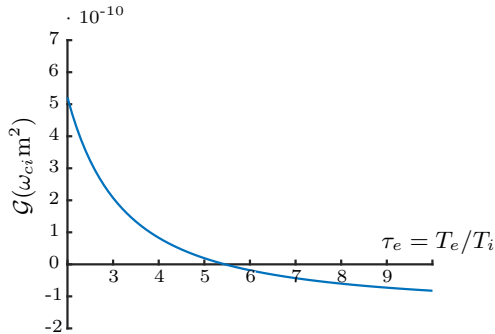
	\caption{Dependency of the dispersion coefficient $\calG$ on the electron-to-ion temperature ratio $\tau_e$ as defined in \eqs\eqref{eq:Analytic-D}-\eqref{eq:Definition-G}. The parameters were chosen as specified in \tab\ref{tab:parameters}, with ion cyclotron frequency $\omega_{ci} \approx 1.82\cdot10^8\,\frac{\text{rad}}{\text{s}}$ and ion Larmor radius $\rho_i/ a_\text{min} = 4.08\cdot10^{-4}$.}
	\label{fig:G-vs-taue}
\end{figure} 

The second term on the right-hand side of \eq\eqref{eq:nlse}, $-\calG/2\,\partial_r^2 \psi$ (for $\calG \neq \calG(r)$), gives the dispersive properties to the GAM dynamics. The nature of the dispersion is determined by the sign of the parameter $\calG$, which, as can be deduced from \eqs\eqref{eq:Analytic-D}-\eqref{eq:Definition-G}, depends solely on the value of the electron to ion temperature ratio $\tau_e$, with $\tau_e \approx 5.45$ marking the boundary between positive and negative values of $\calG$ \cite{Nguyen08}. The corresponding regimes are labelled as follows:
\begin{align}
&\text{Anomalous Dispersion} \hspace{-0.8cm}&& \hspace{-0.8cm}\calG > 0  \hspace{-0.8cm}&& \hspace{-0.8cm} \tau_e \lesssim 5.45, \nonumber\\
&\text{No Dispersion}  \hspace{-0.8cm}&& \hspace{-0.8cm} \calG = 0  \hspace{-0.8cm}&& \hspace{-0.8cm} \tau_e \approx 5.45, \nonumber\\
&\text{Normal Dispersion} \hspace{-0.8cm}&& \hspace{-0.8cm} \calG < 0  \hspace{-0.8cm}&& \hspace{-0.8cm}  \tau_e \gtrsim 5.45.\nonumber
\end{align}

\begin{figure}[h!]
	\centering
	\def\svgwidth{.78\textwidth}
	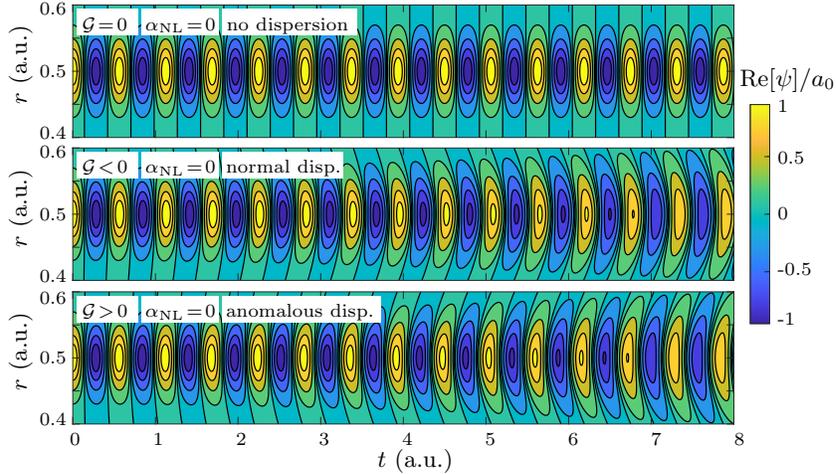
	\caption{NLSE simulations illustrating the isolated impact of the dispersive term on the dynamics. The figure shows the real part $\Rea[\psi]$ of the wavefunction (which for the GAM corresponds to the radial electric field $E_r$) normalized to the maximum amplitude $a_0$ of the Gaussian initial condition. The nonlinear term is disregarded ($\alpha_\text{NL} = 0$). The selected values of $\calF$ and $\calG$ are chosen such that their relative orders of magnitude match the GAM simulations considered in the later sections. It can be observed that the coefficient $\calF$ is responsible for the oscillation of $\Rea[\psi]$, while the dispersive term introduces a curvature in $(r,t)$-space as well as an increase of the Gaussian width as time progresses.}
	\label{fig:Linear-Dispersion}
\end{figure}

The full dependency of $\calG$ on $\tau_e$ is depicted in \fig\ref{fig:G-vs-taue}. The different kinds of dynamics corresponding to the three dispersion regimes are illustrated through NLSE simulations (without the nonlinear term) with an initial Gaussian profile in \fig\ref{fig:Linear-Dispersion}. The dispersive term broadens the width of packets as time progresses and alters the oscillation frequency of the Gaussian flanks compared to the maximum at $r = 0.5\,$(a.u.), resulting in a curvature of the phase front in $(r, t)$-space. In the case of normal dispersion, the flanks oscillate faster than the packet center, leading to convex curvature, and vice versa for anomalous dispersion.

\begin{figure}[h!]
	\centering
	\def\svgwidth{.78\textwidth}
	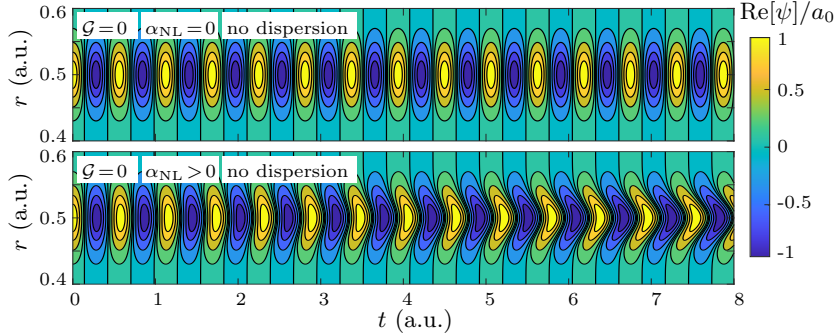
	\caption{NLSE simulations illustrating the nonlinear frequency shift (in the dispersionless regime, $\calG = 0$) for a Gaussian initial condition. Similarly to \fig\ref{fig:Linear-Dispersion} the real part $\Rea[\psi]$ of the wavefunction is illustrated, which corresponds to the GAM radial electric field $E_r$. Comparing the upper simulation (without nonlinearity, $\alpha_\text{NL} = 0$) to the lower one, it is evident to see that the frequency shift as described by \eq\eqref{eq:nl-phaseshift} is most pronounced at the center of the Gaussian at $r=0.5$, since there the amplitude reaches its highest value.}
	\label{fig:Nonlinear-Term}
\end{figure}

The nonlinear term introduces a shift that lowers the frequency (due to $\alpha_\text{NL} > 0$ for GAMs) proportionally to the packet amplitude squared at the location $r$, which for $\calG = 0$ amounts to 
\begin{equation}
\Delta \omega_\text{NL}(r,t) = - \alpha_\text{NL} |\psi|^2(r,t). \label{eq:nl-phaseshift}
\end{equation}
This shift is illustrated in \fig\ref{fig:Nonlinear-Term}. The interaction between the nonlinear phase shift and the dispersive term creates two distinct regimes called self-defocusing regime (when $\calG / \alpha_\text{NL} < 0$, \ie for GAMs for normal dispersion, \ie $\tau_e \gtrsim 5.45$) and self-focusing regime (for $\calG / \alpha_\text{NL} > 0$, \ie vice-versa) \cite{scott2006encyclopedia}. Since the self-focusing regime is deeply connected with the formation of MI, it will be explained in further detail in the next section.

\FloatBarrier

\section{Modulational Instability} \label{sec:mi}
We recall in this section some known results concerning the Modulational Instability (MI). Although the material reported here can be found in textbooks (\eg \Refr\cite{Agrawal}), we present it to put the results of the next sections into context. The MI, also called Benjamin-Feir instability \cite{benjamin67}, is believed to be one of the most ubiquitous instabilities in nature \cite{Zakharov09}. It appears not only in the NLSE, but also in other equations describing nonlinear dispersive waves, \eg in the Withham equation and the Korteweg-de Vries equation. The instability only develops when the nonlinear and dispersive contributions to the oscillation dynamics interact such that the NLSE is self-focusing. As the name suggests, the NLSE dynamics creates a self-focusing effect of maxima in the wave packet, which is explained in further detail in \App\ref{app:qualitative-mi}.

\subsection{Introduction and Behaviour} \label{sec:mi-intro}
\begin{figure}[h!]
	\centering
	\def\svgwidth{.78\textwidth}
	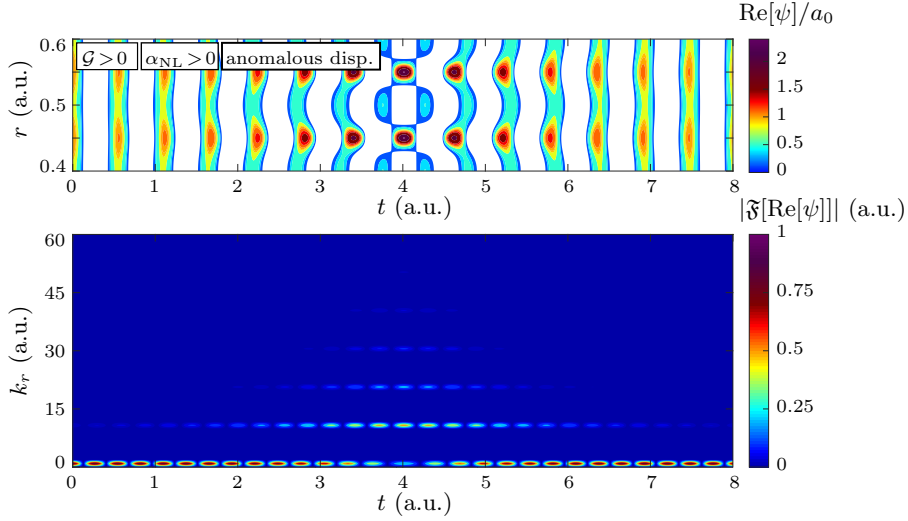
	\caption{Single cycle of an Akhmediev breather where the initial perturbation wavevector was chosen to be $k_r = 10\,$(a.u.). The upper figure illustrates $\Rea[\psi]/a_0$, which for the GAM corresponds to the radial electric field $E_r$. As indicated by the color bar only the positive values are drawn in the figure to emphasize the decoherent phase front at $t=4\,$(a.u.). The bottom figure shows the evolution of the corresponding radial spectrum (\ie the absolute value of the Fourier transform $|\mathfrak{F}\left[\Rea\!\left[\psi\right]\right]|$). The perturbation grows exponentially until $t \approx 2.5$, after which the growth slows down and at $t\approx 4$ the perturbation reaches its maximum value (which can be seen in the spectrum as well as in real space).}
	\label{fig:MI-depiction}
\end{figure}

The MI is usually analyzed for the case of a plane wave $A_0(t)$ with amplitude $a_0$ (which can be considered to be a packet with very large width, $k_r\!\rightarrow\!0$) superimposed with a radially periodic perturbation $A_1$ with a wavevector $k_\text{pert}$
\begin{align}
\psi(r, t) &= [A_0(t) + A_1(r,t)\hspace{.05em}] e^{- \imag \omega_0 t} ,  \label{eq:MI-initial-condition}\\
A_1(r,t=0) &= a_1 \cos(k_\text{pert} r),\quad a_1 \ll a_0.
\end{align}
When the perturbation wavevector $k_\text{pert}$ lies in the unstable range, which will be specified in \sec\ref{sec:mi-conditions}, due to the previously mentioned self-focusing effect the sinusoidal perturbation will grow exponentially (as long as the perturbation amplitude $a_1$ is small compared to the plane wave amplitude $a_0$) at the expense of the plane wave ($k_r = 0$) component of the wave, which in the following will be called the wave background. More details on the growth process are given in \App\ref{app:qualitative-mi}. When the perturbation has grown so large that it dominates the shape of the wave $\psi$ it saturates and acquires a large nonlinear phase shift (as mentioned in \sec\ref{sec:nlse-dynamics}) compared to the background, leading to strongly incoherent phase fronts. In this saturation phase the radial spectrum of the wave contains many high-$k_r$ components. When the nonlinear phase shift is large enough that the perturbation has skipped an entire oscillation compared to the background, the wave front reconnects and the perturbation decreases again. Finally, the initial condition is restored and the MI process can start anew, leading to a cyclic behaviour called Akhmediev Breathers \cite{Akhmediev86}. A single Akhmediev Breather cycle together with the corresponding radial spectrum is shown in \fig\ref{fig:MI-depiction}.

\subsection{Conditions for instability} \label{sec:mi-conditions}
\begin{figure}[h!]
	\centering
	\def\svgwidth{.45\textwidth}
	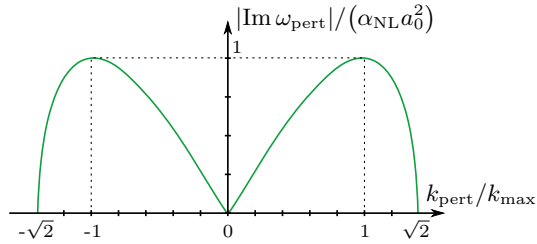
	\caption{Dependency of the MI growth rate $\gamma_\text{MI} := |\Imm\,\omega_\text{pert}|$ on the perturbation wave\-number $k_\text{pert}$, as given by \eq\eqref{eq:mi-growthrate-general}. It is assumed that the first condition, given by \eq\eqref{eq:MI-Condition-1}, is satisfied, \ie dispersion is anomalous.}
	\label{fig:MI-growthcurve}
\end{figure}

The conditions for instability will be briefly summarised in the following (see \eg\Refr\cite{Remoissenet96}). By assuming that the time evolution of the perturbation $A_1(r,t)$ will be of the form
\begin{equation}
    A_1(r,t) = \frac{a_1}{2} e^{\imag (k_\text{pert} r - \omega_\text{pert} t)} + \frac{a_1^*}{2} e^{-\imag (k_\text{pert} r - \omega_\text{pert} t)},
\end{equation}
one finds that $\omega_\text{pert}$ fulfills the dispersion relation
\begin{equation}
    \omega_\text{pert}^2 = \left(k_\text{pert}^2 - 4 \frac{\alpha_\text{NL}}{\calG}a_0^2\right) \frac{\calG^2 k_\text{pert}^2}{4}. \label{eq:MI-disp-rel}
\end{equation}
It is immediate to see that exponential growth occurs (i.e. $\omega_\text{pert}$ is imaginary) when firstly
\begin{equation}
    \hspace{2cm} \frac{\alpha_\text{NL}}{\calG} > 0,  \quad (\Leftrightarrow\text{NLSE is self-focusing)} \label{eq:MI-Condition-1}
\end{equation} 
which due to $\alpha_\text{NL} > 0$ for GAMs is equivalent to requiring anomalous dispersion ($\calG > 0$, $\tau_e \lesssim 5.45$ as described in \sec\ref{sec:nlse-dynamics}), and secondly when the perturbation wavevector is within the range 
\begin{equation}
    |k_\text{pert}| < |k_\text{lim}| = 2 a_0 \sqrt{\frac{\alpha_\text{NL}}{\calG}} = \sqrt{2}k_\text{max}, \label{eq:MI-Condition-2}
\end{equation}
where $k_\text{lim}$ marks the boundary between stable and unstable perturbation wavevectors. The corresponding growth rate, which will be labelled $\gamma_\text{MI} := |\Imm\, \omega_\text{pert}|$ in the following, reaches its maximum value at the wavevector $|k_\text{max}| := a_0 \sqrt{2 \alpha_\text{NL}/\calG}$. From \eq\eqref{eq:MI-disp-rel} one finds that
\begin{equation} \label{eq:mi-growthrate-general}
    \gamma_\text{MI}(k_\text{pert}) = \Bigg|\Imm\left[\frac{\calG k_\text{pert}}{2} \sqrt{ k_\text{pert}^2 - 4 \frac{\alpha_\text{NL}}{\calG} a_0^2}  \,\right]\Bigg|,
\end{equation}
\begin{equation} \label{eq:max-growthrate}
    \gamma_\text{MI}(k_\text{pert}\!=\!k_\text{max}) = \alpha_\text{NL} a_0^2.
\end{equation}
The full dependency of $\gamma_\text{MI}$ on the wavevector of the initial perturbation $k_\text{pert}$ is illustrated in \fig\ref{fig:MI-growthcurve}.

\section{Numerical approach}
\label{sec:numerics}
This section introduces the numerical tools employed in this paper to simulate the GAM gyrokinetically and with the NLSE model. As mentioned in the introduction, \sec\ref{sec:intro}, the gyrokinetic results, which are obtained using the global particle-in-cell code ORB5 \cite{Bottino15, Lanti20}, are assumed to provide an accurate physical description of the GAM dynamics and are thus an adequate reference solution to validate the NLSE results and show shortcomings of this simplified model.

\subsection{Split-step NLSE solver} \label{sec:Numerics-SSS}
The split-step solver (SSS) code was written by G. P. Agrawal \cite{Agrawal} in MatLab and was created to solve the NLSE \eqref{eq:NLSE-Envelope}, which describes the dynamics of the envelope, in the context of optical fibres where it is reported. The code was modified to incorporate the originally missing oscillation term by utilizing the transformation introduced in \eq\eqref{eq:NLSE-Transform} and adapted to make its input and output consistent with ORB5 simulations. 

The split-step method states that, if the chosen numerical time step $\Delta t$ is small enough, the nonlinear and dispersive contributions to the NLSE dynamics (of the wave envelope $\hat{\psi}$) act (mostly) independently from each other \cite{Agrawal,Faou12}. It follows that the solution can be approximated by alternatingly solving the two equations 
\begin{align}
	\hspace{1cm} \imag \frac{\partial \hat{\psi}}{\partial t} &= - \alpha_\text{NL} |\hat{\psi}|^2 \hat{\psi} \label{eq:Split-Step-NL},\hspace{-.5cm} && \hspace{-.5cm} \text{(isolated nonlinear dynamics)}\hspace{1cm} \\
	\hspace{1cm} \imag \frac{\partial \hat{\psi}}{\partial t} &= - \frac{\calG}{2} \frac{\partial^2 \hat{\psi}}{\partial r^2}, \hspace{-.5cm}&&\hspace{-.5cm}\text{(isolated dispersive dynamics)}\hspace{1cm} \label{eq:Split-Step-D}
\end{align}
for each time step. This approach has the significant advantage that the equations for the isolated nonlinear and isolated dispersive dynamics can each be immediately solved from an initial condition $\hat{\psi}(r,t=0)=\hat{\psi}_0$  using the following analytic solutions  
\begin{align}
	\hat{\psi}_\text{NL}(r,t) &= \exp\left(\imag \alpha_\text{NL}|\hat{\psi}_0|^2 t\right)\hat{\psi}_0 =: \varphi_\text{NL}^t[\hat{\psi}_0], \\ 
	 \hat{\psi}_\text{D}(r, t) &= \mathfrak{F}^{-1}\left\lbrace \exp\left(- \imag t \, \frac{\calG}{2} k^2\right) \mathfrak{F}\lbrace\hat{\psi}_0 \rbrace \right\rbrace =: \varphi_\text{D}^t [\hat{\psi}_0], \label{eq:NLSE-solver-dispersive}
\end{align}
where $\mathfrak{F}\left\lbrace.\right\rbrace $ is the Fourier transform which is calculated using the Fast-Fourier-Transform algorithm, and $\varphi_\text{NL}^t\left[.\right]$ and $\varphi_\text{D}^t\left[.\right]$ denote the exact flows (\ie the temporal evolution starting from an initial condition $\hat\psi_0(r)$) for the isolated nonlinear and isolated dispersive equation, respectively.

The flow $\varphi_\text{NLSE}^t$ associated with the complete dynamics of the NLSE is then approximated as follows 
\begin{equation} \label{eq:split-step-Fourier-step}
	 \hat{\psi}(r, t=n\Delta t) = \varphi_\text{NLSE}^{n \Delta t} [\hat{\psi}_0] \approx (\varphi_\text{D}^{\Delta t} \circ \varphi_\text{NL}^{\Delta t})^n [\hat{\psi}_0], 
\end{equation}
also known as the Lie splitting method, where ``$\circ$'' denotes the concatenation of flows, $(\varphi_\text{D}^{\Delta t} \circ \varphi_\text{NL}^{\Delta t}) [\hat{\psi}_0] = \varphi_\text{D}^{\Delta t} [\varphi_\text{NL}^{\Delta t} [\hat{\psi}_0]]$.

The SSS implements a refinement of the split-step method with higher accuracy, called the symmetrised split-step Fourier method or Strang splitting scheme. In this form, the nonlinear dynamics is split into two parts of duration $\Delta t/2$, with the dispersive (and thus smoothing) evolution acting in between \cite{Agrawal, Faou12}.

\subsection{ORB5} \label{sec:Numerics-ORB5}
The gyrokinetic code ORB5 \cite{Bottino15,Lanti20} is a global particle-in-cell (PIC) code that simulates the plasma dynamics inside a tokamak for processes occurring on a time scale slower than that of the ion gyromotion. It can be executed with or without nonlinear plasma interactions.

The gyrokinetic theory reduces the 6D kinetic theory by one dimension by averaging over the gyromotion to obtain a 5D problem describing the dynamics of the guiding centre distribution function $f_s$ of each species $s$. The set of gyrokinetic equations describing the dynamics of $f_s$ can be constructed in different ways. The model implemented in ORB5 is derived from a gyrokinetic Lagrangian describing the particle motion in a magnetic field \cite{BrizardHahm}. The time-symmetric Hamiltonian within the Lagrangian conserves energy automatically, leading to a model which is particularly useful for numerical simulations \cite{Bottino15}. ORB5 provides numeric results that exhibit strong agreement with other gyrokinetic codes \cite{Lanti20}. Its recent application to the dynamics of isolated GAMs is documented in \Refrs\cite{PalermoEPL,Biancalani16, Palermo17,Palermo20,PoP20,Palermo23}. The interested reader is referred to these references for more details about the numerical model. 

In order to create GAMs in the ORB5 simulations, an electric field $E_r$ is initialised in ORB5 via a perturbation $\delta n$ of the ion density $n_0$, which is related to the electric field as
\begin{equation}
    \delta n_i(r) \propto \frac{1}{r} \prti{}{r}\left(r E_r(r)\right). \label{eq:relation-E-ion-density}
\end{equation}
As a result, the GAM is essentially ``dropped into'' the simulations and the formation process is not considered.

\section{Simulation results} \label{sec:results}
\begin{figure}[h!]
    \centering
    \def\svgwidth{.45\textwidth}
	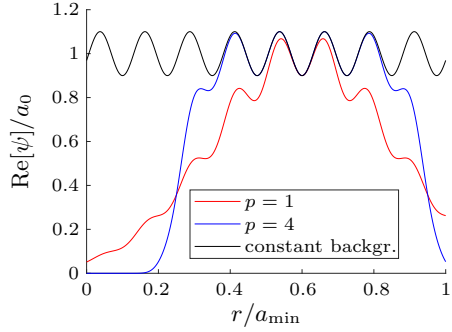    
    \caption{Initial condition of the GAM radial electric field $E_r = \Rea[\psi]$ according to \eq\eqref{eq:Init-cond-plateau} for two values of $p$, compared to the usual constant background initial condition of MI described in \sec\ref{sec:mi-intro}. Unless stated otherwise, $p=4$ will be used for the packet steepness, $\w_0 = 0.35\,a_\text{min}$ for the width and the center will be placed at $r_0 = 0.6\,a_\text{min}$ in subsequent simulations. The perturbation wavevector and amplitude in this example are $k_\text{pert} = 8\,\frac{2\pi}{a_\text{min}}$ and $a_1 = 0.1$, respectively.}
    \label{fig:initial-condition}
\end{figure}

In the following simulations the (axisymmetric) GAM radial electric field is initialised in the following form
\begin{equation}
    E_r(r,t=0) = a_0 \exp\left(-\left[\frac{r-r_0}{\w_0}\right]^{2p}\right) \left(1 + a_1 \cos(k_\text{pert}[r-r_0])\right), \label{eq:Init-cond-plateau}
\end{equation}
which is similar to the initial condition discussed in \eq\eqref{eq:MI-initial-condition} in \sec\ref{sec:mi}, but uses a plateau function instead of a constant background in order to avoid numerical artifacts at the boundaries of the simulation domain, \ie $r = 0$ and $r = a_\text{min}$. This initial condition is depicted for different parameters in \fig\ref{fig:initial-condition}. The exponent $p$ determines the steepness of the plateau edge, where $p = 1$ corresponds to a Gaussian profile and $p = 4$ was chosen in the following simulations. 

The parameters $a_0$ (which is related to the density perturbation amplitude $\delta n/n_0$ as discussed in \App\ref{app:alpha_NL}), $r_0, \w_0, a_1$ for the initial condition and $a_\text{min}, R_0, \tau_e, B_0, q_s, m_i$ for the simulation conditions are chosen as specified in \tab\ref{tab:parameters}, with the ion Larmor radius $\rho_i$ and the perturbation wavevector $k_\text{pert}$ assuming different values in the following sections. Through the choice of $\tau_e = T_e/T_i = 3 < 5.45$ the first MI condition, \eq\eqref{eq:MI-Condition-1}, \ie $\calG>0$ (anomalous dispersion, self-focusing NLSE) is satisfied. The geometry of the tokamak with an aspect ratio of $R_0/a_\text{min} = 10$ has a high cylindricity, which more closely resembles the NLSE model where the geometry is not considered in the equations. 

\begin{table}[h!]
    \caption{Parameters used in the GK (ORB5) and NLSE simulations.}
    \centering
    \begin{tabular}{cccc}
        \toprule
        \multicolumn{2}{c}{Initial Condition} & \multicolumn{2}{c}{Tokamak plasma} \\
        \midrule
        Parameter & Value & Parameter & Value \\
        \midrule 
        $a_0$ & 2 -- 3.4$\,\cdot\,$10$^{\text{-4}}$\,(a.u.) & 0.13\,m & $\w_0$ \\
        $\delta n / n_0$ & 2 -- 3.4$\,\cdot\,$10$^{\text{-4}}$ & $R_0$ & 1.3\,m \\
        $r_0$  & 0.6\,$a_\text{min}$ & $B_0$ & 1.9\,T \\
        $\w_0$ & 0.35\,$a_\text{min}$ & $\tau_e$ & 3  \\
        $p$    & 4 & $m_i$ & 1\,a.m.u. \\
        $a_1$  & 0.1 & $q_s$ & 15 \\
        $k_\text{pert}$ & 4 -- 14\,$\frac{2\pi}{a_\text{min}}$ & $\rho_s/a_\text{min}$ & $\frac{2}{425}$ -- $\frac{2}{325}$ \\
        & & $\rho_i/a_\text{min}$ & 3.842 -- 5.025\,$\cdot10^{-3}$ \\
        \bottomrule
    \end{tabular}
    \label{tab:parameters}
\end{table}

GAM damping is known to decrease with rising safety factor $q_s$ \cite{PalermoEPL}. In order to fulfill the assumption of weak damping (compared to the MI growth rate $\gamma_\text{MI}$, \eq\eqref{eq:max-growthrate}) that was posed in \sec\ref{sec:intro}, $q_s = 15$ is chosen. In this regime the damping term derived by Qiu \etal in \Refr\cite{Qiu09} is applicable as the assumptions of $1/q_s^2 \ll k_r^2 \rho_i^2 \ll 1$ and $\frac{1}{2} \tau_e k_r^2 \rho_i^2 \ll 1$ are satisfied for all of the chosen perturbation wavevectors $k_\text{pert}$ (note that for $\rho_i/ a_\text{min} = 5.025\cdot10^{-3}$ and $k_r = 10\cdot2\pi/a_\text{min}$ the value of $\frac{1}{2}\tau_e k_r^2 \rho_i^2 \approx 0.1$ and for $k_r = 14\cdot2\pi/a_\text{min}$, $\frac{1}{2}\tau_e k_r^2 \rho_i^2 \approx 0.3$). We remark that the damping rate derived by Sugama and Watanabe, see \Refr\cite{SugamaJPP06}, is not applicable due to the approximation $k_r^2 \rho_i^2 \ll 1/q_s^2$ employed there, which is not fulfilled for any of the chosen perturbation wavevectors $k_\text{pert}$. One obtains from \Refr\cite{Qiu09} that the unmodulated packet and wavevectors $k_r < 5\,\frac{2\pi}{a_\text{min}}$ are undamped (see \fig\ref{fig:Damping-term}). However, damping rates of larger wavevectors, \eg $\gamma_\text{Qiu}(k_r = 10\,\frac{2\pi}{a_\text{min}}) = -6.07\cdot10^{-6}\,\omega_{ci}$ are in similar orders of magnitude as the MI growth rate $\gamma_\text{MI}(k_r = 10\,\frac{2\pi}{a_\text{min}}) = 3.11\cdot10^{-5}\,\omega_\text{ci}$, \ie $|\gamma_\text{Qiu}| \approx \frac{1}{5}|\gamma_\text{MI}|$, which leads to the conclusion that damping may slow the growth of MI down but it is not expected to suppress MI.

\FloatBarrier

\subsection{General comparison between NLSE and GK simulations} \label{sec:results-general}
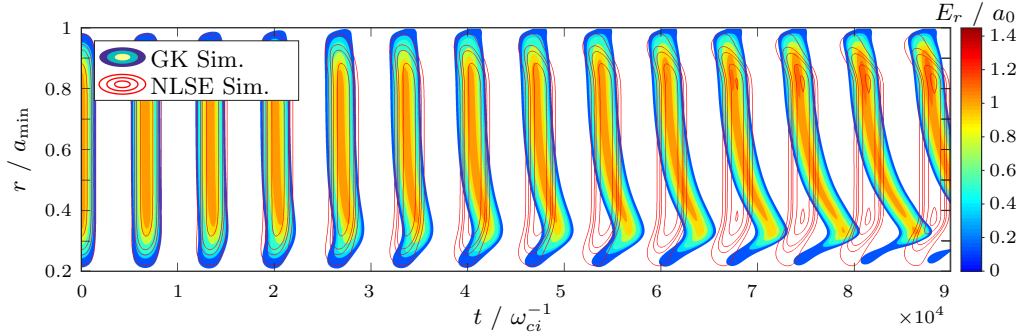
\begin{figure}[h!]
    \centering
    \def\svgwidth{\textwidth}
	\import{./images-simulations/first-comparison}{196-k0-comparison.pdf_tex}
    \caption{GAM radial electric field $E_r$ in an ORB5 gyrokinetic (GK, color contours) and NLSE (red levels) simulation of an initially unmodulated GAM. While the frequency of the oscillation matches well at the packet center and further outward (for $r \geq 0.6\,a_\text{min}$), differences increase when moving to smaller values of $r$. Since the plasma parameters were chosen to be constant across the radial coordinate $r$, these discrepancies can be attributed to an influence of the tokamak geometry on the nonlinear parameter $\alpha_\text{NL}$, which is analyzed in \App\ref{app:alpha_NL}.}
    \label{fig:Comparison-k-0}
\end{figure}

A first comparison between NLSE and gyrokinetic (GK) simulations of the general GAM dynamics (without modulation of the initial condition and thus without MI) was made using the parameters in \tab\ref{tab:parameters}. Specifically, the initial GAM electric field amplitude $a_0 = 2.5\cdot10^{-4}\,$(a.u.), sound Larmor radius $\rho_s/a_\text{min} = 2/375$ and ion Larmor radius $\rho_i/a_\text{min} \approx 4.355\cdot10^{-3}$ (where $\rho_i$ and $\rho_s$ are determined as described in \eqs\eqref{eq:Lx-rhos} and \eqref{eq:rhos-rhoi}) were chosen as a starting point. The dispersion coefficient $\calG$ is obtained according to \eq\eqref{eq:Definition-G} and $\alpha_\text{NL}$ is chosen as described in the \App\ref{app:alpha_NL} (with $r\!=r_0\!=\!0.6\,a_\text{tok}$). Figure \ref{fig:Comparison-k-0} presents the GK and NLSE simulation results.

The comparison in \fig\ref{fig:Comparison-k-0} shows that despite the uniform plasma background and thus (according to the equations introduced in \sec\ref{sec:nlse-introdtion}) constant values of $\calF$ and $\calG$, the frequency of the nonlinear GAM is radially varying in the GK simulations. This major discrepancy between the NLSE and GK simulation results may be explained by a radial dependence of $\alpha_\text{NL}$ that stems from the geometry of the tokamak. This is not included in the NLSE used in this paper since, as mentioned in \sec\ref{sec:nlse-introdtion}, the NLSE model for GAMs has not yet been derived from analytic theory. The dependency is analyzed numerically in more detail in the \App\ref{app:alpha_NL} and will be considered in future work.

Next, an initial condition with a modulated envelope is considered. From the parameters $a_0 = 2.5\cdot10^{-4}\,$(a.u.), $\alpha_\text{NL}$ and $\calG$ one obtains the range of unstable perturbation wavevectors $k_\text{pert}$ through \eq\eqref{eq:MI-Condition-2} as
\begin{equation} \label{eq:results-MI-condition}
    |k_\text{pert}| < |k_\text{lim}| \approx 14.6 \,\frac{2\pi}{a_\text{min}} = \sqrt{2}\cdot k_\text{max} \approx \sqrt{2}\cdot 10.3\,\frac{2\pi}{a_\text{min}} .
\end{equation}
It follows that the wave is unstable to MI when the wavelength of the perturbation $\lambda_\text{pert} = 2\pi/k_\text{pert}$ is larger than approximately 1/15$^\text{th}$ of the minor tokamak radius, and the maximum growth rate from \eq\eqref{eq:max-growthrate} of $\gamma_\text{MI} = 3.05\cdot10^{-5}\,\omega_{ci}$ is achieved when the perturbation wavelength is approximately $a_\text{min}/10$.

\begin{figure}[h!]
    \centering
    \begin{subfigure}{\textwidth}
        \centering
        \def\svgwidth{\textwidth}
        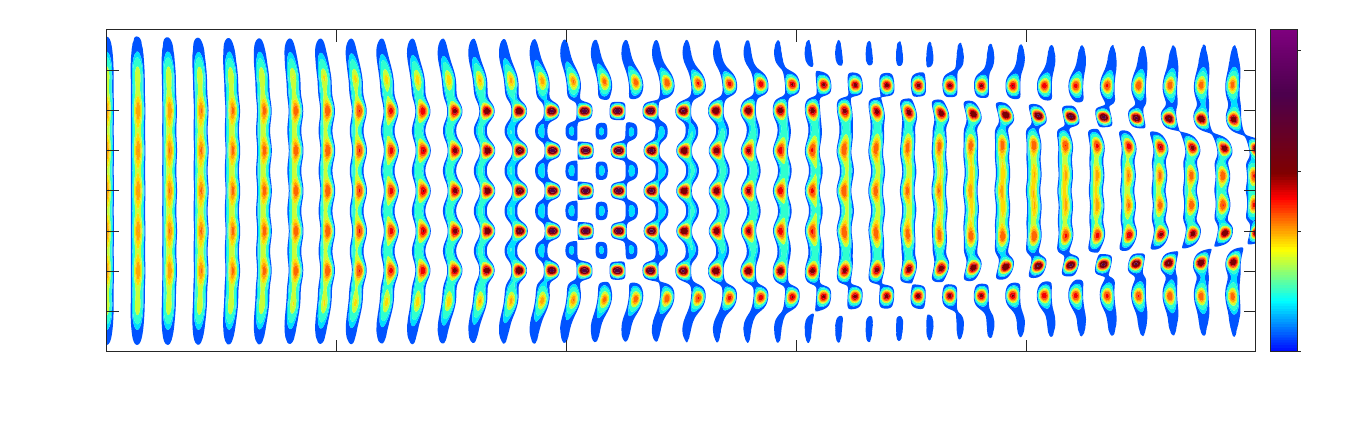
        \caption{NLSE simulation result}    
        \label{fig:comparison-mi-k-10-NLSE}
    \end{subfigure}
    \begin{subfigure}{\textwidth}
        \centering
        \def\svgwidth{\textwidth}
        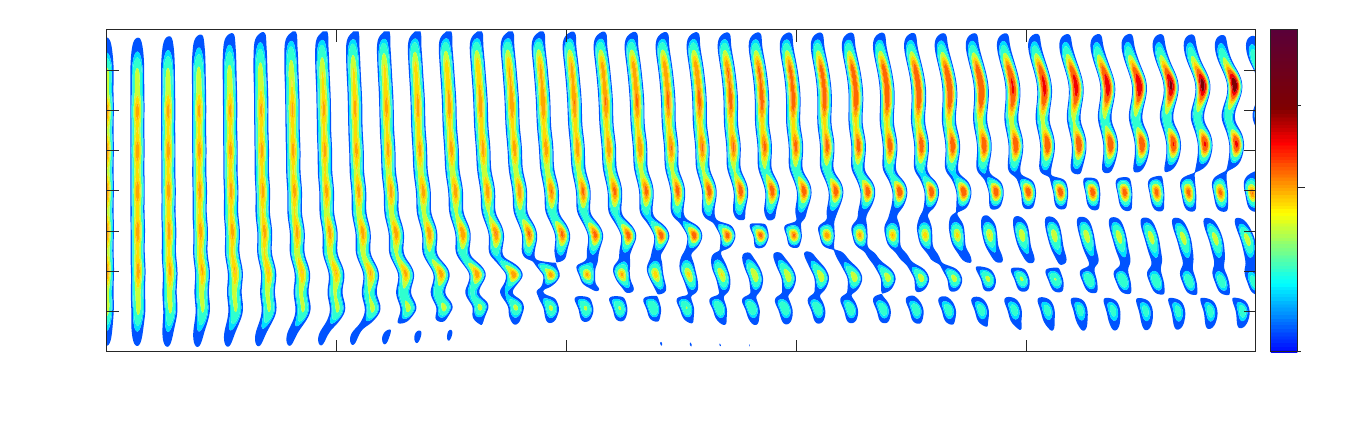
        \caption{Gyrokinetic simulation result}
        \label{fig:comparison-mi-k-10-ORB5}
    \end{subfigure}
\caption{Comparison of a NLSE and a GK simulation where the envelopes are modulated sinusoidally with the perturbation wavevector $k_\text{pert} = 10\,\frac{2\pi}{a_\text{min}} \approx k_\text{max}$. The figures depict the GAM radial electric field $E_r$, which in the NLSE model is the real part of the wavefunction $\Rea[\psi] = E_r$. While the growth of the modulation is observable in both simulations, the growth rate appears to be significantly lower in the GK result compared to the NLSE simulation. One can find further discrepancies, \eg the two individual maxima that form in the NLSE simulation at $r=0.8\,a_\text{min}$ and $r=0.9\,a_\text{min}$ are seemingly merged together in the GK simulation.}
    \label{fig:comparison-mi-k-10}
\end{figure}

A comparison of an NLSE and GK simulation with $k_\text{pert} = 10\,\frac{2\pi}{a_\text{min}} \approx k_\text{max}$ and $a_1 = 0.1$ is shown in \fig\ref{fig:comparison-mi-k-10}. While again significant differences between the NLSE and GK simulations can be observed, the results illustrate that the MI does occur in GK GAM simulations. This is apparent since the maxima of the initial condition start to grow as time progresses and the phase-skipping of the maxima (as described in \sec\ref{sec:mi-intro}) compared to the background is observed \eg for $t\approx 1.5 \cdot 10^5/\omega_{ci}$ at $r = 0.5\,a_\text{min}$ and for $t\approx 2.2 \cdot 10^5/\omega_{ci}$ at $r = 0.6\,a_\text{min}$ in \fig\ref{fig:comparison-mi-k-10-ORB5}. The observed radial difference in growth rate can be explained by the radial dependency of the nonlinear parameter $\alpha_\text{NL}$, as from \eq\eqref{eq:MI-disp-rel} it follows that $\gamma_\text{MI}$ increases monotonously with $\alpha_\text{NL}$, which gets larger for smaller values of $r$ (as determined in the \App\ref{app:alpha_NL}). Another discrepancy is the merging of two maxima at $r=0.8\,a_\text{min}$ and $r=0.9\,a_\text{min}$ in the GK simulation, which stay separated for the NLSE. This is again explained by $\alpha_\text{NL}$ decreasing with $r$, since according to \eq\eqref{eq:MI-Condition-2} the wavevector with the highest growth rate depends as $k_\text{max} \propto \sqrt{\alpha_\text{NL}}$. As a consequence at $r\approx 0.85\,a_\text{min}$ smaller wavevectors (\ie larger structures) will grow faster, thus favoring the merging of structures. Generally, one finds that across the whole radial space the growth rate and maximum amplitude in the GK simulation are significantly smaller compared to the NLSE simulation and theoretic predictions, as can be noticed comparing the color bars in \fig\ref{fig:comparison-mi-k-10}. This observation is further discussed in \secs\ref{sec:results-kpert} and \ref{sec:results-rhoi}.

\begin{figure}[h!]
    \centering
    \begin{subfigure}{.46\textwidth}
        \centering
        \hspace*{-0.12\textwidth}
        \def\svgwidth{\textwidth}
        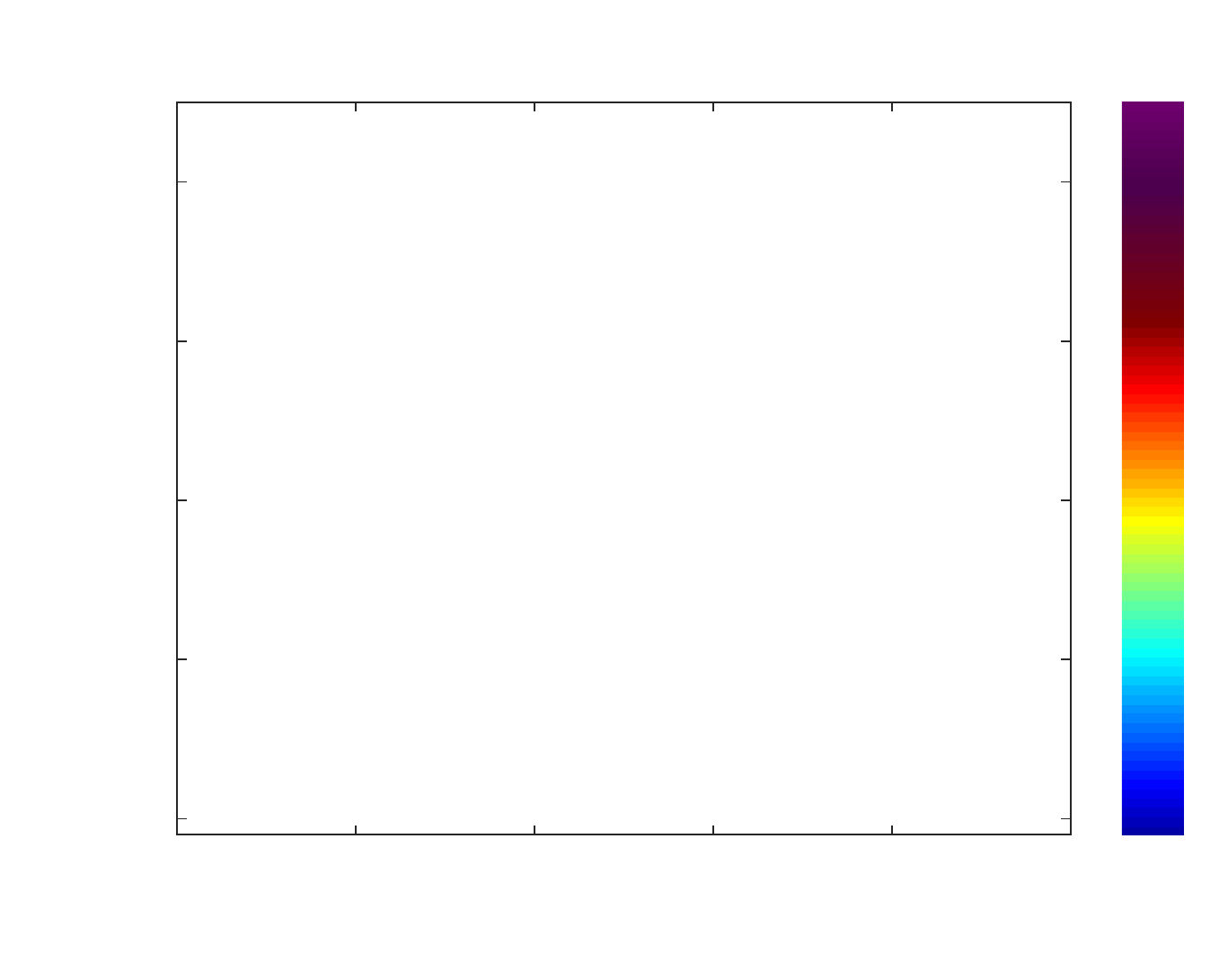
        \caption{NLSE radial spectrum} 
     \label{fig:spectrum-196-k-10-NLSE}
   \end{subfigure}
   \hspace{0.01\textwidth}
    \begin{subfigure}{.46\textwidth}
        \centering
        \def\svgwidth{\textwidth}
        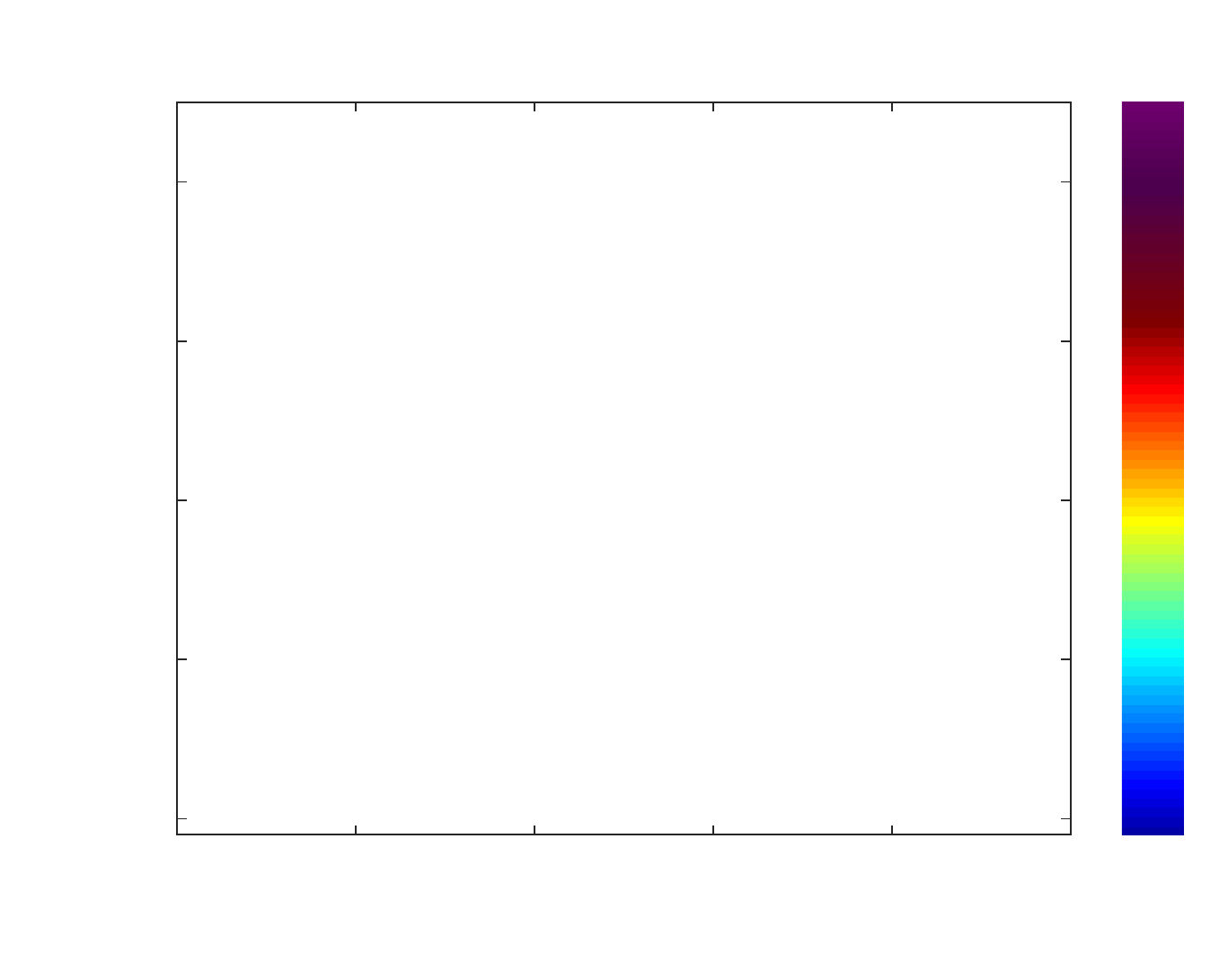
        \caption{Gyrokinetic radial spectrum} 
    \label{fig:spectrum-196-k-10-ORB5}
    \end{subfigure}
\caption{Comparison of the radial spectrum of the GK and NLSE simulations reported in \fig\ref{fig:comparison-mi-k-10}. The figures show the absolute value of the the radial Fourier transform of the GAM radial electric field, $|\mathfrak{F}[E_r]|$. It is apparent that the GK spectrum is in general more restrained to small wavevectors compared to the NLSE spectrum. This is especially noticeable in the saturation phase at $t\approx 1.1  \cdot 10^5/\omega_{ci}$, which as described in \sec\ref{sec:mi} is (according to the NLSE predictions) associated with a spectrum that contains high wavevector components.}
    \label{fig:spectrum-196-k-10}
\end{figure}

To study the aforementioned differences in further detail the radial spectra of the results are compared, as depicted in \fig\ref{fig:spectrum-196-k-10}. The  spectrum of the NLSE simulation contains much larger wavevectors $k_r$ than the GK simulation spectrum, most notably in the nonlinear saturation phase at $t\approx 1.1\cdot10^5/\omega_{ci}$ (see \fig\ref{fig:comparison-mi-k-10-NLSE}). Together with the lower amplitude and growth rate these findings indicate that a damping mechanism acting preferentially on higher wavevectors is present in the GK simulations, which is not contained in the NLSE solver. As a side note it is remarked that the NLSE spectrum extends up to wavevectors larger than $30\,\frac{2\pi}{a_\text{min}}$, where the assumption $k_r^2 \rho_i^2 \ll 1$ posed in \sec\ref{sec:nlse-introdtion} is marginally or no longer satisfied, as \eg $k_r^2\rho_i^2 = 0.2$ corresponds to $k_r \approx 16\,\frac{2\pi}{a_\text{min}}$ for the current ion Larmor radius $\rho_i/a_\text{min} \approx 4.355\cdot10^{-3}$. One can also notice in \fig\ref{fig:spectrum-196-k-10} that higher spectral components develop later in ORB5 simulations as compared to the NLSE solution. This is due to the fact that the instability develops on a slower scale in GK simulations, in particular at larger $r$.

\FloatBarrier

\subsection{Damping term} \label{sec:results-damping}
\begin{figure}[h!]
    \centering
    \def\svgwidth{.45\textwidth}
	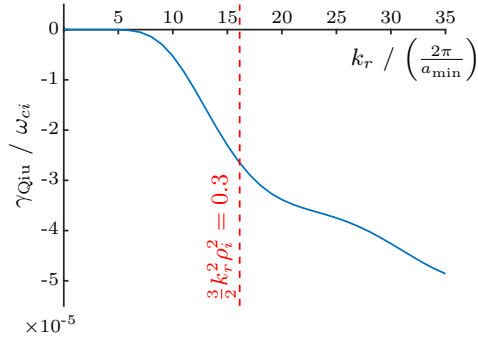
    \caption{Damping term for the parameters given in \tab\ref{tab:parameters}, with $\rho_s/a_\text{min} = 2/375$, $\rho_i/a_\text{min} = 4.355\cdot10^{-3}$. The red line illustrates the point $\frac{3}{2} k_r^2 \rho_i^2 = 0.3$ beyond which the expression may not be applicable anymore. It can be observed that the slope of the damping expression changes roughly where the line is located, which can be an indicator that after this point the behaviour is unphysical. It is remarked that the high end of the perturbation wavevectors unstable to MI, \ie for $k_\text{pert} = 14\,\frac{2\pi}{a_\text{min}}$, is very close to the damping applicability limit at $k_r \approx 16\,\frac{2\pi}{a_\text{min}}$.}
    \label{fig:Damping-term}
\end{figure}

This section gives a short introduction to the damping term derived in \Refr\cite{Qiu09}, which as mentioned in \sec\ref{sec:results} can be applied to the radial GAM spectrum when electrons are considered to be adiabatic, $1/q_s^2 \ll k_r^2\rho_i^2 \ll 1$ and $\frac{1}{2}\tau_e k_r^2\rho_i^2 \ll 1$. The condition $1/q_s^2 \ll k_r^2\rho_i^2$ is always satisfied for the chosen safety factor $q_s = 15$ and the wavevectors $k_\text{pert}$ of the initial perturbation. Furthermore, due to $\tau_e = 3$ the remaining requirements can be merged as follows
\begin{equation}
    k_r^2\rho_i^2 < \frac{\tau_e}{2} k_r^2\rho_i^2 = \frac{3}{2} k_r^2\rho_i^2 \ll 1.
\end{equation}
Taking $0.3\ll 1$ as the boundary value for the validity of this assumption, one finds that for the largest ion Larmor radius used in this paper, $\rho_i/ a_\text{min} = 5.025\cdot10^{-3}$, the applicability of the damping term is limited to wavevectors $k_r \lesssim 14.2\,\frac{2\pi}{a_\text{min}}$. The term is given by the following expression
\begin{align}\nonumber
    \gamma_\text{Qiu} = &- \frac{|\omega_b|}{\sqrt{2}b} \exp\left\lbrace -\sigma\frac{\omega_b}{\omega_\text{dt}} \right\rbrace \left[1 + b \frac{v_{Ti}^2}{\omega_b^2R_0^2}\left(\frac{31}{16}+\frac{9}{4}\tau_e + \tau_e^2\right)\right. \\ \nonumber
    & \left. -b\frac{v_{Ti}^4}{\omega_b^4R_0^4}\left(\frac{747}{32} + \frac{481}{32}\tau_e+\frac{35}{8}\tau_e^2+\frac{1}{2}\tau_e^3\right) - 2\frac{v_{Ti}^4}{\omega_b^4R_0^4q_s^2}\left(\frac{23}{8} + 2 \tau_e + \frac{1}{2}\tau_e^2\right)\right] \\
    &\times\left\lbrace 1 + \frac{1}{24}\omega_b\omega_\text{dt}^2 \left(-\sigma\frac{4}{\omega_\text{dt}^3} + \frac{\omega_b}{\omega_\text{dt}^4}\right) + \sigma \frac{\omega_\text{dt}}{\omega_b}\tau_e + \left(\tau_e^2 + \frac{5}{4} \tau_e + 1\right)\frac{\omega_\text{dt}^2}{\omega_b} - 2b \right\rbrace, \label{eq:Damping-Qiu}
\end{align}
where the components are defined as
\begin{align}
    b &= k_r^2 \rho_i^2 / 2,\\
    v_{Ti} &= \sqrt{2T_i/m_i},\\
    \nonumber\omega_b &= \sqrt{\frac{7}{4} + \tau_e}\, \frac{v_{Ti}}{R_0} \left\lbrace 1 - \frac{b}{2}\left(\frac{31}{16} + \frac{9}{4}\tau_e + \tau_e^2\right)\left(\frac{7}{4} + \tau_e\right)^{-1} \right. \\
    \nonumber&\quad+ \frac{b}{2}\left(\frac{747}{32} + \frac{481}{32}\tau_e + \frac{35}{8}\tau_e^2 + \frac{1}{2}\tau_e^3\right)\left(\frac{7}{4} + \tau_e\right)^{-2} \\
    &\quad\left. + \frac{1}{2q_s^2} \left(\frac{23}{8} + 2\tau_e + \frac{1}{2}\tau_e^2\right)\left(\frac{7}{4}\tau_e\right)^{-2}\right\rbrace,\\
    \omega_\text{dt} &= \frac{v_{Ti}}{R_0} k_r \rho_i, \\
    \omega_\text{tt} &= \frac{v_{Ti}}{R_0 q_s}, \\
    \sigma &= \mathrm{sgn}\left[\frac{\omega_\text{b}}{\omega_\text{dt}}\right].
\end{align}
This damping term describes the collisionless Landau damping of the GAM and can be applied to larger wavevectors $k_r$ than \eg the term derived by Sugama and Watanabe in \Refr\cite{SugamaJPP06}, which is achieved by including higher order harmonics of the ion transit resonances in the derivation. The resulting dependency is depicted in \fig\ref{fig:Damping-term} for the parameters from \tab\ref{tab:parameters} (notably with $\rho_s/a_\text{min} = 2/375$, $\rho_i/a_\text{min} = 4.355\cdot10^{-3}$). It is apparent that the strength is negligible for wavevectors $k_\text{r} < 5\,\frac{2\pi}{a_\text{min}}$, but increases rapidly at $k\approx 10\,\frac{2\pi}{a_\text{min}}$.

\FloatBarrier

\subsection{Role of the perturbation wavelength}\label{sec:results-kpert}
This section analyzes the influence of the perturbation wavelength $\lambda_\text{pert} = 2\pi/k_\text{pert}$ chosen in the initial condition of the GAM radial electric field $\delta E_r$ on MI growth (as predicted by \eq\eqref{eq:mi-growthrate-general}, see \fig\ref{fig:MI-growthcurve}). Additionally the damping mechanism introduced in the previous section is taken into account. Similarly to the previous section the parameters from \sec\ref{sec:results-general} are used and consequently all perturbation wavevectors $k_\text{pert} < 14.6\,\frac{2\pi}{a_\text{min}} = k_\text{lim}$ should be susceptible to MI. Thus, simulations with modulation wavevectors in the range $4 - 14\,\frac{2\pi}{a_\text{min}}$ are analyzed in this section.

\begin{figure}[h!]
    \centering
    \begin{subfigure}{\textwidth}
        \centering
        \def\svgwidth{\textwidth}
        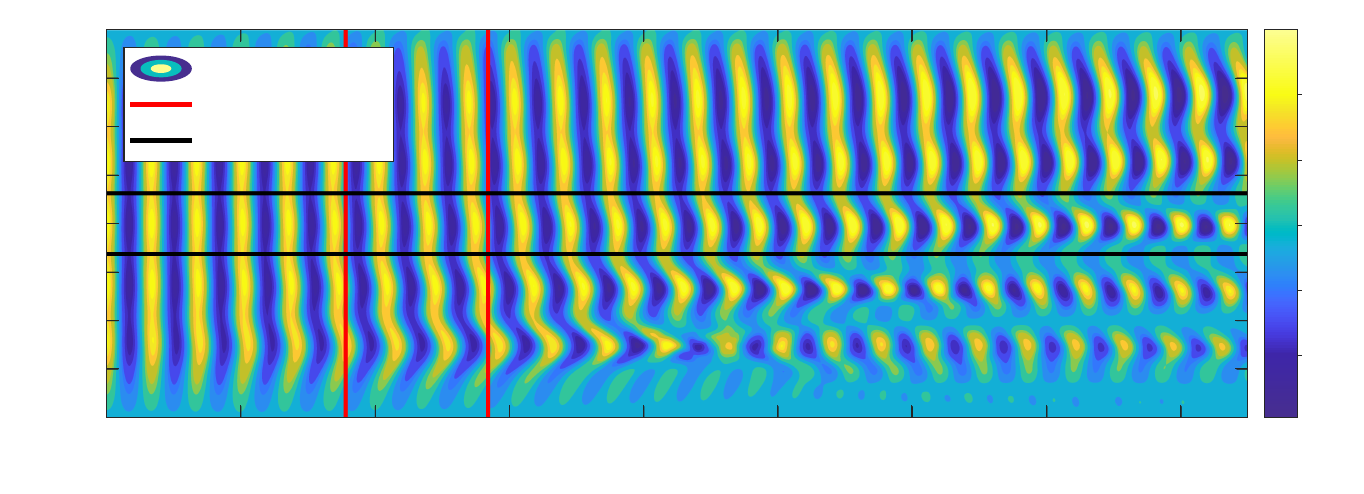  
        \caption{GAM radial electric field $E_r$ from a gyrokinetic simulation with perturbation wavevector $k_\text{pert} = 8\,\frac{2\pi}{a_\text{min}}$. The red lines show the time window where the fit was applied, the black lines indicate the radial region that was included in the calculation of the Fourier coefficient.}    
    \end{subfigure}
    \begin{subfigure}{\textwidth}
        \centering
        \def\svgwidth{\textwidth}
        \vspace*{.3em}
        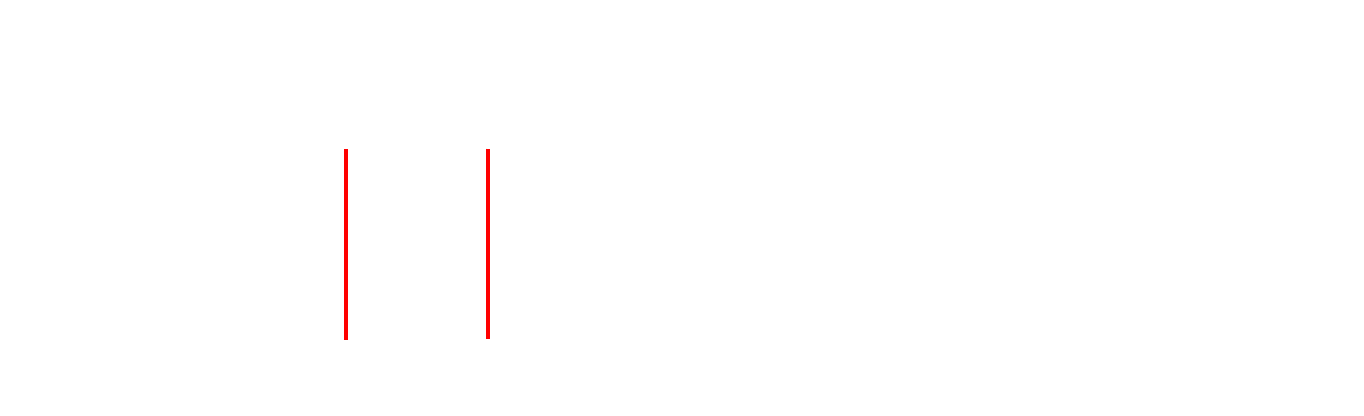 \caption{Time evolution of the absolute value of the Fourier coefficient of the perturbation wavevector $k_\text{pert} = 8\,\frac{2\pi}{a_\text{min}}$. The red lines indicate the time window where the fit was applied.}    
    \end{subfigure}
\caption{Scheme for determining the growth rate $\gamma$ of the perturbation in the GK GAM simulations. The upper picture depicts the radial electric field of the case with $k_\text{pert} = 8\,\frac{2\pi}{a_\text{min}}$.  The bottom picture depicts the absolute value of the Fourier coefficient at $|\mathfrak{F}[E_r](k_r = 8\,\frac{2\pi}{a_\text{min}},t)|$ and the envelope of the coefficient. The exponential fit is applied only to the region where the growth rate is highest, as for the first oscillation cycles the GAM electric field is experiencing an initial transient where higher GAM harmonics that were excited by the initial ``drop-in'' are still fading away, and for higher values of $t$ the assumption that the perturbation amplitude $a_1$ is small compared to the plateau amplitude $a_0$ is not fulfilled anymore.}
    \label{fig:gamma-scheme}
\end{figure}

In order to obtain a quantitative measure of the growth (or damping) $\gamma$ of the sinusoidal perturbation, the radial Fourier transforms of the GK simulation results, $\mathfrak{F}[E_r](k_r, t)$, are calculated using the Fast-Fourier-Transform algorithm. The Fourier coefficient corresponding to the perturbation wavevector $k_\text{pert}$, $\mathfrak{F}[E_r](k_r = k_\text{pert}, t)$, is then extracted and an exponential fit is applied to the region of exponential MI growth or exponential decay. Due to the fact that the growth rate depends on $\alpha_\text{NL}$, which in turn depends on the radial position $r$, the Fourier coefficient of the whole packet would return a complex mixture of the different growth stages at the different radial positions. 
As a consequence, before the Fourier transform is performed, a mask is applied such that only the simulation data around each maximum of the wavefront is included. This scheme is illustrated for the example of $k_\text{pert} = 8\,\frac{2\pi}{a_\text{min}}$ in \fig\ref{fig:gamma-scheme}.

\begin{figure}[h!]
    \centering
    \begin{subfigure}{0.48\textwidth}
    \def\svgwidth{\textwidth}
    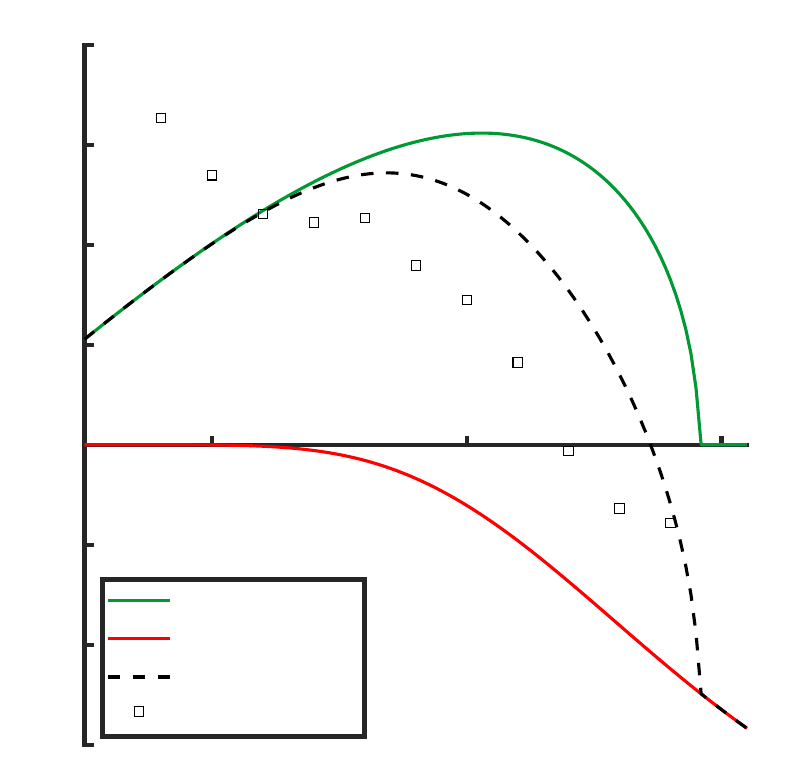
    \caption{Unmodified damping according to $\gamma_\text{Qiu}$}
    \label{fig:results-k-scan-unmodified}
    \end{subfigure}
    \begin{subfigure}{0.48\textwidth}
    \def\svgwidth{\textwidth}
    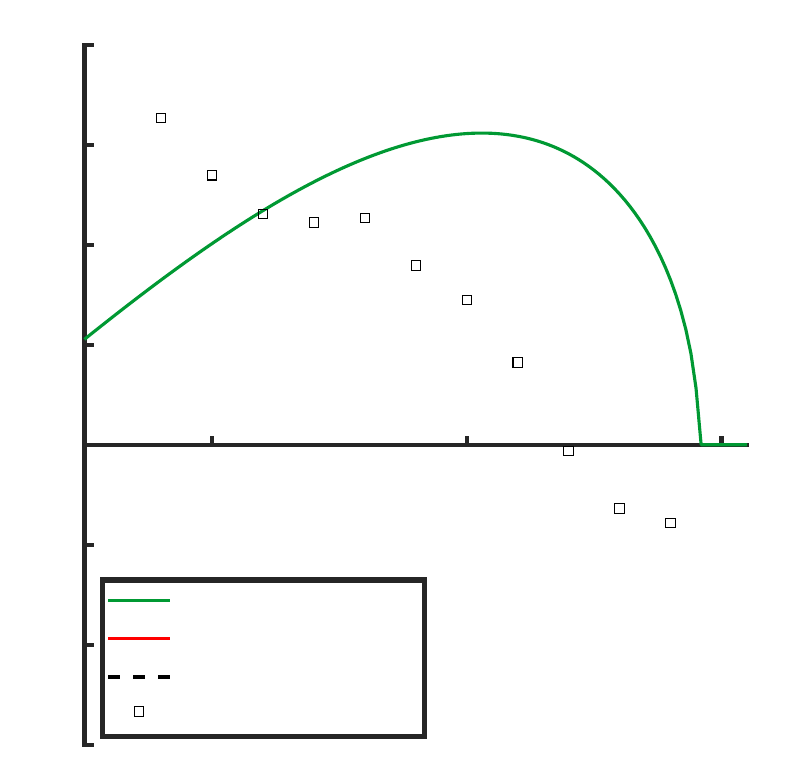
    \caption{Damping multiplied by a factor of 2.5}
    \label{fig:results-k-scan-adjusted}
    \end{subfigure}
\caption{GAM perturbation growth and damping rates $\gamma$ in GK simulations where the initial condition was modulated with different perturbation wavevectors $k_\text{pert}$. The GK results were obtained via the method illustrated in \fig\ref{fig:gamma-scheme}. The left figure shows the comparison of the simulation results to the theoretic predictions for the analytic MI growth rate $\gamma_\text{MI}$, \eq\eqref{eq:MI-disp-rel}, and the damping $\gamma_\text{Qiu}$, \eq\eqref{eq:Damping-Qiu}. The theoretic predictions are found to overestimate the growth rate significantly, as seen in \fig\ref{fig:results-k-scan-unmodified}. The right figure establishes that, when the damping term is amplified by a factor of 2.5, data for the wavevectors in the region $6\,\frac{2\pi}{a_\text{min}} \leq k_\text{pert} \leq 11\,\frac{2\pi}{a_\text{min}}$ more closely matches the theoretic predictions.}
    \label{fig:results-k-scan}
\end{figure}

\newpage
The aforementioned scheme, illustrated in \fig\ref{fig:gamma-scheme}, is applied to all wavevectors $k = 4, 5, \dots, 14\,\,\frac{2\pi}{a_\text{min}}$. In \fig\ref{fig:results-k-scan}, the MI growth rates obtained in this way are compared to the theoretical results of \sec\ref{sec:mi} (see \fig\ref{fig:MI-growthcurve} and \eq\eqref{eq:mi-growthrate-general}), where a damping term is included according to \eq\eqref{eq:Damping-Qiu}. The NLSE results clearly overestimate the GK MI growth rate $\gamma$ for $k_\text{pert} > 6\,\frac{2\pi}{a_\text{min}}$, however, after adjusting the strength of the damping amplitude by multiplying it with a factor of 2.5, the simulation results show good agreement with the theoretic predictions, most notably for wavevectors in the domain $6\,\frac{2\pi}{a_\text{min}} \leq k_\text{pert} \leq 11\,\frac{2\pi}{a_\text{min}}$. This assumption is justified by the benchmark in \Refr\cite{Biancalani17} (see \fig 4b in this reference), where in a comparison of the GAM damping $\gamma_\text{Qiu}$ to numeric simulations a similar difference was observed. 

In contrast to the well-matching growth rates at medium wavevectors, the matching becomes poorer at larger and smaller $k_\text{pert}$, as is apparent in \fig\ref{fig:results-k-scan-adjusted}. For wavevectors $k_\text{pert} > 11\,\frac{2\pi}{a_\text{min}}$, the MI is damped in GK simulations and its time behaviour is more prone to fitting errors due to the initial transient process. Additionally, these values are close to the applicability limit $\frac{3}{2} k_r^2\rho_i^2 \ll 1$ of the damping term. On the other hand, the growth rates obtained from the GK simulations at the lower end, with $k_\text{pert} < 6\,\frac{2\pi}{a_\text{min}}$, are too high compared to theory. Further investigations regarding this discrepancy are needed.

\FloatBarrier

\subsection{Role of the ion Larmor radius} \label{sec:results-rhoi}
The last section established that the small scales (high wavevectors) are more strongly affected by damping and that the analytical term from \Refr\cite{Qiu09}, when adjusted by a factor 2.5, gives a good approximation to the observations. This section tests this hypothesis further by analyzing the impact of a change of the ion Larmor radius $\rho_i$ on the damping scale, where the values $\rho_{i1}/a_\text{min} = 3.842\cdot10^{-3}$, $\rho_{i2}/a_\text{min} = 4.355\cdot10^{-3}$ (same value as in the previous sections) and $\rho_{i3}/a_\text{min} = 5.025\cdot10^{-3}$ were chosen for the comparison. The changes of $\rho_i$ are achieved by adjusting the ion thermal velocity $v_{Ti}$, which in the simulations is determined according to \eqs\eqref{eq:Lx-rhos} and \eqref{eq:rhos-rhoi}. The analytic damping term $\gamma_\text{Qiu}$ predicts that, for smaller Larmor radii, smaller scales \ie higher wavevectors $k_\text{pert}$ should be less damped.

\begin{figure}[h!]
	\centering
	\def\svgwidth{.65\textwidth}
	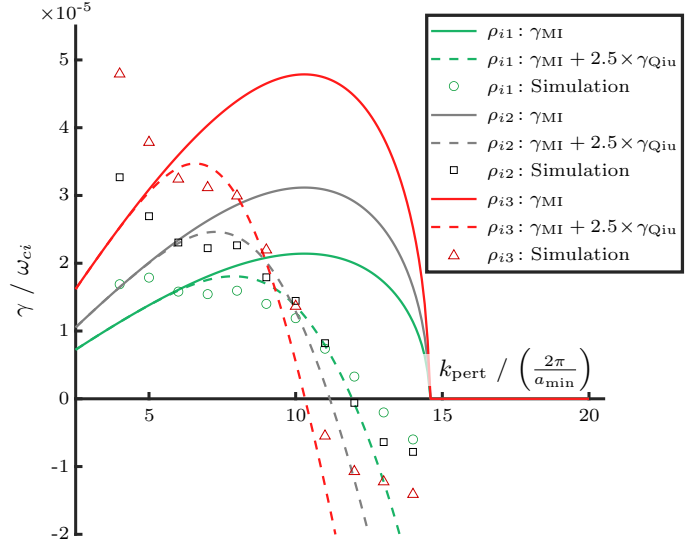
	\caption{GAM MI growth and damping rates in GK simulations with different perturbation wavevectors $k_\text{pert}$ and ion Larmor radii $\rho_i$ ($\rho_{i1}/a_\text{min} = 3.842\cdot10^{-3}$, $\rho_{i2}/a_\text{min} = 4.355\cdot10^{-3}$ and $\rho_{i3}/a_\text{min} = 5.025\cdot10^{-3}$) compared to the theoretic predictions from the analytic MI  growth rate $\gamma_\text{MI}$, \eq\eqref{eq:MI-disp-rel} and the analytic GAM damping $\gamma_\text{Qiu}$, \eq\eqref{eq:Damping-Qiu}. The difference in the maxima of the growth rates stems mainly from the different packet background amplitudes that were chosen according to \eq\eqref{eq:Dependency-growthrate-rhoi}.}
	\label{fig:Growthrate-Lx}
\end{figure}

A change of $\rho_i$ (via $v_{Ti}$) influences both $\calG$ and $\alpha_\text{NL}$, meaning that the range of unstable wavevectors $k_\text{pert} < k_\text{lim} = 2 a_0 \sqrt{\alpha_\text{NL}/\calG}$ from \eq\eqref{eq:MI-Condition-2} may change. To decorrelate the change of $\rho_i$ from the change of the unstable MI wavevectors, the parameter $a_0$ is selected depending on $\rho_i$ to ensure that the range of the unstable MI wavevectors stays the same. Precisely, due to $\calG \propto \rho_i^3$ (see \eq\eqref{eq:Definition-G} with $v_{Ti} \propto \rho_i$) and $\alpha_\text{NL} \propto 1/\rho_i$ (see \App\ref{app:alpha_NL}) one finds from \eq\eqref{eq:MI-Condition-2} that in order to keep the unstable range constant $a_0$ needs to be adjusted as
\begin{equation}
    k_\text{lim} = 2 a_0 \sqrt{\frac{\alpha_\text{NL}}{\calG}} \propto \sqrt{\frac{1}{\rho_i^4}} \, a_0 \stackrel{!}{=} \text{const.} \quad \Rightarrow \quad a_0 \propto \rho_i^2. \label{eq:Dependency-growthrate-rhoi}
\end{equation}
In order to obtain the same $k_\text{lim} = 14.5\,\frac{2\pi}{a_\text{min}}$ as in the previous sections the corresponding amplitudes are $a_{01} \approx 1.99\cdot10^{-4}\,$(a.u.), $a_{02} \approx 2.56\cdot10^{-4}\,$(a.u.) and $a_{03} \approx 3.40\cdot10^{-4}\,$(a.u.) for $\rho_{i1}, \rho_{i2}$ and $\rho_{i3}$, respectively. The simulations were analyzed with the scheme introduced in the last section with the corresponding growth and damping rates presented in \fig\ref{fig:Growthrate-Lx}. 

The results confirm the findings of the previous section that the adjustment by a factor of 2.5 of the amplitude of the damping term does reproduce the observed damping rate. Additionally, the theoretical predictions for the change of growth rates from $\gamma_\text{MI}$ due to the changing amplitude is in good agreement with the simulation results. However, similarly to \fig\ref{fig:results-k-scan-adjusted} the damping rate is overestimated in the region where the simulations are damped and the growth rate results for $k_\text{pert} < 6\,\frac{2\pi}{a_\text{min}}$ are again higher than the theoretical prediction. 
 
\newpage
\subsection{NLSE simulations including damping} \label{sec:results-damped-NLSE}
We now include the findings regarding the damping term of the previous two sections in the NLSE solver and compare a damped NLSE simulation to a GK simulation. The damping given by \eq\eqref{eq:Damping-Qiu} is amplified by a factor 2.5 and included in the numerical solver at the point where the dispersive term is applied, \ie \eq\eqref{eq:NLSE-solver-dispersive} in \sec\ref{sec:Numerics-SSS}. However, as shown in \fig\ref{fig:spectrum-196-k-10-NLSE}, during the MI saturation phase large wavevectors ($k_r > 20\,\frac{2\pi}{a_\text{min}}$) appear in the spectrum which, due to the condition $\frac{\tau_e}{2} k_r^2 \rho_i^2 \ll 1$ from \sec\ref{sec:results-damping}, lie outside of the applicability range of $\gamma_\text{Qiu}$. From the GK simulations, see \fig\ref{fig:spectrum-196-k-10-ORB5}, it is clear that these high wavevectors should be strongly suppressed, which was achieved by applying $\gamma = -4\cdot10^{-4}\,\omega_{ci}$ to all wavevectors fulfilling $\frac{\tau_e}{2} k_r^2 \rho_i^2 \geq 0.3$.

\begin{figure}[h!]
    \centering
    \def\svgwidth{\textwidth}
    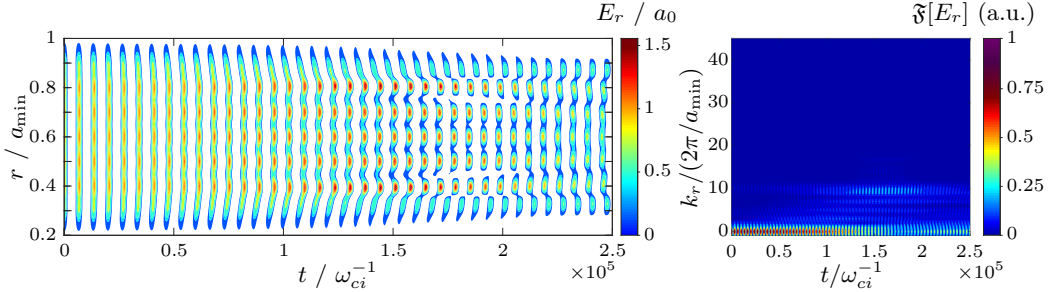
    \caption{Repetition of the NLSE simulation reported in \fig\ref{fig:comparison-mi-k-10-NLSE}, where now the damping scheme described at the beginning of this section included. The figure shows the time evolution of the GAM radial electric field $E_r$ (left) as well as the corresponding radial spectrum (right), \ie the absolute value of the the radial Fourier transform of the GAM radial electric field, $|\mathfrak{F}[E_r]|$.}
    \label{fig:results-damped-NLSE}
\end{figure}

The (undamped) NLSE simulation presented in \fig\ref{fig:comparison-mi-k-10-NLSE} is now repeated with the inclusion of the above described damping scheme. The damped result is reported in \fig\ref{fig:results-damped-NLSE} together with the corresponding spectrum. A clear improvement in regards to multiple aspects can be discerned when comparing damped and undamped NLSE simulations to the related GK result from \fig\ref{fig:comparison-mi-k-10-ORB5}. First of all, the moment where MI saturation phase appears in the simulations matches much better, with $t^\text{NLSE damp.}_\text{sat.} \approx 1.8\cdot10^5\,1/\omega_i $ and $t^\text{GK}_\text{sat.} \approx 2.2\cdot10^5\,1/\omega_i$ (at the packet center $r = r_0 = 0.6\,a_\text{tok}$), compared to the undamped NLSE simulation from \fig\ref{fig:comparison-mi-k-10-NLSE} with $t^\text{NLSE undamp.}_\text{sat.} \approx 1.1\cdot10^5\,1/\omega_i$. Furthermore, the maximal amplitude at the center $r=r_0$ is in better agreement, with $E^\text{NLSE damp.}_\text{max} = 1.31\,a_0$ and $E^\text{GK}_\text{max} \approx 1.26\,a_0$, compared to $E^\text{NLSE undamp.}_\text{max} = 2.43\,a_0$. The growth rate is $\gamma^\text{NLSE damp.} \approx 1.48\cdot10^{-5}\,\omega_{ci}$, which is very similar to the value obtained for the GK simulation with $\gamma^\text{GK} \approx 1.41\cdot10^{-5}\,\omega_{ci}$. However, a spectral comparison between \figs\ref{fig:results-damped-NLSE} and \ref{fig:spectrum-196-k-10-ORB5} illustrates significant differences, notably that the GK spectrum is much more complex, with more interactions of the different wavevectors. This may in part stem from the radial dependence of the value of $\alpha_\text{NL}$, which leads to incoherent phase fronts in the GK simulations and is not included in the NLSE simulation (see also discussion in \ref{sec:results-general}), on the other hand it may also be a consequence of the generally reduced complexity of the NLSE model.

\subsection{Self-focusing of Gaussian packets} \label{sec:results-self-focusing}
\begin{figure}[h!]
    \begin{subfigure}{\textwidth}
        \centering
        \def\svgwidth{.85\textwidth}
    	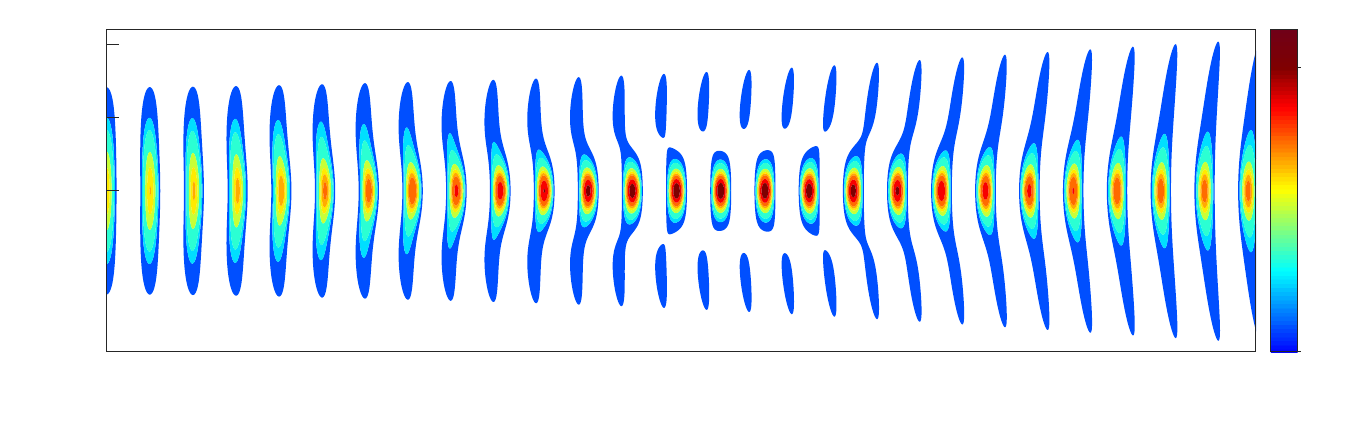
        \caption{NLSE simulation result}
    \end{subfigure}
    \begin{subfigure}{\textwidth}
    	\centering
        \def\svgwidth{.85\textwidth}
    	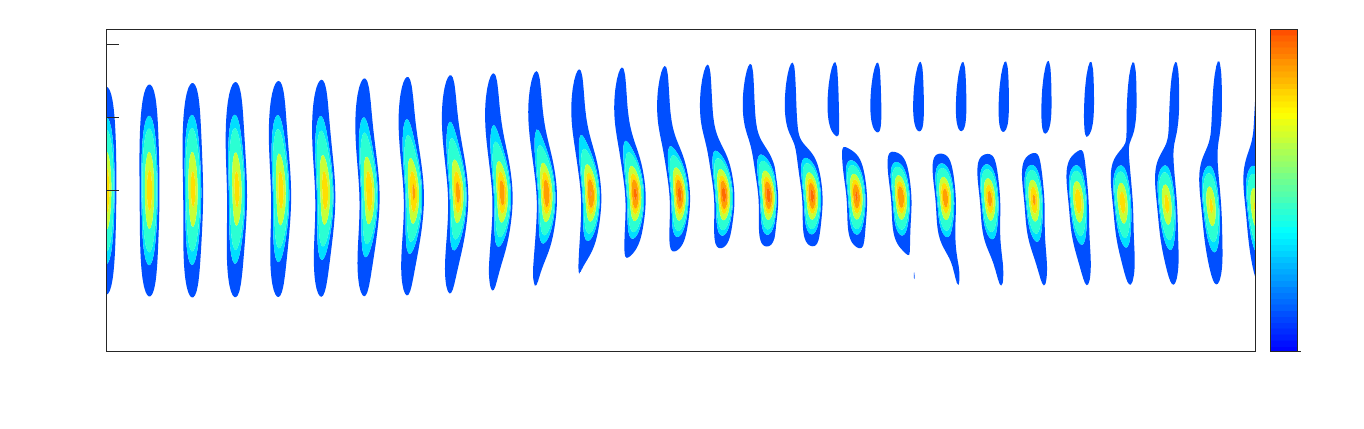
        \caption{GK simulation result}
    \end{subfigure}
	\begin{subfigure}{\textwidth}
    	\centering
        \def\svgwidth{.85\textwidth}
    	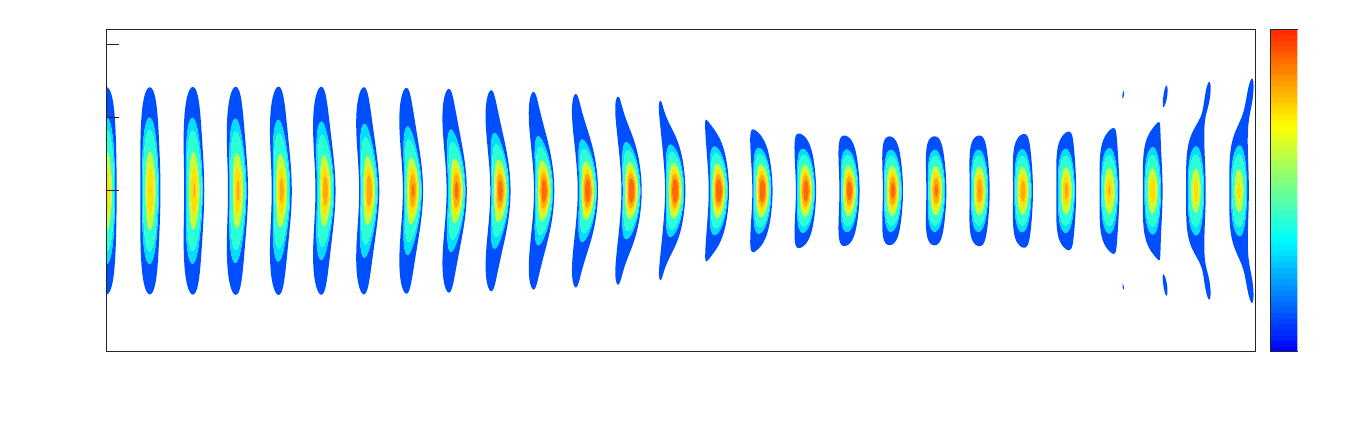
        \caption{Damped NLSE simulation result}
    \end{subfigure}
	\caption{Comparison of the evolution of the GAM radial electric field $E_r$ with an unperturbed Gaussian initial condition according to the undamped NLSE, damped NLSE and GK theory. The plasma background parameters of the simulations are $\tau_e = 2$, $q_s = 11$, $\rho_i/a_\text{min} = 3.784\cdot10^{-3}$, the initial condition is given by \eq\eqref{eq:Init-cond-plateau} with $a_0 = 3\cdot10^{-4}\,$(a.u.), $a_1 = 0$, $\w_0 = 0.1\,a_\text{min}$ and $p = 1$.}
    \label{fig:self-focusing}
\end{figure}

This section illustrates the self-focusing effect of the NLSE on an unmodulated initial condition where for the envelope a Gaussian packet is chosen. As detailed in \sec\ref{sec:nlse-dynamics}, self-focusing is observed for anomalous dispersion, \ie $\calG > 0$ and $\tau_e \lesssim 5.45$. The resulting comparison between a GK, an undamped NLSE and a damped NLSE (as specified in \sec\ref{sec:results-damping}) simulation is presented in \fig\ref{fig:self-focusing}. 

One can observe that in all three simulations the Gaussian packet experiences self-focusing, resulting in an increase of the maximal amplitude and a phase skip of the Gaussian center compared to the packet edges. It is apparent that this behaviour is very similar to the MI, while it in contrast does not require an initial modulation of the envelope. Similarly to \sec\ref{sec:results-damping} it is observed that, with damping included in the NLSE model, the maximum amplitude of the GK simulation is more closely reproduced. Furthermore one finds better agreement for the width and steepness of the focused packet, \ie in the GK and damped NLSE simulations at $t \approx 1.1\cdot10^5\,1/\omega_i$ compared to the undamped NLSE at $t \approx 0.8\cdot10^5\,1/\omega_i$. The radial asymmetry that develops in the GK simulation is not found in the NLSE simulations and can be explained by the radial dependence of $\alpha_\text{NL}$ that was introduced in \sec\ref{sec:results-general} and is explored further in \App\ref{app:alpha_NL}.

\subsection{Breather simulations} \label{sec:results-breather}
\begin{figure}[h!]
    \centering
    \def\svgwidth{\textwidth}
    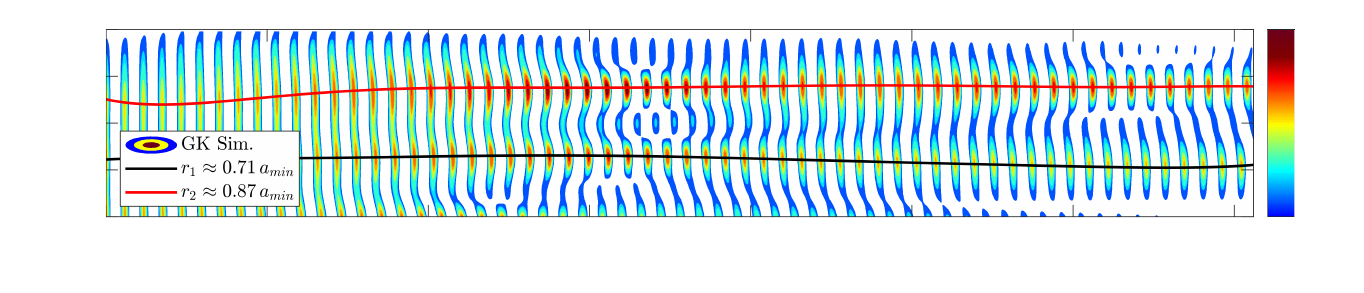
    \centering
    \def\svgwidth{\textwidth}
    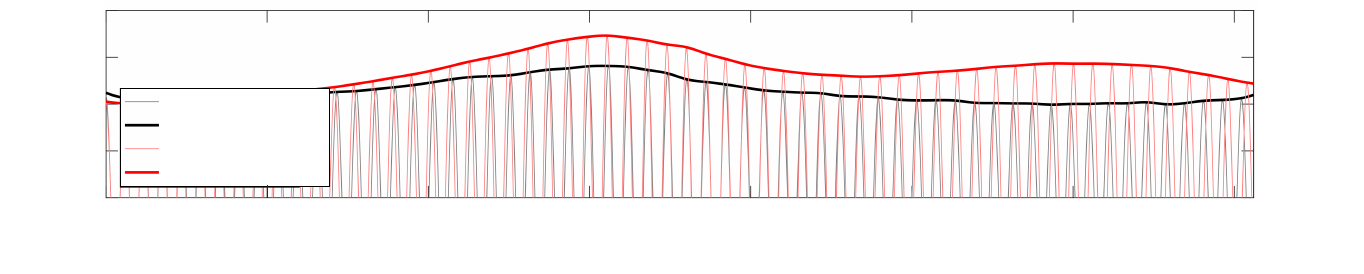
    \caption{Gyrokinetic simulation of a GAM which shows the breather behaviour of the MI. The upper figure shows the radial electric field $E_r$ with its radial shape, the bottom figure shows the value of $E_r$ along the lines shown in the upper figure, which follow the maxima. The red curve shows two MI saturation phases, the first at $t\approx 1.7\cdot10^5\,1/\omega_{ci}$ and the second one at $t\approx 3.3\cdot10^5\,1/\omega_{ci}$. The black curve shows the start of a second MI growth phase at the end of the simulation. The parameters are chosen as specified in \tab\ref{tab:parameters}, with $\rho_{i3}/a_\text{min} = 5.025\cdot10^{-3}$, $a_0 = 3.4\cdot10^{-4}$ and $k_\text{pert} = 8\,\frac{2\pi}{a_\text{min}}$.}
    \label{fig:Breather}
\end{figure}

In this section the phenomenon of the Akhmediev breather is studied in GK simulations. Akhmediev Breathers (ABs) \cite{Akhmediev86} are special types of MI solutions to the NLSE which predict that after the saturation phase, the MI initial condition is restored, as illustrated in \fig\ref{fig:MI-depiction}, and new MI growth should be observable. This phenomenon is hard to observe in GK simulations as \eg the dependency of $\alpha_\text{NL}$ breaks the packet apart and hinders the return to the initial condition after the first saturation phase. As a consequence we concentrate on the region $r> 0.6\,a_\text{min}$, where the results from \App\ref{app:alpha_NL} predict only small changes of $\alpha_\text{NL}$. A GK simulation with a breather solution is depicted in \fig\ref{fig:Breather}.

The maximum at $r_2\approx 0.87\,a_\text{min}$ experiences two saturation phases, the first one at $t\approx 1.5\cdot10^5\,\omega_{ci}^{-1}$ and the second at $t\approx 3.3\cdot10^5\,\omega_{ci}^{-1}$, thus demonstrating the AB phenomenon for a GAM packet in a GK simulation. Comparable behaviour happens for the maximum at $r\approx0.71$, however the growth is slower and not as pronounced. The impact of the damping mechanism and the differences in the nonlinear coefficient $\alpha_\text{NL}$ are again present in the simulations: For example, the second saturation phase at $t\approx 3.3\cdot10^5\,\omega_{ci}^{-1}$ of the maximum centered around $r_2$ is observed to have a maximum amplitude of $1.44\,a_0$, which is significantly lower than the value of $1.74\,a_0$ found in the first saturation phase, whereas an undamped AB would under ideal conditions yield the same maximum value in both saturation phases.

While more complex patterns of the AB are possible under ideal conditions, such as the appearance of growth-phases of other unstable wavevectors in-between the saturation phases of the initial perturbation wavevector $k_\text{pert}$, see \eg \Refr\cite{Copie2020}, the deviations from a pure NLSE-like behaviour due to the mechanisms introduced in the previous sections make the possibility of such observations highly unlikely.

\section{Summary and conclusions}
\label{sec:concl}
The results of this paper show that geodesic-acoustic oscillations (GAMs) are susceptible to modulational instability (MI) under the conditions predicted by the nonlinear Schr\"{o}dinger equation (NLSE) model. However, the high wavevectors that are part of the nonlinear saturation phase and important for the MI cycle (and MI breathers, see \sec\ref{sec:results-breather}) do not develop in gyrokinetic (GK) simulations due to the Landau damping which was characterized in detail in \secs\ref{sec:results-damping} -- \ref{sec:results-damped-NLSE}. For the parameters considered in this study the aforementioned damping effect hinders the MI process significantly from developing to its full extent and is strong enough to stabilize some of the (according to the undamped NLSE model) unstable wavevectors, as was illustrated in \fig\ref{fig:Growthrate-Lx} for $k \gtrsim 12\,\frac{2\pi}{a_\text{min}}$. A second significant shortcoming of the NLSE model was the assumption that $\alpha_\text{NL}$ is independent of the radial coordinate $r$. The GK simulations of this paper establish the radial variation of the nonlinear coefficient $\alpha_\text{NL} = \alpha_\text{NL}(r)$ of the NLSE model for GAMs, which was not evident in the prior study on this topic, \Refr\cite{PoP21}, due to the relatively narrow Gaussian packets that were considered there.

One can conclude from the theoretical descriptions of the MI and the damping mechanism (\eqs\eqref{eq:mi-growthrate-general} and \eqref{eq:Damping-Qiu}, respectively) that GAM MI is more likely to be observable for high safety factors and small Larmor radii (\ie low ion temperatures $T_i$ and thermal velocities $v_{Ti}$). Theory further suggests that the tokamak aspect ratio $A = R_0 / a_\text{min}$ may have an impact on the GAM MI growth rate, however further investigations on the dependency of the nonlinear coefficient $\alpha_\text{NL}$ on the geometric parameters are needed to confirm this prediction. 

The observed significance of damping in the gyrokinetic simulation results might seem surprising as the initial spectrum of the packet lies in a range for which theory predicts negligible damping. This enhanced damping observed in the GK simulations is akin to the phenomenon analyzed in detail in \Refr\cite{PalermoEPL,Biancalani16}. There, the generation of higher wavevectors through linear processes due to the presence of radially nonuniform profiles was found to increase GAM damping significantly, leading to the inclusion of a ``phase-mixing damping adjustment''. Similarly to these findings, as was discussed in \sec\ref{sec:mi-intro}, the MI process is associated with the (in our case nonlinear) generation of high wavevector components in the radial GAM spectrum. This interpretation of the simulation results is confirmed by the good agreement they exhibit with an NLSE model corrected with the inclusion of the damping rate derived in \Refr\cite{Qiu09}, as is presented in \sec\ref{sec:results-damped-NLSE}. The overall correction factor adopted here to achieve a quantitative matching with the GK simulations is in line with previous findings \cite{Biancalani17}.

While the NLSE model proved to give accurate predictions of the general GAM behaviour, the exact dynamics shows significant differences. An improvement of the model, putting it on a more firm theoretical ground, requires a derivation of the NLSE equation for GAMs from first principles, which should be addressed in the near future. 

Predictive capability to other parameter regimes is limited and will only improve with further studies on the unknown variable $\alpha_\text{NL}$. 

Altogether, the results indicate that the possibility and impact of an MI on the GAM dynamics will be small. This conclusion becomes more evident by the fact that in this study, electrons were treated adiabatically in simulations and in the damping term, while recent research suggests that a kinetic treatment of the electrons will increase damping, thus decreasing the likeliness of MI even further \cite{Zhang10,Ehrlacher18}. On the other hand, the self-focusing behaviour associated with the MI formation process is omnipresent in the regime of anomalous dispersion ($\calG > 0$, $\tau_e \lesssim 5.45$) and may be observable in other simulations similarly to the case presented in \sec\ref{sec:results-self-focusing}. The main reason behind this is the size of the involved structures and unstable wavevectors $k_r$: while the self-focusing effect only requires that a local maximum is present in the packet envelope, which can be of any size (or even simply the packet itself), MI formation requires that the maximum is ``on top of'' a relatively constant packet, thus demanding much finer structures and higher wavevectors. This is not only a much more unlikely initial condition to spontaneously develop in a tokamak, but is also much more significantly affected by the damping process illustrated in \secs\ref{sec:results-damping}-\ref{sec:results-rhoi}.

\newpage 
\paragraph{Acknowledgments}\mbox{}\\
This work has been carried out within the framework of the EUROfusion Consortium, funded by the European Union via the Euratom Research and Training Programme (Grant Agreement No 101052200 — EUROfusion). Views and opinions expressed are however those of the author(s) only and do not necessarily reflect those of the European Union or the European Commission. Neither the European Union nor the European Commission can be held responsible for them.

We thank Prof.~Dr.~Zhiyong Qiu (University of Zhejiang) for his support and fruitful discussions concerning the damping term derived in \Refr\cite{Qiu09}. We thank Prof.~Dr.~Fulvio Zonca for helpful discussions.

The data that support the findings of this study are available from the corresponding author upon reasonable request.

\appendix
\section{Determination of $\alpha_\text{NL}$} 
\setcounter{equation}{0}
\renewcommand{\theequation}{\thesection.\arabic{equation}}
\label{app:alpha_NL}

In order to make meaningful predictions about the MI behaviour and growth rates $\gamma_\text{MI}$ (see \eq\eqref{eq:mi-growthrate-general}), it is necessary to assess the magnitude of the strength $\alpha_\text{NL}$ of the nonlinear term. Since an analytical expression for this parameter has currently not yet been derived, as was already mentioned in \sec\ref{sec:nlse-introdtion}, its value is in this study determined by comparing NLSE results to gyrokinetic simulations of isolated GAMs obtained with the code ORB5. The axisymmetric component of the GAM radial electric field is initialised as described in \sec\ref{sec:Numerics-ORB5}, where the NLSE evolves an initial electric field perturbation, while in ORB5 the electric field is generated from an initial perturbation $\delta n$ of the ion density.

To obtain the exact value of the initial electric field perturbation from the initial density perturbation amplitude $\delta n / n_0$ one would have to evaluate the proportionality constant of the relation in \eq\eqref{eq:relation-E-ion-density}. Since the exact physical values of the electric field do not matter for the results of this study and to simplify comparisons between the NLSE and gyrokinetic simulations, we define a normalized amplitude of the electric field simply as $a_0 = \delta n / n_0$. As a consequence, the parameter $\alpha_\text{NL}$ is given here in units related to the amplitude of the relative density perturbation $\delta n / n_0$. 

\begin{figure}[h!]
	\centering
	\def\svgwidth{.5\textwidth}
	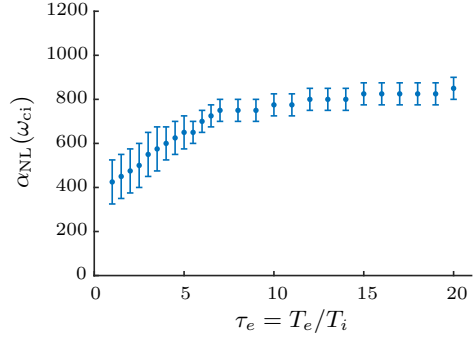
	\caption{Dependency of the nonlinear parameter $\alpha_\text{NL}$ on the electron-to-ion temperature ratio $\tau_e$. The remaining parameters were chosen as specified in \tab\ref{tab:parameters}, with $\rho_i/a_\text{min} = 3.842\cdot10^{-3}$, $q_s = 5$ and $r_0 = 0.5\,a_\text{min}$.}
	\label{fig:alpha-NL-scan-taue}
\end{figure}

\begin{figure}[h!]
	\centering
	\def\svgwidth{.5\textwidth}
	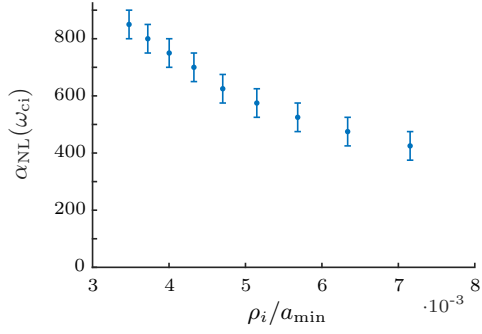
	\caption{Dependency of the nonlinear parameter $\alpha_\text{NL}$ on the ion Larmor radius $\rho_i$. The remaining parameters were chosen as specified in \tab\ref{tab:parameters}, with $\tau_e = 4$, $q_s = 5$ and $r_0 = 0.5\,a_\text{min}$.}
	\label{fig:alpha-NL-scan-rhoi}
\end{figure}

\begin{figure}[h!]
	\centering
	\def\svgwidth{.5\textwidth}
	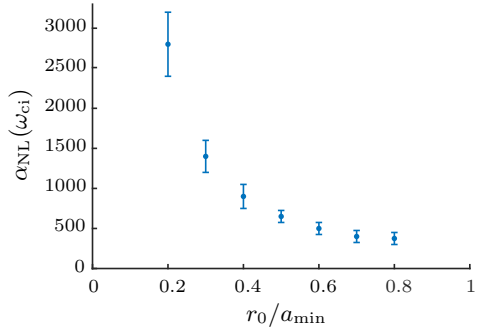
	\caption{Dependency of the nonlinear parameter $\alpha_\text{NL}$ on the position $r_0$ of the GAM in the minor tokamak radius. This dependency was not included in the NLSE simulation dynamics. The remaining parameters were chosen as specified in \tab\ref{tab:parameters}, with $\tau_e = 4$, $\rho_i/a_\text{min} = 3.842\cdot10^{-3}$ and $q_s = 5$.}
	\label{fig:alpha-NL-scan-r0}
\end{figure}

For the comparison of NLSE and gyrokinetic simulations an unperturbed Gaussian envelope (\ie $a_1 = 0$, $p = 1$ in \eq\eqref{eq:Init-cond-plateau}) was chosen for the initial condition. The dependency of $\alpha_\text{NL}$ on the electron-to-ion temperature ratio $\tau_e = T_e/T_i$, the safety factor $q_s$, the radial position $r_0$ of the GAM and the ion Larmor radius $\rho_i$ was determined. The ion Larmor radius is set in ORB5 by the choice of the parameter $L_x$, which is defined through the following relations
\begin{equation} \label{eq:Lx-rhos}
    L_x = 2\frac{a_\text{min}}{\rho_s},
\end{equation}
\begin{equation} \label{eq:rhos-rhoi}
    \rho_s = \frac{c_\text{s}}{\omega_{ci}} = \frac{\sqrt{T_e/m_i}}{\omega_{ci}} = \sqrt{\frac{\tau_e}{2}} \frac{\sqrt{2\,T_i/m_i}}{\omega_{ci}} = \sqrt{\frac{\tau_e}{2}}\rho_i.
\end{equation}
This subsequently also affects the value of the ion temperature. The remaining parameters $a_\text{min}, R_0, B_0, m_i$ were chosen as specified in \tab\ref{tab:parameters}. The results for the dependencies on $\tau_e$, $\rho_i$ (\ie $L_x$) and $r_0$ are depicted in figures \ref{fig:alpha-NL-scan-taue} -- \ref{fig:alpha-NL-scan-r0}, respectively. The respective error bars illustrate the range of values of $\alpha_\text{NL}$ in which the NLSE simulations matched GK results within one quarter of the oscillation period $1/4\,T_\text{GAM}$ at the end of the simulation.

From \fig\ref{fig:alpha-NL-scan-taue} one finds that $\alpha_\text{NL}(\tau_e)$ increases nearly linearly with $\tau_e$, but with different slopes in the regime of anomalous ($\tau_e \lesssim 5.45$) and normal ($\tau_e \gtrsim 5.45$) dispersion. The error bars in the regime of anomalous dispersion are higher due to the self-focusing effect of Gaussian packets, which is discussed in \sec\ref{sec:results-self-focusing} and complicated comparisons. From \fig\ref{fig:alpha-NL-scan-rhoi} as a first approximation the proportionality $\alpha_\text{NL} \propto 1/\rho_i$ is obtained. The radial position of the GAM is found to heavily influence the strength of the nonlinear parameter as seen in \fig\ref{fig:alpha-NL-scan-r0}, most notably when $r_0$ is below $0.4\,a_\text{min}$. For the safety factor $q_s$ it was found that the value of $\alpha_\text{NL}$ does not change significantly in the range $q_s = 1 \dots 15$ considered in this paper. 

\section{Qualitative picture for the self-focusing NLSE and Modulational Instability} \label{app:qualitative-mi}
This section presents a qualitative explanation for the MI formation process in terms of an interaction of the between the effects of the nonlinear and the dispersive term, which were introduced in \sec\ref{sec:nlse-dynamics}. Although the MI is meanwhile a well-known phenomenon, we believe that the summary of this simple interpretation might be useful, in particular to interpret the results of this paper.  

\begin{figure}[h!]
	\centering
	\def\svgwidth{.78\textwidth}
	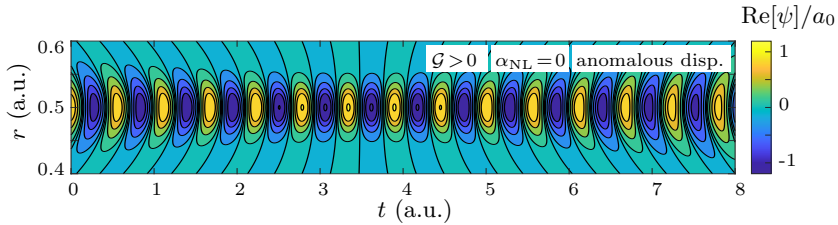
	\caption{NLSE simulation of the GAM electric field $E_r = \Rea[\psi]$ for a Gaussian with an initial convex curvature in $(r,t)$-space. The anomalous dispersion reduces the width of the Gaussian while increasing its amplitude until the phase front is flat at $t\approx 3.5$. After this point one observes the usual dispersive broadening and increase of concave curvature.}
	\label{fig:MI-understanding-0}
\end{figure}

As depicted in \fig\ref{fig:Linear-Dispersion}, the effect of the dispersive term (without nonlinearity, $\alpha_\text{NL} = 0$) on a ``flat'' phase front at $t=0$ is an increase in width and the appearance of (in the case of anomalous dispersion concave) curvature as time progresses. When considering a packet with initial curvature opposite to what is generated by the dispersive term, one can observe that during the process of reducing the phase-front curvature, the packet width decreases until the phase front is flat, as shown in \fig\ref{fig:MI-understanding-0}. After this point one finds the usual dispersive broadening. This is the well-known behaviour of a Gaussian pulse in optics (but with the roles of time and space reversed), see \eg \Refr\cite{Hecht12}. One can conclude that, in the regime of anomalous dispersion, as long as the packet curvature is convex in $(r,t)$-space, the width and curvature of the packet will decrease while the amplitude of the maximum increases. For normal dispersion one will observe the same effect for an initially concave packet.

As mentioned in \sec\ref{sec:nlse-dynamics} the nonlinear term introduces a phase shift, described by \eq\eqref{eq:nl-phaseshift}, in regions where the amplitude is higher, which gives maxima in the packet a convex curvature, as seen in \fig\ref{fig:Nonlinear-Term}. It follows that when both the nonlinear and (anomalous) dispersive contributions to the dynamics are considered, the dispersive term acts to reduce the curvature from the nonlinear phase shift and decreases the width of the packet. As long as the strength of the nonlinear phase shift is stronger than the effect of the dispersive term, the packet stays convexly curved. As a result, the width of the region around the maximum will decrease while its amplitude rises. This competition between the nonlinear and the dispersive term is represented mathematically by second condition for MI growth stated in the previous section, \eq\eqref{eq:MI-Condition-2}, which can be rewritten to
\begin{equation}
    \frac{\calG k_\text{pert}^2}{2} < 2 a_0^2 \alpha_\text{NL}.
\end{equation}
Here, the left-hand side is the strength of the phase shift of a local sinusoidal maximum due to the dispersive term, while the \rhs\ is the nonlinear phase-shift of the maximum relative to its surrounding background.

\bibliography{./biblio}
\bibliographystyle{ieeetr}

\end{document}

%% file: images-theory/G-vs-taue-thick-Lx-400.pdf_tex
\begingroup%
  \makeatletter%
  \providecommand\color[2][]{%
    \errmessage{(Inkscape) Color is used for the text in Inkscape, but the package 'color.sty' is not loaded}%
    \renewcommand\color[2][]{}%
  }%
  \providecommand\transparent[1]{%
    \errmessage{(Inkscape) Transparency is used (non-zero) for the text in Inkscape, but the package 'transparent.sty' is not loaded}%
    \renewcommand\transparent[1]{}%
  }%
  \providecommand\rotatebox[2]{#2}%
  \newcommand*\fsize{\dimexpr\f@size pt\relax}%
  \newcommand*\lineheight[1]{\fontsize{\fsize}{#1\fsize}\selectfont}%
  \ifx\svgwidth\undefined%
    \setlength{\unitlength}{367.63320923bp}%
    \ifx\svgscale\undefined%
      \relax%
    \else%
      \setlength{\unitlength}{\unitlength * \real{\svgscale}}%
    \fi%
  \else%
    \setlength{\unitlength}{\svgwidth}%
  \fi%
  \global\let\svgwidth\undefined%
  \global\let\svgscale\undefined%
  \makeatother%
  \begin{picture}(1,0.76998364)%
    \lineheight{1}%
    \setlength\tabcolsep{0pt}%
    \put(0,0){\includegraphics[width=\unitlength,page=1]{G-vs-taue-thick-Lx-400.pdf}}%
    \put(0.87220494,0.20710276){\color[rgb]{0,0,0}\makebox(0,0)[lt]{\lineheight{1.25}\smash{\begin{tabular}[t]{l}\SVGlabelsize $\tau_e = T_e / T_i$\end{tabular}}}}%
    \put(0.06483142,0.39101925){\color[rgb]{0,0,0}\makebox(0,0)[lt]{\lineheight{1.25}\smash{\begin{tabular}[t]{l}\SVGunitsize 3\end{tabular}}}}%
    \put(0.06501776,0.46876883){\color[rgb]{0,0,0}\makebox(0,0)[lt]{\lineheight{1.25}\smash{\begin{tabular}[t]{l}\SVGunitsize 4\end{tabular}}}}%
    \put(0.06474267,0.54651838){\color[rgb]{0,0,0}\makebox(0,0)[lt]{\lineheight{1.25}\smash{\begin{tabular}[t]{l}\SVGunitsize 5\end{tabular}}}}%
    \put(0.0642391,0.62426795){\color[rgb]{0,0,0}\makebox(0,0)[lt]{\lineheight{1.25}\smash{\begin{tabular}[t]{l}\SVGunitsize 6\end{tabular}}}}%
    \put(0.06475846,0.70201752){\color[rgb]{0,0,0}\makebox(0,0)[lt]{\lineheight{1.25}\smash{\begin{tabular}[t]{l}\SVGunitsize 7\end{tabular}}}}%
    \put(0.10509256,0.75138388){\color[rgb]{0,0,0}\makebox(0,0)[lt]{\lineheight{1.25}\smash{\begin{tabular}[t]{l}\SVGunitsize $\cdot$ 10$^\text{-10}$\end{tabular}}}}%
    \put(0.06470392,0.31326946){\color[rgb]{0,0,0}\makebox(0,0)[lt]{\lineheight{1.25}\smash{\begin{tabular}[t]{l}\SVGunitsize 2\end{tabular}}}}%
    \put(0.21001409,0.12303516){\color[rgb]{0,0,0}\makebox(0,0)[lt]{\lineheight{1.25}\smash{\begin{tabular}[t]{l}\SVGunitsize 3\end{tabular}}}}%
    \put(0.32068828,0.1226646){\color[rgb]{0,0,0}\makebox(0,0)[lt]{\lineheight{1.25}\smash{\begin{tabular}[t]{l}\SVGunitsize 4\end{tabular}}}}%
    \put(0.43136245,0.12303516){\color[rgb]{0,0,0}\makebox(0,0)[lt]{\lineheight{1.25}\smash{\begin{tabular}[t]{l}\SVGunitsize 5\end{tabular}}}}%
    \put(0.65271078,0.1226646){\color[rgb]{0,0,0}\makebox(0,0)[lt]{\lineheight{1.25}\smash{\begin{tabular}[t]{l}\SVGunitsize 7\end{tabular}}}}%
    \put(0.76338495,0.12307102){\color[rgb]{0,0,0}\makebox(0,0)[lt]{\lineheight{1.25}\smash{\begin{tabular}[t]{l}\SVGunitsize 8\end{tabular}}}}%
    \put(0.87405909,0.12301124){\color[rgb]{0,0,0}\makebox(0,0)[lt]{\lineheight{1.25}\smash{\begin{tabular}[t]{l}\SVGunitsize 9\end{tabular}}}}%
    \put(0.01859976,0.28549514){\color[rgb]{0,0,0}\rotatebox{90}{\makebox(0,0)[lt]{\lineheight{1.25}\smash{\begin{tabular}[t]{l}\SVGlabelsize $\calG (\omega_{ci}\text{m}^2)$\end{tabular}}}}}%
    \put(0.06613878,0.23551992){\color[rgb]{0,0,0}\makebox(0,0)[lt]{\lineheight{1.25}\smash{\begin{tabular}[t]{l}\SVGunitsize 1\end{tabular}}}}%
    \put(0.06491846,0.15777032){\color[rgb]{0,0,0}\makebox(0,0)[lt]{\lineheight{1.25}\smash{\begin{tabular}[t]{l}\SVGunitsize 0\end{tabular}}}}%
    \put(0.0558508,0.08002078){\color[rgb]{0,0,0}\makebox(0,0)[lt]{\lineheight{1.25}\smash{\begin{tabular}[t]{l}\SVGunitsize -1\end{tabular}}}}%
    \put(0.0558508,0.00227118){\color[rgb]{0,0,0}\makebox(0,0)[lt]{\lineheight{1.25}\smash{\begin{tabular}[t]{l}\SVGunitsize -2\end{tabular}}}}%
    \put(0,0){\includegraphics[width=\unitlength,page=2]{G-vs-taue-thick-Lx-400.pdf}}%
    \put(0.54203662,0.12303516){\color[rgb]{0,0,0}\makebox(0,0)[lt]{\lineheight{1.25}\smash{\begin{tabular}[t]{l}\SVGunitsize 6\end{tabular}}}}%
  \end{picture}%
\endgroup%

%% file: images-theory/Linear-Dispersion-Comparison-2-v2-altered-edge.pdf_tex
\begingroup%
  \makeatletter%
  \providecommand\color[2][]{%
    \errmessage{(Inkscape) Color is used for the text in Inkscape, but the package 'color.sty' is not loaded}%
    \renewcommand\color[2][]{}%
  }%
  \providecommand\transparent[1]{%
    \errmessage{(Inkscape) Transparency is used (non-zero) for the text in Inkscape, but the package 'transparent.sty' is not loaded}%
    \renewcommand\transparent[1]{}%
  }%
  \providecommand\rotatebox[2]{#2}%
  \newcommand*\fsize{\dimexpr\f@size pt\relax}%
  \newcommand*\lineheight[1]{\fontsize{\fsize}{#1\fsize}\selectfont}%
  \ifx\svgwidth\undefined%
    \setlength{\unitlength}{1350.28945082bp}%
    \ifx\svgscale\undefined%
      \relax%
    \else%
      \setlength{\unitlength}{\unitlength * \real{\svgscale}}%
    \fi%
  \else%
    \setlength{\unitlength}{\svgwidth}%
  \fi%
  \global\let\svgwidth\undefined%
  \global\let\svgscale\undefined%
  \makeatother%
  \begin{picture}(1,0.58309348)%
    \lineheight{1}%
    \setlength\tabcolsep{0pt}%
    \put(0,0){\includegraphics[width=\unitlength,page=1]{Linear-Dispersion-Comparison-2-v2-altered-edge.pdf}}%
    \put(0.08922924,0.54687799){\color[rgb]{0,0,0}\makebox(0,0)[lt]{\lineheight{1.25}\smash{\begin{tabular}[t]{l}{\SVGmathsize $\calG\!=\!0$}\end{tabular}}}}%
    \put(0,0){\includegraphics[width=\unitlength,page=2]{Linear-Dispersion-Comparison-2-v2-altered-edge.pdf}}%
    \put(0.16786288,0.54687798){\color[rgb]{0,0,0}\makebox(0,0)[lt]{\lineheight{1.25}\smash{\begin{tabular}[t]{l}{\SVGmathsize $\alpha_\text{NL}\!=\!0$}\end{tabular}}}}%
    \put(0,0){\includegraphics[width=\unitlength,page=3]{Linear-Dispersion-Comparison-2-v2-altered-edge.pdf}}%
    \put(0.26930388,0.54687799){\color[rgb]{0,0,0}\makebox(0,0)[lt]{\lineheight{1.25}\smash{\begin{tabular}[t]{l}{\SVGunitsize no dispersion}\end{tabular}}}}%
    \put(0,0){\includegraphics[width=\unitlength,page=4]{Linear-Dispersion-Comparison-2-v2-altered-edge.pdf}}%
    \put(0.26835981,0.36892715){\color[rgb]{0,0,0}\makebox(0,0)[lt]{\lineheight{1.25}\smash{\begin{tabular}[t]{l}{\SVGunitsize normal disp.}\end{tabular}}}}%
    \put(0,0){\includegraphics[width=\unitlength,page=5]{Linear-Dispersion-Comparison-2-v2-altered-edge.pdf}}%
    \put(0.2683598,0.18985075){\color[rgb]{0,0,0}\makebox(0,0)[lt]{\lineheight{1.25}\smash{\begin{tabular}[t]{l}{\SVGunitsize anomalous disp.}\end{tabular}}}}%
    \put(0,0){\includegraphics[width=\unitlength,page=6]{Linear-Dispersion-Comparison-2-v2-altered-edge.pdf}}%
    \put(0.08922924,0.36892715){\color[rgb]{0,0,0}\makebox(0,0)[lt]{\lineheight{1.25}\smash{\begin{tabular}[t]{l}{\SVGmathsize $\calG\!<\!0$}\end{tabular}}}}%
    \put(0,0){\includegraphics[width=\unitlength,page=7]{Linear-Dispersion-Comparison-2-v2-altered-edge.pdf}}%
    \put(0.16786288,0.36892715){\color[rgb]{0,0,0}\makebox(0,0)[lt]{\lineheight{1.25}\smash{\begin{tabular}[t]{l}{\SVGmathsize $\alpha_\text{NL}\!=\!0$}\end{tabular}}}}%
    \put(0,0){\includegraphics[width=\unitlength,page=8]{Linear-Dispersion-Comparison-2-v2-altered-edge.pdf}}%
    \put(0.08922924,0.18985075){\color[rgb]{0,0,0}\makebox(0,0)[lt]{\lineheight{1.25}\smash{\begin{tabular}[t]{l}{\SVGmathsize $\calG\!>\!0$}\end{tabular}}}}%
    \put(0,0){\includegraphics[width=\unitlength,page=9]{Linear-Dispersion-Comparison-2-v2-altered-edge.pdf}}%
    \put(0.16786288,0.18985075){\color[rgb]{0,0,0}\makebox(0,0)[lt]{\lineheight{1.25}\smash{\begin{tabular}[t]{l}{\SVGmathsize $\alpha_\text{NL}\!=\!0$}\end{tabular}}}}%
    \put(0.03641842,0.41225012){\color[rgb]{0,0,0}\makebox(0,0)[lt]{\lineheight{1.25}\smash{\begin{tabular}[t]{l}{\SVGunitsize 0.4}\end{tabular}}}}%
    \put(0.03641842,0.48897167){\color[rgb]{0,0,0}\makebox(0,0)[lt]{\lineheight{1.25}\smash{\begin{tabular}[t]{l}{\SVGunitsize 0.5}\end{tabular}}}}%
    \put(0.03641842,0.56621346){\color[rgb]{0,0,0}\makebox(0,0)[lt]{\lineheight{1.25}\smash{\begin{tabular}[t]{l}{\SVGunitsize 0.6}\end{tabular}}}}%
    \put(0.03641842,0.23368136){\color[rgb]{0,0,0}\makebox(0,0)[lt]{\lineheight{1.25}\smash{\begin{tabular}[t]{l}\SVGunitsize 0.4\end{tabular}}}}%
    \put(0.03641842,0.31040298){\color[rgb]{0,0,0}\makebox(0,0)[lt]{\lineheight{1.25}\smash{\begin{tabular}[t]{l}{\SVGunitsize 0.5}\end{tabular}}}}%
    \put(0.03641842,0.38764479){\color[rgb]{0,0,0}\makebox(0,0)[lt]{\lineheight{1.25}\smash{\begin{tabular}[t]{l}{\SVGunitsize 0.6}\end{tabular}}}}%
    \put(0.03641842,0.05487994){\color[rgb]{0,0,0}\makebox(0,0)[lt]{\lineheight{1.25}\smash{\begin{tabular}[t]{l}\SVGunitsize 0.4\end{tabular}}}}%
    \put(0.07416002,0.03074976){\color[rgb]{0,0,0}\makebox(0,0)[lt]{\lineheight{1.25}\smash{\begin{tabular}[t]{l}{\SVGunitsize 0}\end{tabular}}}}%
    \put(0.17761099,0.03074976){\color[rgb]{0,0,0}\makebox(0,0)[lt]{\lineheight{1.25}\smash{\begin{tabular}[t]{l}{\SVGunitsize 1}\end{tabular}}}}%
    \put(0.28074617,0.03074976){\color[rgb]{0,0,0}\makebox(0,0)[lt]{\lineheight{1.25}\smash{\begin{tabular}[t]{l}{\SVGunitsize 2}\end{tabular}}}}%
    \put(0.38406426,0.03074976){\color[rgb]{0,0,0}\makebox(0,0)[lt]{\lineheight{1.25}\smash{\begin{tabular}[t]{l}{\SVGunitsize 3}\end{tabular}}}}%
    \put(0.48750337,0.03074976){\color[rgb]{0,0,0}\makebox(0,0)[lt]{\lineheight{1.25}\smash{\begin{tabular}[t]{l}{\SVGunitsize 4}\end{tabular}}}}%
    \put(0.95453854,0.44468467){\color[rgb]{0,0,0}\makebox(0,0)[lt]{\lineheight{1.25}\smash{\begin{tabular}[t]{l}{\SVGunitsize 1}\end{tabular}}}}%
    \put(0.59057473,0.03074976){\color[rgb]{0,0,0}\makebox(0,0)[lt]{\lineheight{1.25}\smash{\begin{tabular}[t]{l}{\SVGunitsize 5}\end{tabular}}}}%
    \put(0.69394867,0.03074976){\color[rgb]{0,0,0}\makebox(0,0)[lt]{\lineheight{1.25}\smash{\begin{tabular}[t]{l}{\SVGunitsize 6}\end{tabular}}}}%
    \put(0.79723135,0.03074976){\color[rgb]{0,0,0}\makebox(0,0)[lt]{\lineheight{1.25}\smash{\begin{tabular}[t]{l}{\SVGunitsize 7}\end{tabular}}}}%
    \put(0.95453854,0.17961649){\color[rgb]{0,0,0}\makebox(0,0)[lt]{\lineheight{1.25}\smash{\begin{tabular}[t]{l}{\SVGunitsize -1}\end{tabular}}}}%
    \put(0.95453854,0.24209912){\color[rgb]{0,0,0}\makebox(0,0)[lt]{\lineheight{1.25}\smash{\begin{tabular}[t]{l}{\SVGunitsize -0.5}\end{tabular}}}}%
    \put(0.95453854,0.31157546){\color[rgb]{0,0,0}\makebox(0,0)[lt]{\lineheight{1.25}\smash{\begin{tabular}[t]{l}{\SVGunitsize 0}\end{tabular}}}}%
    \put(0.95453854,0.38108841){\color[rgb]{0,0,0}\makebox(0,0)[lt]{\lineheight{1.25}\smash{\begin{tabular}[t]{l}{\SVGunitsize 0.5}\end{tabular}}}}%
    \put(0.9005528,0.03074976){\color[rgb]{0,0,0}\makebox(0,0)[lt]{\lineheight{1.25}\smash{\begin{tabular}[t]{l}{\SVGunitsize 8}\end{tabular}}}}%
    \put(0.03641842,0.13160147){\color[rgb]{0,0,0}\makebox(0,0)[lt]{\lineheight{1.25}\smash{\begin{tabular}[t]{l}\SVGunitsize 0.5\end{tabular}}}}%
    \put(0.02034934,0.0981503){\color[rgb]{0,0,0}\rotatebox{90}{\makebox(0,0)[lt]{\lineheight{1.25}\smash{\begin{tabular}[t]{l}\SVGlabelsize $r$ (a.u.)\end{tabular}}}}}%
    \put(0.45716504,0.00338913){\color[rgb]{0,0,0}\makebox(0,0)[lt]{\lineheight{1.25}\smash{\begin{tabular}[t]{l}\SVGlabelsize $t$ (a.u.)\end{tabular}}}}%
    \put(0.90774912,0.47716422){\color[rgb]{0,0,0}\makebox(0,0)[lt]{\lineheight{1.25}\smash{\begin{tabular}[t]{l}{\SVGlabelsize $\text{Re}[\psi]$/$a_0$}\end{tabular}}}}%
    \put(0.02034934,0.27695172){\color[rgb]{0,0,0}\rotatebox{90}{\makebox(0,0)[lt]{\lineheight{1.25}\smash{\begin{tabular}[t]{l}\SVGlabelsize $r$ (a.u.)\end{tabular}}}}}%
    \put(0.02034934,0.45552051){\color[rgb]{0,0,0}\rotatebox{90}{\makebox(0,0)[lt]{\lineheight{1.25}\smash{\begin{tabular}[t]{l} \SVGlabelsize $r$ (a.u.)\end{tabular}}}}}%
    \put(0.03641842,0.20662161){\color[rgb]{0,0,0}\makebox(0,0)[lt]{\lineheight{1.25}\smash{\begin{tabular}[t]{l}\SVGunitsize 0.6\end{tabular}}}}%
    \put(0,0){\includegraphics[width=\unitlength,page=10]{Linear-Dispersion-Comparison-2-v2-altered-edge.pdf}}%
  \end{picture}%
\endgroup%

%% file: images-theory/Nonlinear-Term-Comparison-alter-v2.pdf_tex
\begingroup%
  \makeatletter%
  \providecommand\color[2][]{%
    \errmessage{(Inkscape) Color is used for the text in Inkscape, but the package 'color.sty' is not loaded}%
    \renewcommand\color[2][]{}%
  }%
  \providecommand\transparent[1]{%
    \errmessage{(Inkscape) Transparency is used (non-zero) for the text in Inkscape, but the package 'transparent.sty' is not loaded}%
    \renewcommand\transparent[1]{}%
  }%
  \providecommand\rotatebox[2]{#2}%
  \newcommand*\fsize{\dimexpr\f@size pt\relax}%
  \newcommand*\lineheight[1]{\fontsize{\fsize}{#1\fsize}\selectfont}%
  \ifx\svgwidth\undefined%
    \setlength{\unitlength}{1350.29758243bp}%
    \ifx\svgscale\undefined%
      \relax%
    \else%
      \setlength{\unitlength}{\unitlength * \real{\svgscale}}%
    \fi%
  \else%
    \setlength{\unitlength}{\svgwidth}%
  \fi%
  \global\let\svgwidth\undefined%
  \global\let\svgscale\undefined%
  \makeatother%
  \begin{picture}(1,0.4044838)%
    \lineheight{1}%
    \setlength\tabcolsep{0pt}%
    \put(0,0){\includegraphics[width=\unitlength,page=1]{Nonlinear-Term-Comparison-alter-v2.pdf}}%
    \put(0.08922479,0.36849032){\color[rgb]{0,0,0}\makebox(0,0)[lt]{\lineheight{1.25}\smash{\begin{tabular}[t]{l}{\SVGmathsize $\calG\!=\!0$}\end{tabular}}}}%
    \put(0,0){\includegraphics[width=\unitlength,page=2]{Nonlinear-Term-Comparison-alter-v2.pdf}}%
    \put(0.16785796,0.36849045){\color[rgb]{0,0,0}\makebox(0,0)[lt]{\lineheight{1.25}\smash{\begin{tabular}[t]{l}{\SVGmathsize $\alpha_\text{NL}\!=\!0$}\end{tabular}}}}%
    \put(0,0){\includegraphics[width=\unitlength,page=3]{Nonlinear-Term-Comparison-alter-v2.pdf}}%
    \put(0.2692982,0.36849044){\color[rgb]{0,0,0}\makebox(0,0)[lt]{\lineheight{1.25}\smash{\begin{tabular}[t]{l}{\SVGunitsize no dispersion}\end{tabular}}}}%
    \put(0,0){\includegraphics[width=\unitlength,page=4]{Nonlinear-Term-Comparison-alter-v2.pdf}}%
    \put(0.08924233,0.18988455){\color[rgb]{0,0,0}\makebox(0,0)[lt]{\lineheight{1.25}\smash{\begin{tabular}[t]{l}{\SVGmathsize $\calG\!=\!0$}\end{tabular}}}}%
    \put(0,0){\includegraphics[width=\unitlength,page=5]{Nonlinear-Term-Comparison-alter-v2.pdf}}%
    \put(0.16787545,0.18988453){\color[rgb]{0,0,0}\makebox(0,0)[lt]{\lineheight{1.25}\smash{\begin{tabular}[t]{l}{\SVGmathsize $\alpha_\text{NL}\!>\!0$}\end{tabular}}}}%
    \put(0,0){\includegraphics[width=\unitlength,page=6]{Nonlinear-Term-Comparison-alter-v2.pdf}}%
    \put(0.26931583,0.18988454){\color[rgb]{0,0,0}\makebox(0,0)[lt]{\lineheight{1.25}\smash{\begin{tabular}[t]{l}{\SVGunitsize no dispersion}\end{tabular}}}}%
    \put(0,0){\includegraphics[width=\unitlength,page=7]{Nonlinear-Term-Comparison-alter-v2.pdf}}%
    \put(0.03642425,0.23364148){\color[rgb]{0,0,0}\makebox(0,0)[lt]{\lineheight{1.25}\smash{\begin{tabular}[t]{l}{\SVGunitsize 0.4}\end{tabular}}}}%
    \put(0.03642425,0.31036256){\color[rgb]{0,0,0}\makebox(0,0)[lt]{\lineheight{1.25}\smash{\begin{tabular}[t]{l}{\SVGunitsize 0.5}\end{tabular}}}}%
    \put(0.03642425,0.38760388){\color[rgb]{0,0,0}\makebox(0,0)[lt]{\lineheight{1.25}\smash{\begin{tabular}[t]{l}{\SVGunitsize 0.6}\end{tabular}}}}%
    \put(0.03642425,0.0550738){\color[rgb]{0,0,0}\makebox(0,0)[lt]{\lineheight{1.25}\smash{\begin{tabular}[t]{l}\SVGunitsize 0.4\end{tabular}}}}%
    \put(0.03642425,0.13179497){\color[rgb]{0,0,0}\makebox(0,0)[lt]{\lineheight{1.25}\smash{\begin{tabular}[t]{l}{\SVGunitsize 0.5}\end{tabular}}}}%
    \put(0.03642425,0.2090363){\color[rgb]{0,0,0}\makebox(0,0)[lt]{\lineheight{1.25}\smash{\begin{tabular}[t]{l}{\SVGunitsize 0.6}\end{tabular}}}}%
    \put(0.07416566,0.03075061){\color[rgb]{0,0,0}\makebox(0,0)[lt]{\lineheight{1.25}\smash{\begin{tabular}[t]{l}{\SVGunitsize 0}\end{tabular}}}}%
    \put(0.17761597,0.03075061){\color[rgb]{0,0,0}\makebox(0,0)[lt]{\lineheight{1.25}\smash{\begin{tabular}[t]{l}{\SVGunitsize 1}\end{tabular}}}}%
    \put(0.28075053,0.03075061){\color[rgb]{0,0,0}\makebox(0,0)[lt]{\lineheight{1.25}\smash{\begin{tabular}[t]{l}{\SVGunitsize 2}\end{tabular}}}}%
    \put(0.38406801,0.03075061){\color[rgb]{0,0,0}\makebox(0,0)[lt]{\lineheight{1.25}\smash{\begin{tabular}[t]{l}{\SVGunitsize 3}\end{tabular}}}}%
    \put(0.48750651,0.03075061){\color[rgb]{0,0,0}\makebox(0,0)[lt]{\lineheight{1.25}\smash{\begin{tabular}[t]{l}{\SVGunitsize 4}\end{tabular}}}}%
    \put(0.95453884,0.35383456){\color[rgb]{0,0,0}\makebox(0,0)[lt]{\lineheight{1.25}\smash{\begin{tabular}[t]{l}{\SVGunitsize 1}\end{tabular}}}}%
    \put(0.59057723,0.03075061){\color[rgb]{0,0,0}\makebox(0,0)[lt]{\lineheight{1.25}\smash{\begin{tabular}[t]{l}{\SVGunitsize 5}\end{tabular}}}}%
    \put(0.69395055,0.03075061){\color[rgb]{0,0,0}\makebox(0,0)[lt]{\lineheight{1.25}\smash{\begin{tabular}[t]{l}{\SVGunitsize 6}\end{tabular}}}}%
    \put(0.79723261,0.03075061){\color[rgb]{0,0,0}\makebox(0,0)[lt]{\lineheight{1.25}\smash{\begin{tabular}[t]{l}{\SVGunitsize 7}\end{tabular}}}}%
    \put(0.95453884,0.08876761){\color[rgb]{0,0,0}\makebox(0,0)[lt]{\lineheight{1.25}\smash{\begin{tabular}[t]{l}{\SVGunitsize -1}\end{tabular}}}}%
    \put(0.95453884,0.15124985){\color[rgb]{0,0,0}\makebox(0,0)[lt]{\lineheight{1.25}\smash{\begin{tabular}[t]{l}{\SVGunitsize -0.5}\end{tabular}}}}%
    \put(0.95453884,0.22072581){\color[rgb]{0,0,0}\makebox(0,0)[lt]{\lineheight{1.25}\smash{\begin{tabular}[t]{l}{\SVGunitsize 0}\end{tabular}}}}%
    \put(0.95453884,0.29023831){\color[rgb]{0,0,0}\makebox(0,0)[lt]{\lineheight{1.25}\smash{\begin{tabular}[t]{l}{\SVGunitsize 0.5}\end{tabular}}}}%
    \put(0.90055343,0.03075061){\color[rgb]{0,0,0}\makebox(0,0)[lt]{\lineheight{1.25}\smash{\begin{tabular}[t]{l}{\SVGunitsize 8}\end{tabular}}}}%
    \put(0.45716834,0.00338992){\color[rgb]{0,0,0}\makebox(0,0)[lt]{\lineheight{1.25}\smash{\begin{tabular}[t]{l}\SVGlabelsize $t$ (a.u.)\end{tabular}}}}%
    \put(0.90774968,0.38631337){\color[rgb]{0,0,0}\makebox(0,0)[lt]{\lineheight{1.25}\smash{\begin{tabular}[t]{l}{\SVGlabelsize $\text{Re}[\psi]$/$a_0$}\end{tabular}}}}%
    \put(0.02035525,0.09834389){\color[rgb]{0,0,0}\rotatebox{90}{\makebox(0,0)[lt]{\lineheight{1.25}\smash{\begin{tabular}[t]{l}\SVGlabelsize $r$ (a.u.)\end{tabular}}}}}%
    \put(0.02035525,0.27691161){\color[rgb]{0,0,0}\rotatebox{90}{\makebox(0,0)[lt]{\lineheight{1.25}\smash{\begin{tabular}[t]{l} \SVGlabelsize $r$ (a.u.)\end{tabular}}}}}%
  \end{picture}%
\endgroup%

%% file: images-theory/MI-breather-new-with-FFT4.pdf_tex
\begingroup%
  \makeatletter%
  \providecommand\color[2][]{%
    \errmessage{(Inkscape) Color is used for the text in Inkscape, but the package 'color.sty' is not loaded}%
    \renewcommand\color[2][]{}%
  }%
  \providecommand\transparent[1]{%
    \errmessage{(Inkscape) Transparency is used (non-zero) for the text in Inkscape, but the package 'transparent.sty' is not loaded}%
    \renewcommand\transparent[1]{}%
  }%
  \providecommand\rotatebox[2]{#2}%
  \newcommand*\fsize{\dimexpr\f@size pt\relax}%
  \newcommand*\lineheight[1]{\fontsize{\fsize}{#1\fsize}\selectfont}%
  \ifx\svgwidth\undefined%
    \setlength{\unitlength}{1352.02807617bp}%
    \ifx\svgscale\undefined%
      \relax%
    \else%
      \setlength{\unitlength}{\unitlength * \real{\svgscale}}%
    \fi%
  \else%
    \setlength{\unitlength}{\svgwidth}%
  \fi%
  \global\let\svgwidth\undefined%
  \global\let\svgscale\undefined%
  \makeatother%
  \begin{picture}(1,0.62966489)%
    \lineheight{1}%
    \setlength\tabcolsep{0pt}%
    \put(0,0){\includegraphics[width=\unitlength,page=1]{MI-breather-new-with-FFT4.pdf}}%
    \put(0.03637157,0.42340499){\color[rgb]{0,0,0}\makebox(0,0)[lt]{\lineheight{1.25}\smash{\begin{tabular}[t]{l}\SVGunitsize 0.4\end{tabular}}}}%
    \put(0.06962687,0.39930586){\color[rgb]{0,0,0}\makebox(0,0)[lt]{\lineheight{1.25}\smash{\begin{tabular}[t]{l}{\SVGunitsize 0}\end{tabular}}}}%
    \put(0.17294485,0.39930586){\color[rgb]{0,0,0}\makebox(0,0)[lt]{\lineheight{1.25}\smash{\begin{tabular}[t]{l}{\SVGunitsize 1}\end{tabular}}}}%
    \put(0.27594734,0.39930586){\color[rgb]{0,0,0}\makebox(0,0)[lt]{\lineheight{1.25}\smash{\begin{tabular}[t]{l}{\SVGunitsize 2}\end{tabular}}}}%
    \put(0.37913257,0.39930586){\color[rgb]{0,0,0}\makebox(0,0)[lt]{\lineheight{1.25}\smash{\begin{tabular}[t]{l}{\SVGunitsize 3}\end{tabular}}}}%
    \put(0.48243869,0.39930586){\color[rgb]{0,0,0}\makebox(0,0)[lt]{\lineheight{1.25}\smash{\begin{tabular}[t]{l}{\SVGunitsize 4}\end{tabular}}}}%
    \put(0.5853774,0.39930586){\color[rgb]{0,0,0}\makebox(0,0)[lt]{\lineheight{1.25}\smash{\begin{tabular}[t]{l}{\SVGunitsize 5}\end{tabular}}}}%
    \put(0.68861845,0.39930586){\color[rgb]{0,0,0}\makebox(0,0)[lt]{\lineheight{1.25}\smash{\begin{tabular}[t]{l}{\SVGunitsize 6}\end{tabular}}}}%
    \put(0.7917683,0.39930586){\color[rgb]{0,0,0}\makebox(0,0)[lt]{\lineheight{1.25}\smash{\begin{tabular}[t]{l}{\SVGunitsize 7}\end{tabular}}}}%
    \put(0.89495691,0.39930586){\color[rgb]{0,0,0}\makebox(0,0)[lt]{\lineheight{1.25}\smash{\begin{tabular}[t]{l}{\SVGunitsize 8}\end{tabular}}}}%
    \put(0,0){\includegraphics[width=\unitlength,page=2]{MI-breather-new-with-FFT4.pdf}}%
    \put(0.03637158,0.50002786){\color[rgb]{0,0,0}\makebox(0,0)[lt]{\lineheight{1.25}\smash{\begin{tabular}[t]{l}\SVGunitsize 0.5\end{tabular}}}}%
    \put(0,0){\includegraphics[width=\unitlength,page=3]{MI-breather-new-with-FFT4.pdf}}%
    \put(0.2680147,0.55820224){\color[rgb]{0,0,0}\makebox(0,0)[lt]{\lineheight{1.25}\smash{\begin{tabular}[t]{l}{\SVGunitsize anomalous disp.}\end{tabular}}}}%
    \put(0,0){\includegraphics[width=\unitlength,page=4]{MI-breather-new-with-FFT4.pdf}}%
    \put(0.08911449,0.55820223){\color[rgb]{0,0,0}\makebox(0,0)[lt]{\lineheight{1.25}\smash{\begin{tabular}[t]{l}{\SVGmathsize $\calG\!>\!0$}\end{tabular}}}}%
    \put(0,0){\includegraphics[width=\unitlength,page=5]{MI-breather-new-with-FFT4.pdf}}%
    \put(0.16764701,0.55820223){\color[rgb]{0,0,0}\makebox(0,0)[lt]{\lineheight{1.25}\smash{\begin{tabular}[t]{l}{\SVGmathsize $\alpha_\text{NL}\!>\!0$}\end{tabular}}}}%
    \put(0.02032316,0.46661971){\color[rgb]{0,0,0}\rotatebox{90}{\makebox(0,0)[lt]{\lineheight{1.25}\smash{\begin{tabular}[t]{l}\SVGlabelsize $r$ (a.u.)\end{tabular}}}}}%
    \put(0.45657721,0.37198042){\color[rgb]{0,0,0}\makebox(0,0)[lt]{\lineheight{1.25}\smash{\begin{tabular}[t]{l}\SVGlabelsize $t$ (a.u.)\end{tabular}}}}%
    \put(0.90658182,0.61280653){\color[rgb]{0,0,0}\makebox(0,0)[lt]{\lineheight{1.25}\smash{\begin{tabular}[t]{l}{\SVGlabelsize $\text{Re}[\psi]$/$a_0$}\end{tabular}}}}%
    \put(0.03637157,0.57495155){\color[rgb]{0,0,0}\makebox(0,0)[lt]{\lineheight{1.25}\smash{\begin{tabular}[t]{l}\SVGunitsize 0.6\end{tabular}}}}%
    \put(0,0){\includegraphics[width=\unitlength,page=6]{MI-breather-new-with-FFT4.pdf}}%
    \put(0.95331106,0.55787478){\color[rgb]{0,0,0}\makebox(0,0)[lt]{\lineheight{1.25}\smash{\begin{tabular}[t]{l}{\SVGunitsize 2}\end{tabular}}}}%
    \put(0,0){\includegraphics[width=\unitlength,page=7]{MI-breather-new-with-FFT4.pdf}}%
    \put(0.95331106,0.5233211){\color[rgb]{0,0,0}\makebox(0,0)[lt]{\lineheight{1.25}\smash{\begin{tabular}[t]{l}{\SVGunitsize 1.5}\end{tabular}}}}%
    \put(0,0){\includegraphics[width=\unitlength,page=8]{MI-breather-new-with-FFT4.pdf}}%
    \put(0.95331106,0.48876742){\color[rgb]{0,0,0}\makebox(0,0)[lt]{\lineheight{1.25}\smash{\begin{tabular}[t]{l}{\SVGunitsize 1}\end{tabular}}}}%
    \put(0,0){\includegraphics[width=\unitlength,page=9]{MI-breather-new-with-FFT4.pdf}}%
    \put(0.95331106,0.45421374){\color[rgb]{0,0,0}\makebox(0,0)[lt]{\lineheight{1.25}\smash{\begin{tabular}[t]{l}{\SVGunitsize 0.5}\end{tabular}}}}%
    \put(0,0){\includegraphics[width=\unitlength,page=10]{MI-breather-new-with-FFT4.pdf}}%
    \put(0.95331106,0.41966006){\color[rgb]{0,0,0}\makebox(0,0)[lt]{\lineheight{1.25}\smash{\begin{tabular}[t]{l}{\SVGunitsize 0}\end{tabular}}}}%
    \put(0,0){\includegraphics[width=\unitlength,page=11]{MI-breather-new-with-FFT4.pdf}}%
    \put(0.04255345,0.05480757){\color[rgb]{0,0,0}\makebox(0,0)[lt]{\lineheight{1.25}\smash{\begin{tabular}[t]{l}\SVGunitsize \phantom{0}0\end{tabular}}}}%
    \put(0.06962687,0.03070847){\color[rgb]{0,0,0}\makebox(0,0)[lt]{\lineheight{1.25}\smash{\begin{tabular}[t]{l}{\SVGunitsize 0}\end{tabular}}}}%
    \put(0.17294483,0.03070847){\color[rgb]{0,0,0}\makebox(0,0)[lt]{\lineheight{1.25}\smash{\begin{tabular}[t]{l}{\SVGunitsize 1}\end{tabular}}}}%
    \put(0.27594734,0.03070847){\color[rgb]{0,0,0}\makebox(0,0)[lt]{\lineheight{1.25}\smash{\begin{tabular}[t]{l}{\SVGunitsize 2}\end{tabular}}}}%
    \put(0.37913254,0.03070847){\color[rgb]{0,0,0}\makebox(0,0)[lt]{\lineheight{1.25}\smash{\begin{tabular}[t]{l}{\SVGunitsize 3}\end{tabular}}}}%
    \put(0.48243866,0.03070847){\color[rgb]{0,0,0}\makebox(0,0)[lt]{\lineheight{1.25}\smash{\begin{tabular}[t]{l}{\SVGunitsize 4}\end{tabular}}}}%
    \put(0.5853774,0.03070847){\color[rgb]{0,0,0}\makebox(0,0)[lt]{\lineheight{1.25}\smash{\begin{tabular}[t]{l}{\SVGunitsize 5}\end{tabular}}}}%
    \put(0.68861845,0.03070847){\color[rgb]{0,0,0}\makebox(0,0)[lt]{\lineheight{1.25}\smash{\begin{tabular}[t]{l}{\SVGunitsize 6}\end{tabular}}}}%
    \put(0.7917683,0.03070847){\color[rgb]{0,0,0}\makebox(0,0)[lt]{\lineheight{1.25}\smash{\begin{tabular}[t]{l}{\SVGunitsize 7}\end{tabular}}}}%
    \put(0.89495691,0.03070847){\color[rgb]{0,0,0}\makebox(0,0)[lt]{\lineheight{1.25}\smash{\begin{tabular}[t]{l}{\SVGunitsize 8}\end{tabular}}}}%
    \put(0.04255345,0.19385599){\color[rgb]{0,0,0}\makebox(0,0)[lt]{\lineheight{1.25}\smash{\begin{tabular}[t]{l}\SVGunitsize 30\end{tabular}}}}%
    \put(0.04255345,0.2662531){\color[rgb]{0,0,0}\makebox(0,0)[lt]{\lineheight{1.25}\smash{\begin{tabular}[t]{l}\SVGunitsize 45\end{tabular}}}}%
    \put(0.04255345,0.12145888){\color[rgb]{0,0,0}\makebox(0,0)[lt]{\lineheight{1.25}\smash{\begin{tabular}[t]{l}\SVGunitsize 15\end{tabular}}}}%
    \put(0,0){\includegraphics[width=\unitlength,page=12]{MI-breather-new-with-FFT4.pdf}}%
    \put(0.02032316,0.2014565){\color[rgb]{0,0,0}\rotatebox{90}{\makebox(0,0)[t]{\lineheight{1.25}\smash{\begin{tabular}[t]{c}\SVGlabelsize $k_r$ (a.u.)\end{tabular}}}}}%
    \put(0.45657718,0.00338303){\color[rgb]{0,0,0}\makebox(0,0)[lt]{\lineheight{1.25}\smash{\begin{tabular}[t]{l}\SVGlabelsize $t$ (a.u.)\end{tabular}}}}%
    \put(0.90658172,0.36957812){\color[rgb]{0,0,0}\makebox(0,0)[lt]{\lineheight{1.25}\smash{\begin{tabular}[t]{l}{\SVGlabelsize $|\mathfrak{F}[\Rea\!\left[\psi\right]]|$ (a.u.)}\end{tabular}}}}%
    \put(0.04255345,0.33172313){\color[rgb]{0,0,0}\makebox(0,0)[lt]{\lineheight{1.25}\smash{\begin{tabular}[t]{l}\SVGunitsize 60\end{tabular}}}}%
    \put(0,0){\includegraphics[width=\unitlength,page=13]{MI-breather-new-with-FFT4.pdf}}%
    \put(0.95331104,0.34233379){\color[rgb]{0,0,0}\makebox(0,0)[lt]{\lineheight{1.25}\smash{\begin{tabular}[t]{l}{\SVGunitsize 1}\end{tabular}}}}%
    \put(0.95331104,0.2699349){\color[rgb]{0,0,0}\makebox(0,0)[lt]{\lineheight{1.25}\smash{\begin{tabular}[t]{l}{\SVGunitsize 0.75}\end{tabular}}}}%
    \put(0,0){\includegraphics[width=\unitlength,page=14]{MI-breather-new-with-FFT4.pdf}}%
    \put(0.95331104,0.19753598){\color[rgb]{0,0,0}\makebox(0,0)[lt]{\lineheight{1.25}\smash{\begin{tabular}[t]{l}{\SVGunitsize 0.5}\end{tabular}}}}%
    \put(0,0){\includegraphics[width=\unitlength,page=15]{MI-breather-new-with-FFT4.pdf}}%
    \put(0.95331104,0.1251371){\color[rgb]{0,0,0}\makebox(0,0)[lt]{\lineheight{1.25}\smash{\begin{tabular}[t]{l}{\SVGunitsize 0.25}\end{tabular}}}}%
    \put(0,0){\includegraphics[width=\unitlength,page=16]{MI-breather-new-with-FFT4.pdf}}%
    \put(0.95331104,0.05273818){\color[rgb]{0,0,0}\makebox(0,0)[lt]{\lineheight{1.25}\smash{\begin{tabular}[t]{l}{\SVGunitsize 0}\end{tabular}}}}%
    \put(0,0){\includegraphics[width=\unitlength,page=17]{MI-breather-new-with-FFT4.pdf}}%
  \end{picture}%
\endgroup%

%% file: images-theory/Growthrate-Curve-paper2.pdf_tex
\begingroup%
  \makeatletter%
  \providecommand\color[2][]{%
    \errmessage{(Inkscape) Color is used for the text in Inkscape, but the package 'color.sty' is not loaded}%
    \renewcommand\color[2][]{}%
  }%
  \providecommand\transparent[1]{%
    \errmessage{(Inkscape) Transparency is used (non-zero) for the text in Inkscape, but the package 'transparent.sty' is not loaded}%
    \renewcommand\transparent[1]{}%
  }%
  \providecommand\rotatebox[2]{#2}%
  \newcommand*\fsize{\dimexpr\f@size pt\relax}%
  \newcommand*\lineheight[1]{\fontsize{\fsize}{#1\fsize}\selectfont}%
  \ifx\svgwidth\undefined%
    \setlength{\unitlength}{953.25004037bp}%
    \ifx\svgscale\undefined%
      \relax%
    \else%
      \setlength{\unitlength}{\unitlength * \real{\svgscale}}%
    \fi%
  \else%
    \setlength{\unitlength}{\svgwidth}%
  \fi%
  \global\let\svgwidth\undefined%
  \global\let\svgscale\undefined%
  \makeatother%
  \begin{picture}(1,0.50118015)%
    \lineheight{1}%
    \setlength\tabcolsep{0pt}%
    \put(0,0){\includegraphics[width=\unitlength,page=1]{Growthrate-Curve-paper2.pdf}}%
    \put(0.49314591,0.41035722){\color[rgb]{0,0,0}\makebox(0,0)[lt]{\lineheight{1.25}\smash{\begin{tabular}[t]{l}\SVGunitsize 1\end{tabular}}}}%
    \put(0.47342072,0.01300491){\color[rgb]{0,0,0}\makebox(0,0)[lt]{\lineheight{1.25}\smash{\begin{tabular}[t]{l}\SVGunitsize 0\end{tabular}}}}%
    \put(0.76881003,0.01300491){\color[rgb]{0,0,0}\makebox(0,0)[lt]{\lineheight{1.25}\smash{\begin{tabular}[t]{l}\SVGunitsize 1\end{tabular}}}}%
    \put(0.85925683,0.01377905){\color[rgb]{0,0,0}\makebox(0,0)[lt]{\lineheight{1.25}\smash{\begin{tabular}[t]{l}\SVGunitsize $\sqrt{\text{2}}$\end{tabular}}}}%
    \put(0.17456318,0.01300491){\color[rgb]{0,0,0}\makebox(0,0)[lt]{\lineheight{1.25}\smash{\begin{tabular}[t]{l}\SVGunitsize -1\end{tabular}}}}%
    \put(0.03326803,0.01377947){\color[rgb]{0,0,0}\makebox(0,0)[lt]{\lineheight{1.25}\smash{\begin{tabular}[t]{l}\SVGunitsize -$\sqrt{\text{2}}$\end{tabular}}}}%
    \put(0.91135351,0.0901125){\color[rgb]{0,0,0}\makebox(0,0)[lt]{\lineheight{1.25}\smash{\begin{tabular}[t]{l}\SVGlabelsize $k_\text{pert}/k_\text{max}$\end{tabular}}}}%
    \put(0.50103825,0.47106614){\color[rgb]{0,0,0}\makebox(0,0)[lt]{\lineheight{1.25}\smash{\begin{tabular}[t]{l}\SVGlabelsize $|\Imm\,\omega_\text{pert}|/\!\left(\alpha_\text{NL} a_0^2\right)$\end{tabular}}}}%
    \put(0,0){\includegraphics[width=\unitlength,page=2]{Growthrate-Curve-paper2.pdf}}%
  \end{picture}%
\endgroup%

%% file: images-simulations/initial-condition-simulations.pdf_tex
\begingroup%
  \makeatletter%
  \providecommand\color[2][]{%
    \errmessage{(Inkscape) Color is used for the text in Inkscape, but the package 'color.sty' is not loaded}%
    \renewcommand\color[2][]{}%
  }%
  \providecommand\transparent[1]{%
    \errmessage{(Inkscape) Transparency is used (non-zero) for the text in Inkscape, but the package 'transparent.sty' is not loaded}%
    \renewcommand\transparent[1]{}%
  }%
  \providecommand\rotatebox[2]{#2}%
  \newcommand*\fsize{\dimexpr\f@size pt\relax}%
  \newcommand*\lineheight[1]{\fontsize{\fsize}{#1\fsize}\selectfont}%
  \ifx\svgwidth\undefined%
    \setlength{\unitlength}{299.29481506bp}%
    \ifx\svgscale\undefined%
      \relax%
    \else%
      \setlength{\unitlength}{\unitlength * \real{\svgscale}}%
    \fi%
  \else%
    \setlength{\unitlength}{\svgwidth}%
  \fi%
  \global\let\svgwidth\undefined%
  \global\let\svgscale\undefined%
  \makeatother%
  \begin{picture}(1,0.70008649)%
    \lineheight{1}%
    \setlength\tabcolsep{0pt}%
    \put(0,0){\includegraphics[width=\unitlength,page=1]{initial-condition-simulations.pdf}}%
    \put(0.14674528,0.66962416){\makebox(0,0)[rt]{\lineheight{1.25}\smash{\begin{tabular}[t]{r}\svgunitsize 1.2\end{tabular}}}}%
    \put(0.1466004,0.57472658){\color[rgb]{0,0,0}\makebox(0,0)[rt]{\lineheight{1.25}\smash{\begin{tabular}[t]{r}\svgunitsize 1\end{tabular}}}}%
    \put(0.14615671,0.47982899){\color[rgb]{0,0,0}\makebox(0,0)[rt]{\lineheight{1.25}\smash{\begin{tabular}[t]{r}\svgunitsize 0.8\end{tabular}}}}%
    \put(0.1460571,0.38493145){\color[rgb]{0,0,0}\makebox(0,0)[rt]{\lineheight{1.25}\smash{\begin{tabular}[t]{r}\svgunitsize 0.6\end{tabular}}}}%
    \put(0.14674528,0.19513632){\color[rgb]{0,0,0}\makebox(0,0)[rt]{\lineheight{1.25}\smash{\begin{tabular}[t]{r}\svgunitsize 0.2\end{tabular}}}}%
    \put(0.14593033,0.2900339){\color[rgb]{0,0,0}\makebox(0,0)[rt]{\lineheight{1.25}\smash{\begin{tabular}[t]{r}\svgunitsize 0.4\end{tabular}}}}%
    \put(0.03636774,0.39694471){\color[rgb]{0,0,0}\rotatebox{90}{\makebox(0,0)[t]{\lineheight{1.25}\smash{\begin{tabular}[t]{c}\svglabelsize $\Rea[\psi]$/$a_0$\end{tabular}}}}}%
    \put(0.55034692,0.00945584){\color[rgb]{0,0,0}\makebox(0,0)[t]{\lineheight{1.25}\smash{\begin{tabular}[t]{c}\svglabelsize $r$/$a_\text{min}$\end{tabular}}}}%
    \put(0.50391635,0.2542689){\color[rgb]{0,0,0}\makebox(0,0)[lt]{\lineheight{1.25}\smash{\begin{tabular}[t]{l}\svglegendsize $p = 1$\end{tabular}}}}%
    \put(0.50391635,0.20572307){\color[rgb]{0,0,0}\makebox(0,0)[lt]{\lineheight{1.25}\smash{\begin{tabular}[t]{l}\svglegendsize $p = 4$\end{tabular}}}}%
    \put(0.5039164,0.15717724){\color[rgb]{0,0,0}\makebox(0,0)[lt]{\lineheight{1.25}\smash{\begin{tabular}[t]{l}\svglegendsize constant backgr.\end{tabular}}}}%
    \put(0.14612048,0.10023876){\color[rgb]{0,0,0}\makebox(0,0)[rt]{\lineheight{1.25}\smash{\begin{tabular}[t]{r}\svgunitsize 0\end{tabular}}}}%
    \put(0.16303295,0.06436022){\color[rgb]{0,0,0}\makebox(0,0)[t]{\lineheight{1.25}\smash{\begin{tabular}[t]{c}\svgunitsize 0\end{tabular}}}}%
    \put(0.3127184,0.06436022){\color[rgb]{0,0,0}\makebox(0,0)[t]{\lineheight{1.25}\smash{\begin{tabular}[t]{c}\svgunitsize 0.2\end{tabular}}}}%
    \put(0.47062583,0.06436022){\color[rgb]{0,0,0}\makebox(0,0)[t]{\lineheight{1.25}\smash{\begin{tabular}[t]{c}\svgunitsize 0.4\end{tabular}}}}%
    \put(0.62934816,0.06436022){\color[rgb]{0,0,0}\makebox(0,0)[t]{\lineheight{1.25}\smash{\begin{tabular}[t]{c}\svgunitsize 0.6\end{tabular}}}}%
    \put(0.78794374,0.06436022){\color[rgb]{0,0,0}\makebox(0,0)[t]{\lineheight{1.25}\smash{\begin{tabular}[t]{c}\svgunitsize 0.8\end{tabular}}}}%
    \put(0.93759303,0.06436022){\color[rgb]{0,0,0}\makebox(0,0)[t]{\lineheight{1.25}\smash{\begin{tabular}[t]{c}\svgunitsize 1\end{tabular}}}}%
  \end{picture}%
\endgroup%

%% file: images-simulations/first-comparison/196-k0-Comparison.pdf_tex
\begingroup%
  \makeatletter%
  \providecommand\color[2][]{%
    \errmessage{(Inkscape) Color is used for the text in Inkscape, but the package 'color.sty' is not loaded}%
    \renewcommand\color[2][]{}%
  }%
  \providecommand\transparent[1]{%
    \errmessage{(Inkscape) Transparency is used (non-zero) for the text in Inkscape, but the package 'transparent.sty' is not loaded}%
    \renewcommand\transparent[1]{}%
  }%
  \providecommand\rotatebox[2]{#2}%
  \newcommand*\fsize{\dimexpr\f@size pt\relax}%
  \newcommand*\lineheight[1]{\fontsize{\fsize}{#1\fsize}\selectfont}%
  \ifx\svgwidth\undefined%
    \setlength{\unitlength}{654.37339783bp}%
    \ifx\svgscale\undefined%
      \relax%
    \else%
      \setlength{\unitlength}{\unitlength * \real{\svgscale}}%
    \fi%
  \else%
    \setlength{\unitlength}{\svgwidth}%
  \fi%
  \global\let\svgwidth\undefined%
  \global\let\svgscale\undefined%
  \makeatother%
  \begin{picture}(1,0.31395694)%
    \lineheight{1}%
    \setlength\tabcolsep{0pt}%
    \put(0,0){\includegraphics[width=\unitlength,page=1]{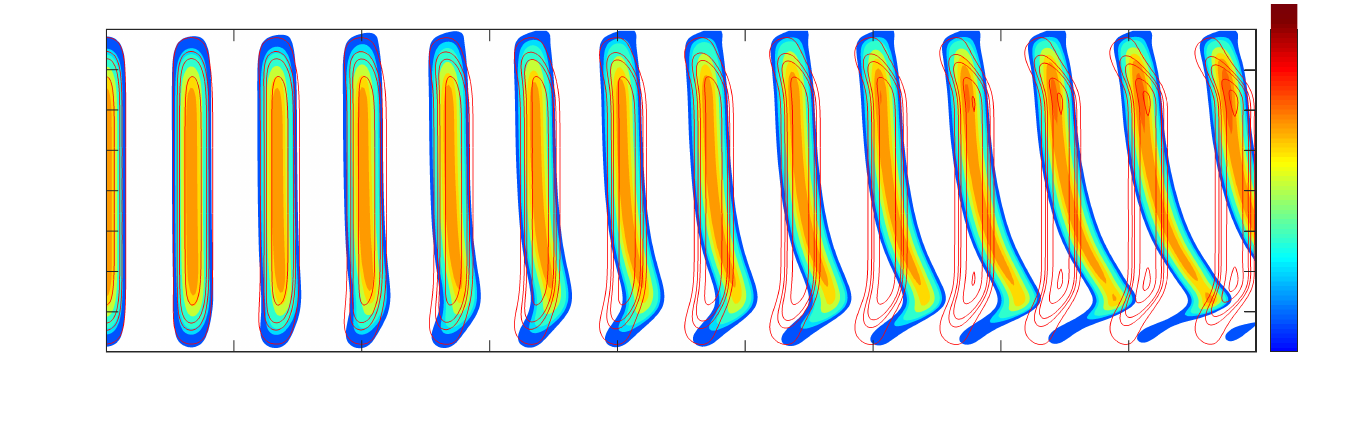}}%
    \put(0.06900446,0.28619287){\color[rgb]{0,0,0}\makebox(0,0)[rt]{\lineheight{1.25}\smash{\begin{tabular}[t]{r}\svgunitsize 1\end{tabular}}}}%
    \put(0.06885848,0.22714003){\color[rgb]{0,0,0}\makebox(0,0)[rt]{\lineheight{1.25}\smash{\begin{tabular}[t]{r}\svgunitsize 0.8\end{tabular}}}}%
    \put(0.0688257,0.16808718){\color[rgb]{0,0,0}\makebox(0,0)[rt]{\lineheight{1.25}\smash{\begin{tabular}[t]{r}\svgunitsize 0.6\end{tabular}}}}%
    \put(0.06878398,0.1090343){\color[rgb]{0,0,0}\makebox(0,0)[rt]{\lineheight{1.25}\smash{\begin{tabular}[t]{r}\svgunitsize 0.4\end{tabular}}}}%
    \put(0.06905213,0.05004885){\color[rgb]{0,0,0}\makebox(0,0)[rt]{\lineheight{1.25}\smash{\begin{tabular}[t]{r}\svgunitsize 0.2\end{tabular}}}}%
    \put(0.07875069,0.03078308){\color[rgb]{0,0,0}\makebox(0,0)[t]{\lineheight{1.25}\smash{\begin{tabular}[t]{c}\svgunitsize 0\end{tabular}}}}%
    \put(0.49721293,0.00432485){\color[rgb]{0,0,0}\makebox(0,0)[t]{\lineheight{1.25}\smash{\begin{tabular}[t]{c}\svglabelsize $t$ / $\omega_{ci}^{-1}$\end{tabular}}}}%
    \put(0,0){\includegraphics[width=\unitlength,page=2]{196-k0-Comparison.pdf}}%
    \put(0.90642952,0.30002426){\color[rgb]{0,0,0}\makebox(0,0)[lt]{\lineheight{1.25}\smash{\begin{tabular}[t]{l}\svglabelsize $E_r$ / $a_0$\end{tabular}}}}%
    \put(0.02358262,0.17462148){\color[rgb]{0,0,0}\rotatebox{90}{\makebox(0,0)[t]{\lineheight{1.25}\smash{\begin{tabular}[t]{c}\svglabelsize $r$ / $a_\text{min}$\end{tabular}}}}}%
    \put(0.17186221,0.03078308){\color[rgb]{0,0,0}\makebox(0,0)[t]{\lineheight{1.25}\smash{\begin{tabular}[t]{c}\svgunitsize 1\end{tabular}}}}%
    \put(0.26448387,0.03078308){\color[rgb]{0,0,0}\makebox(0,0)[t]{\lineheight{1.25}\smash{\begin{tabular}[t]{c}\svgunitsize 2\end{tabular}}}}%
    \put(0.35747112,0.03078308){\color[rgb]{0,0,0}\makebox(0,0)[t]{\lineheight{1.25}\smash{\begin{tabular}[t]{c}\svgunitsize 3\end{tabular}}}}%
    \put(0.45048492,0.03078308){\color[rgb]{0,0,0}\makebox(0,0)[t]{\lineheight{1.25}\smash{\begin{tabular}[t]{c}\svgunitsize 4\end{tabular}}}}%
    \put(0.5433057,0.03078308){\color[rgb]{0,0,0}\makebox(0,0)[t]{\lineheight{1.25}\smash{\begin{tabular}[t]{c}\svgunitsize 5\end{tabular}}}}%
    \put(0.6365295,0.03078308){\color[rgb]{0,0,0}\makebox(0,0)[t]{\lineheight{1.25}\smash{\begin{tabular}[t]{c}\svgunitsize 6\end{tabular}}}}%
    \put(0.72914015,0.03078308){\color[rgb]{0,0,0}\makebox(0,0)[t]{\lineheight{1.25}\smash{\begin{tabular}[t]{c}\svgunitsize 7\end{tabular}}}}%
    \put(0.82222869,0.03078308){\color[rgb]{0,0,0}\makebox(0,0)[t]{\lineheight{1.25}\smash{\begin{tabular}[t]{c}\svgunitsize 8\end{tabular}}}}%
    \put(0.91578041,0.03078416){\color[rgb]{0,0,0}\makebox(0,0)[t]{\lineheight{1.25}\smash{\begin{tabular}[t]{c}\svgunitsize 9\end{tabular}}}}%
    \put(0,0){\includegraphics[width=\unitlength,page=3]{196-k0-Comparison.pdf}}%
    \put(0.95946082,0.27906843){\color[rgb]{0,0,0}\makebox(0,0)[lt]{\lineheight{1.25}\smash{\begin{tabular}[t]{l}\svgunitsize 1.4\end{tabular}}}}%
    \put(0,0){\includegraphics[width=\unitlength,page=4]{196-k0-Comparison.pdf}}%
    \put(0.95979151,0.24646452){\color[rgb]{0,0,0}\makebox(0,0)[lt]{\lineheight{1.25}\smash{\begin{tabular}[t]{l}\svgunitsize 1.2\end{tabular}}}}%
    \put(0,0){\includegraphics[width=\unitlength,page=5]{196-k0-Comparison.pdf}}%
    \put(0.95946077,0.08344496){\color[rgb]{0,0,0}\makebox(0,0)[lt]{\lineheight{1.25}\smash{\begin{tabular}[t]{l}\svgunitsize 0.2\end{tabular}}}}%
    \put(0,0){\includegraphics[width=\unitlength,page=6]{196-k0-Comparison.pdf}}%
    \put(0.95946077,0.11604888){\color[rgb]{0,0,0}\makebox(0,0)[lt]{\lineheight{1.25}\smash{\begin{tabular}[t]{l}\svgunitsize 0.4\end{tabular}}}}%
    \put(0,0){\includegraphics[width=\unitlength,page=7]{196-k0-Comparison.pdf}}%
    \put(0.95946077,0.1812567){\color[rgb]{0,0,0}\makebox(0,0)[lt]{\lineheight{1.25}\smash{\begin{tabular}[t]{l}\svgunitsize 0.8\end{tabular}}}}%
    \put(0,0){\includegraphics[width=\unitlength,page=8]{196-k0-Comparison.pdf}}%
    \put(0.95946082,0.21386061){\color[rgb]{0,0,0}\makebox(0,0)[lt]{\lineheight{1.25}\smash{\begin{tabular}[t]{l}\svgunitsize 1\end{tabular}}}}%
    \put(0,0){\includegraphics[width=\unitlength,page=9]{196-k0-Comparison.pdf}}%
    \put(0.95946082,0.14865279){\color[rgb]{0,0,0}\makebox(0,0)[lt]{\lineheight{1.25}\smash{\begin{tabular}[t]{l}\svgunitsize 0.6\end{tabular}}}}%
    \put(0,0){\includegraphics[width=\unitlength,page=10]{196-k0-Comparison.pdf}}%
    \put(0.95946082,0.05084105){\color[rgb]{0,0,0}\makebox(0,0)[lt]{\lineheight{1.25}\smash{\begin{tabular}[t]{l}\svgunitsize 0\end{tabular}}}}%
    \put(0.89341652,0.00511456){\color[rgb]{0,0,0}\makebox(0,0)[t]{\lineheight{1.25}\smash{\begin{tabular}[t]{c}\svgunitsize $\times$10$^\text{4}$\end{tabular}}}}%
    \put(0,0){\includegraphics[width=\unitlength,page=11]{196-k0-Comparison.pdf}}%
    \put(0.14527064,0.25406261){\color[rgb]{0,0,0}\makebox(0,0)[lt]{\lineheight{1.25}\smash{\begin{tabular}[t]{l}\svglabelsize GK Sim.\end{tabular}}}}%
    \put(0,0){\includegraphics[width=\unitlength,page=12]{196-k0-Comparison.pdf}}%
    \put(0.14514149,0.22770469){\color[rgb]{0,0,0}\makebox(0,0)[lt]{\lineheight{1.25}\smash{\begin{tabular}[t]{l}\svglabelsize NLSE Sim.\end{tabular}}}}%
    \put(0,0){\includegraphics[width=\unitlength,page=13]{196-k0-Comparison.pdf}}%
  \end{picture}%
\endgroup%

%% file: images-simulations/first-comparison/196-k10-long-NLSE.pdf_tex
\begingroup%
  \makeatletter%
  \providecommand\color[2][]{%
    \errmessage{(Inkscape) Color is used for the text in Inkscape, but the package 'color.sty' is not loaded}%
    \renewcommand\color[2][]{}%
  }%
  \providecommand\transparent[1]{%
    \errmessage{(Inkscape) Transparency is used (non-zero) for the text in Inkscape, but the package 'transparent.sty' is not loaded}%
    \renewcommand\transparent[1]{}%
  }%
  \providecommand\rotatebox[2]{#2}%
  \newcommand*\fsize{\dimexpr\f@size pt\relax}%
  \newcommand*\lineheight[1]{\fontsize{\fsize}{#1\fsize}\selectfont}%
  \ifx\svgwidth\undefined%
    \setlength{\unitlength}{654.37339783bp}%
    \ifx\svgscale\undefined%
      \relax%
    \else%
      \setlength{\unitlength}{\unitlength * \real{\svgscale}}%
    \fi%
  \else%
    \setlength{\unitlength}{\svgwidth}%
  \fi%
  \global\let\svgwidth\undefined%
  \global\let\svgscale\undefined%
  \makeatother%
  \begin{picture}(1,0.31395694)%
    \lineheight{1}%
    \setlength\tabcolsep{0pt}%
    \put(0,0){\includegraphics[width=\unitlength,page=1]{196-k10-long-NLSE.pdf}}%
    \put(0.06900446,0.28619286){\color[rgb]{0,0,0}\makebox(0,0)[rt]{\lineheight{1.25}\smash{\begin{tabular}[t]{r}\svgunitsize 1\end{tabular}}}}%
    \put(0.06885848,0.22714004){\color[rgb]{0,0,0}\makebox(0,0)[rt]{\lineheight{1.25}\smash{\begin{tabular}[t]{r}\svgunitsize 0.8\end{tabular}}}}%
    \put(0.06882569,0.16808719){\color[rgb]{0,0,0}\makebox(0,0)[rt]{\lineheight{1.25}\smash{\begin{tabular}[t]{r}\svgunitsize 0.6\end{tabular}}}}%
    \put(0.06878398,0.10903432){\color[rgb]{0,0,0}\makebox(0,0)[rt]{\lineheight{1.25}\smash{\begin{tabular}[t]{r}\svgunitsize 0.4\end{tabular}}}}%
    \put(0.06905212,0.05004885){\color[rgb]{0,0,0}\makebox(0,0)[rt]{\lineheight{1.25}\smash{\begin{tabular}[t]{r}\svgunitsize 0.2\end{tabular}}}}%
    \put(0.07875069,0.03078308){\color[rgb]{0,0,0}\makebox(0,0)[t]{\lineheight{1.25}\smash{\begin{tabular}[t]{c}\svgunitsize 0\end{tabular}}}}%
    \put(0.49721292,0.00432485){\color[rgb]{0,0,0}\makebox(0,0)[t]{\lineheight{1.25}\smash{\begin{tabular}[t]{c}\svglabelsize $t$ / $\omega_{ci}^{-1}$\end{tabular}}}}%
    \put(0.90642954,0.30002426){\color[rgb]{0,0,0}\makebox(0,0)[lt]{\lineheight{1.25}\smash{\begin{tabular}[t]{l}\svglabelsize $E_r$ / $a_0$\end{tabular}}}}%
    \put(0.02358261,0.17462149){\color[rgb]{0,0,0}\rotatebox{90}{\makebox(0,0)[t]{\lineheight{1.25}\smash{\begin{tabular}[t]{c}\svglabelsize $r$ / $a_\text{min}$\end{tabular}}}}}%
    \put(0.24685493,0.03078308){\color[rgb]{0,0,0}\makebox(0,0)[t]{\lineheight{1.25}\smash{\begin{tabular}[t]{c}\svgunitsize 0.5\end{tabular}}}}%
    \put(0.41546908,0.03078308){\color[rgb]{0,0,0}\makebox(0,0)[t]{\lineheight{1.25}\smash{\begin{tabular}[t]{c}\svgunitsize 1\end{tabular}}}}%
    \put(0.58294589,0.03078308){\color[rgb]{0,0,0}\makebox(0,0)[t]{\lineheight{1.25}\smash{\begin{tabular}[t]{c}\svgunitsize 1.5\end{tabular}}}}%
    \put(0.75088153,0.03078308){\color[rgb]{0,0,0}\makebox(0,0)[t]{\lineheight{1.25}\smash{\begin{tabular}[t]{c}\svgunitsize 2\end{tabular}}}}%
    \put(0.91873808,0.03078308){\color[rgb]{0,0,0}\makebox(0,0)[t]{\lineheight{1.25}\smash{\begin{tabular}[t]{c}\svgunitsize 2.5\end{tabular}}}}%
    \put(0.95946085,0.27165148){\color[rgb]{0,0,0}\makebox(0,0)[lt]{\lineheight{1.25}\smash{\begin{tabular}[t]{l}\svgunitsize 2.5\end{tabular}}}}%
    \put(0,0){\includegraphics[width=\unitlength,page=2]{196-k10-long-NLSE.pdf}}%
    \put(0.95946085,0.22465044){\color[rgb]{0,0,0}\makebox(0,0)[lt]{\lineheight{1.25}\smash{\begin{tabular}[t]{l}\svgunitsize 2\end{tabular}}}}%
    \put(0.95946085,0.18329037){\color[rgb]{0,0,0}\makebox(0,0)[lt]{\lineheight{1.25}\smash{\begin{tabular}[t]{l}\svgunitsize 1.5\end{tabular}}}}%
    \put(0.95946085,0.13913363){\color[rgb]{0,0,0}\makebox(0,0)[lt]{\lineheight{1.25}\smash{\begin{tabular}[t]{l}\svgunitsize 1\end{tabular}}}}%
    \put(0,0){\includegraphics[width=\unitlength,page=3]{196-k10-long-NLSE.pdf}}%
    \put(0.95946085,0.09499723){\color[rgb]{0,0,0}\makebox(0,0)[lt]{\lineheight{1.25}\smash{\begin{tabular}[t]{l}\svgunitsize 0.5\end{tabular}}}}%
    \put(0.95946085,0.05084103){\color[rgb]{0,0,0}\makebox(0,0)[lt]{\lineheight{1.25}\smash{\begin{tabular}[t]{l}\svgunitsize 0\end{tabular}}}}%
    \put(0.89341655,0.00511453){\color[rgb]{0,0,0}\makebox(0,0)[t]{\lineheight{1.25}\smash{\begin{tabular}[t]{c}\svgunitsize $\times$10$^\text{5}$\end{tabular}}}}%
  \end{picture}%
\endgroup%

%% file: images-simulations/first-comparison/196-k10-long-ORB5.pdf_tex
\begingroup%
  \makeatletter%
  \providecommand\color[2][]{%
    \errmessage{(Inkscape) Color is used for the text in Inkscape, but the package 'color.sty' is not loaded}%
    \renewcommand\color[2][]{}%
  }%
  \providecommand\transparent[1]{%
    \errmessage{(Inkscape) Transparency is used (non-zero) for the text in Inkscape, but the package 'transparent.sty' is not loaded}%
    \renewcommand\transparent[1]{}%
  }%
  \providecommand\rotatebox[2]{#2}%
  \newcommand*\fsize{\dimexpr\f@size pt\relax}%
  \newcommand*\lineheight[1]{\fontsize{\fsize}{#1\fsize}\selectfont}%
  \ifx\svgwidth\undefined%
    \setlength{\unitlength}{654.37339783bp}%
    \ifx\svgscale\undefined%
      \relax%
    \else%
      \setlength{\unitlength}{\unitlength * \real{\svgscale}}%
    \fi%
  \else%
    \setlength{\unitlength}{\svgwidth}%
  \fi%
  \global\let\svgwidth\undefined%
  \global\let\svgscale\undefined%
  \makeatother%
  \begin{picture}(1,0.31395694)%
    \lineheight{1}%
    \setlength\tabcolsep{0pt}%
    \put(0,0){\includegraphics[width=\unitlength,page=1]{196-k10-long-ORB5.pdf}}%
    \put(0.06900447,0.28619287){\color[rgb]{0,0,0}\makebox(0,0)[rt]{\lineheight{1.25}\smash{\begin{tabular}[t]{r}\svgunitsize 1\end{tabular}}}}%
    \put(0.06885848,0.22714003){\color[rgb]{0,0,0}\makebox(0,0)[rt]{\lineheight{1.25}\smash{\begin{tabular}[t]{r}\svgunitsize 0.8\end{tabular}}}}%
    \put(0.06882569,0.16808719){\color[rgb]{0,0,0}\makebox(0,0)[rt]{\lineheight{1.25}\smash{\begin{tabular}[t]{r}\svgunitsize 0.6\end{tabular}}}}%
    \put(0.06878399,0.10903434){\color[rgb]{0,0,0}\makebox(0,0)[rt]{\lineheight{1.25}\smash{\begin{tabular}[t]{r}\svgunitsize 0.4\end{tabular}}}}%
    \put(0.06905212,0.05004885){\color[rgb]{0,0,0}\makebox(0,0)[rt]{\lineheight{1.25}\smash{\begin{tabular}[t]{r}\svgunitsize 0.2\end{tabular}}}}%
    \put(0.0787507,0.03078308){\color[rgb]{0,0,0}\makebox(0,0)[t]{\lineheight{1.25}\smash{\begin{tabular}[t]{c}\svgunitsize 0\end{tabular}}}}%
    \put(0.49721288,0.00432485){\color[rgb]{0,0,0}\makebox(0,0)[t]{\lineheight{1.25}\smash{\begin{tabular}[t]{c}\svglabelsize $t$ / $\omega_{ci}^{-1}$\end{tabular}}}}%
    \put(0.90642954,0.30002426){\color[rgb]{0,0,0}\makebox(0,0)[lt]{\lineheight{1.25}\smash{\begin{tabular}[t]{l}\svglabelsize $E_r$ / $a_0$\end{tabular}}}}%
    \put(0.02358262,0.17462149){\color[rgb]{0,0,0}\rotatebox{90}{\makebox(0,0)[t]{\lineheight{1.25}\smash{\begin{tabular}[t]{c}\svglabelsize $r$ / $a_\text{min}$\end{tabular}}}}}%
    \put(0.24685492,0.03078308){\color[rgb]{0,0,0}\makebox(0,0)[t]{\lineheight{1.25}\smash{\begin{tabular}[t]{c}\svgunitsize 0.5\end{tabular}}}}%
    \put(0.41546908,0.03078308){\color[rgb]{0,0,0}\makebox(0,0)[t]{\lineheight{1.25}\smash{\begin{tabular}[t]{c}\svgunitsize 1\end{tabular}}}}%
    \put(0.58294589,0.03078308){\color[rgb]{0,0,0}\makebox(0,0)[t]{\lineheight{1.25}\smash{\begin{tabular}[t]{c}\svgunitsize 1.5\end{tabular}}}}%
    \put(0.75088153,0.03078308){\color[rgb]{0,0,0}\makebox(0,0)[t]{\lineheight{1.25}\smash{\begin{tabular}[t]{c}\svgunitsize 2\end{tabular}}}}%
    \put(0.91873808,0.03078308){\color[rgb]{0,0,0}\makebox(0,0)[t]{\lineheight{1.25}\smash{\begin{tabular}[t]{c}\svgunitsize 2.5\end{tabular}}}}%
    \put(0.95946091,0.23126336){\color[rgb]{0,0,0}\makebox(0,0)[lt]{\lineheight{1.25}\smash{\begin{tabular}[t]{l}\svgunitsize 1.5\end{tabular}}}}%
    \put(0.95946091,0.17112409){\color[rgb]{0,0,0}\makebox(0,0)[lt]{\lineheight{1.25}\smash{\begin{tabular}[t]{l}\svgunitsize 1\end{tabular}}}}%
    \put(0,0){\includegraphics[width=\unitlength,page=2]{196-k10-long-ORB5.pdf}}%
    \put(0.95946091,0.11094182){\color[rgb]{0,0,0}\makebox(0,0)[lt]{\lineheight{1.25}\smash{\begin{tabular}[t]{l}\svgunitsize 0.5\end{tabular}}}}%
    \put(0.95946091,0.05084103){\color[rgb]{0,0,0}\makebox(0,0)[lt]{\lineheight{1.25}\smash{\begin{tabular}[t]{l}\svgunitsize 0\end{tabular}}}}%
    \put(0.89341661,0.00511453){\color[rgb]{0,0,0}\makebox(0,0)[t]{\lineheight{1.25}\smash{\begin{tabular}[t]{c}\svgunitsize $\times$10$^\text{5}$\end{tabular}}}}%
  \end{picture}%
\endgroup%

%% file: images-simulations/first-comparison/196-k10-long-NLSE-FFT-new2.pdf_tex
\begingroup%
  \makeatletter%
  \providecommand\color[2][]{%
    \errmessage{(Inkscape) Color is used for the text in Inkscape, but the package 'color.sty' is not loaded}%
    \renewcommand\color[2][]{}%
  }%
  \providecommand\transparent[1]{%
    \errmessage{(Inkscape) Transparency is used (non-zero) for the text in Inkscape, but the package 'transparent.sty' is not loaded}%
    \renewcommand\transparent[1]{}%
  }%
  \providecommand\rotatebox[2]{#2}%
  \newcommand*\fsize{\dimexpr\f@size pt\relax}%
  \newcommand*\lineheight[1]{\fontsize{\fsize}{#1\fsize}\selectfont}%
  \ifx\svgwidth\undefined%
    \setlength{\unitlength}{655.95140076bp}%
    \ifx\svgscale\undefined%
      \relax%
    \else%
      \setlength{\unitlength}{\unitlength * \real{\svgscale}}%
    \fi%
  \else%
    \setlength{\unitlength}{\svgwidth}%
  \fi%
  \global\let\svgwidth\undefined%
  \global\let\svgscale\undefined%
  \makeatother%
  \begin{picture}(1,0.79803813)%
    \lineheight{1}%
    \setlength\tabcolsep{0pt}%
    \put(0,0){\includegraphics[width=\unitlength,page=1]{196-k10-long-NLSE-FFT-new2.pdf}}%
    \put(0.0831364,0.1144941){\color[rgb]{0,0,0}\makebox(0,0)[lt]{\lineheight{1.25}\smash{\begin{tabular}[t]{l}\SVGunitsize \phantom{0}0\end{tabular}}}}%
    \put(0.14450434,0.06710845){\color[rgb]{0,0,0}\makebox(0,0)[t]{\lineheight{1.25}\smash{\begin{tabular}[t]{c}{\SVGunitsize 0}\end{tabular}}}}%
    \put(0.03960271,0.41639473){\color[rgb]{0,0,0}\rotatebox{90}{\makebox(0,0)[t]{\lineheight{1.25}\smash{\begin{tabular}[t]{c}\SVGlabelsize $k_r / (2\pi/a_\text{min})$ \end{tabular}}}}}%
    \put(0.51271199,0.01078613){\color[rgb]{0,0,0}\makebox(0,0)[t]{\lineheight{1.25}\smash{\begin{tabular}[t]{c}\SVGlabelsize $t/\omega_{ci}^{-1}$ \end{tabular}}}}%
    \put(0.82498125,0.7632901){\color[rgb]{0,0,0}\makebox(0,0)[lt]{\lineheight{1.25}\smash{\begin{tabular}[t]{l}{\SVGlabelsize $\mathfrak{F}[E_r]$ (a.u.)}\end{tabular}}}}%
    \put(0,0){\includegraphics[width=\unitlength,page=2]{196-k10-long-NLSE-FFT-new2.pdf}}%
    \put(0.980302,0.70713485){\color[rgb]{0,0,0}\makebox(0,0)[lt]{\lineheight{1.25}\smash{\begin{tabular}[t]{l}{\SVGunitsize 1}\end{tabular}}}}%
    \put(0.980302,0.55790839){\color[rgb]{0,0,0}\makebox(0,0)[lt]{\lineheight{1.25}\smash{\begin{tabular}[t]{l}{\SVGunitsize 0.75}\end{tabular}}}}%
    \put(0,0){\includegraphics[width=\unitlength,page=3]{196-k10-long-NLSE-FFT-new2.pdf}}%
    \put(0.980302,0.40868184){\color[rgb]{0,0,0}\makebox(0,0)[lt]{\lineheight{1.25}\smash{\begin{tabular}[t]{l}{\SVGunitsize 0.5}\end{tabular}}}}%
    \put(0,0){\includegraphics[width=\unitlength,page=4]{196-k10-long-NLSE-FFT-new2.pdf}}%
    \put(0.980302,0.25945524){\color[rgb]{0,0,0}\makebox(0,0)[lt]{\lineheight{1.25}\smash{\begin{tabular}[t]{l}{\SVGunitsize 0.25}\end{tabular}}}}%
    \put(0,0){\includegraphics[width=\unitlength,page=5]{196-k10-long-NLSE-FFT-new2.pdf}}%
    \put(0.980302,0.11022861){\color[rgb]{0,0,0}\makebox(0,0)[lt]{\lineheight{1.25}\smash{\begin{tabular}[t]{l}{\SVGunitsize 0}\end{tabular}}}}%
    \put(0.28996838,0.06710845){\color[rgb]{0,0,0}\makebox(0,0)[t]{\lineheight{1.25}\smash{\begin{tabular}[t]{c}{\SVGunitsize 0.5}\end{tabular}}}}%
    \put(0.43539708,0.06710845){\color[rgb]{0,0,0}\makebox(0,0)[t]{\lineheight{1.25}\smash{\begin{tabular}[t]{c}{\SVGunitsize 1}\end{tabular}}}}%
    \put(0.58089851,0.06710845){\color[rgb]{0,0,0}\makebox(0,0)[t]{\lineheight{1.25}\smash{\begin{tabular}[t]{c}{\SVGunitsize 1.5}\end{tabular}}}}%
    \put(0.72642965,0.06710845){\color[rgb]{0,0,0}\makebox(0,0)[t]{\lineheight{1.25}\smash{\begin{tabular}[t]{c}{\SVGunitsize 2}\end{tabular}}}}%
    \put(0.87197871,0.06710845){\color[rgb]{0,0,0}\makebox(0,0)[t]{\lineheight{1.25}\smash{\begin{tabular}[t]{c}{\SVGunitsize 2.5}\end{tabular}}}}%
    \put(0.0831364,0.24422323){\color[rgb]{0,0,0}\makebox(0,0)[lt]{\lineheight{1.25}\smash{\begin{tabular}[t]{l}\SVGunitsize 10\end{tabular}}}}%
    \put(0.0831364,0.37393272){\color[rgb]{0,0,0}\makebox(0,0)[lt]{\lineheight{1.25}\smash{\begin{tabular}[t]{l}\SVGunitsize 20\end{tabular}}}}%
    \put(0.0831364,0.5036817){\color[rgb]{0,0,0}\makebox(0,0)[lt]{\lineheight{1.25}\smash{\begin{tabular}[t]{l}\SVGunitsize 30\end{tabular}}}}%
    \put(0.0831364,0.63345151){\color[rgb]{0,0,0}\makebox(0,0)[lt]{\lineheight{1.25}\smash{\begin{tabular}[t]{l}\SVGunitsize 40\end{tabular}}}}%
    \put(0,0){\includegraphics[width=\unitlength,page=6]{196-k10-long-NLSE-FFT-new2.pdf}}%
    \put(0.87417065,0.00810592){\color[rgb]{0,0,0}\makebox(0,0)[t]{\lineheight{1.25}\smash{\begin{tabular}[t]{c}\svgunitsize $\times$10$^\text{5}$\end{tabular}}}}%
  \end{picture}%
\endgroup%

%% file: images-simulations/first-comparison/196-k10-long-ORB5-FFT-new2.pdf_tex
\begingroup%
  \makeatletter%
  \providecommand\color[2][]{%
    \errmessage{(Inkscape) Color is used for the text in Inkscape, but the package 'color.sty' is not loaded}%
    \renewcommand\color[2][]{}%
  }%
  \providecommand\transparent[1]{%
    \errmessage{(Inkscape) Transparency is used (non-zero) for the text in Inkscape, but the package 'transparent.sty' is not loaded}%
    \renewcommand\transparent[1]{}%
  }%
  \providecommand\rotatebox[2]{#2}%
  \newcommand*\fsize{\dimexpr\f@size pt\relax}%
  \newcommand*\lineheight[1]{\fontsize{\fsize}{#1\fsize}\selectfont}%
  \ifx\svgwidth\undefined%
    \setlength{\unitlength}{655.95140076bp}%
    \ifx\svgscale\undefined%
      \relax%
    \else%
      \setlength{\unitlength}{\unitlength * \real{\svgscale}}%
    \fi%
  \else%
    \setlength{\unitlength}{\svgwidth}%
  \fi%
  \global\let\svgwidth\undefined%
  \global\let\svgscale\undefined%
  \makeatother%
  \begin{picture}(1,0.79803813)%
    \lineheight{1}%
    \setlength\tabcolsep{0pt}%
    \put(0,0){\includegraphics[width=\unitlength,page=1]{196-k10-long-ORB5-FFT-new2.pdf}}%
    \put(0.08313648,0.1144941){\color[rgb]{0,0,0}\makebox(0,0)[lt]{\lineheight{1.25}\smash{\begin{tabular}[t]{l}\SVGunitsize \phantom{0}0\end{tabular}}}}%
    \put(0.14450438,0.06710845){\color[rgb]{0,0,0}\makebox(0,0)[t]{\lineheight{1.25}\smash{\begin{tabular}[t]{c}{\SVGunitsize 0}\end{tabular}}}}%
    \put(0.03960276,0.41639473){\color[rgb]{0,0,0}\rotatebox{90}{\makebox(0,0)[t]{\lineheight{1.25}\smash{\begin{tabular}[t]{c}\SVGlabelsize $k_r / (2\pi/a_\text{min})$ \end{tabular}}}}}%
    \put(0.51271202,0.01078613){\color[rgb]{0,0,0}\makebox(0,0)[t]{\lineheight{1.25}\smash{\begin{tabular}[t]{c}\SVGlabelsize $t/\omega_{ci}^{-1}$ \end{tabular}}}}%
    \put(0.82498134,0.7632901){\color[rgb]{0,0,0}\makebox(0,0)[lt]{\lineheight{1.25}\smash{\begin{tabular}[t]{l}{\SVGlabelsize $\mathfrak{F}[E_r]$ (a.u.)}\end{tabular}}}}%
    \put(0,0){\includegraphics[width=\unitlength,page=2]{196-k10-long-ORB5-FFT-new2.pdf}}%
    \put(0.980302,0.70713489){\color[rgb]{0,0,0}\makebox(0,0)[lt]{\lineheight{1.25}\smash{\begin{tabular}[t]{l}{\SVGunitsize 1}\end{tabular}}}}%
    \put(0.980302,0.55790839){\color[rgb]{0,0,0}\makebox(0,0)[lt]{\lineheight{1.25}\smash{\begin{tabular}[t]{l}{\SVGunitsize 0.75}\end{tabular}}}}%
    \put(0,0){\includegraphics[width=\unitlength,page=3]{196-k10-long-ORB5-FFT-new2.pdf}}%
    \put(0.980302,0.40868184){\color[rgb]{0,0,0}\makebox(0,0)[lt]{\lineheight{1.25}\smash{\begin{tabular}[t]{l}{\SVGunitsize 0.5}\end{tabular}}}}%
    \put(0,0){\includegraphics[width=\unitlength,page=4]{196-k10-long-ORB5-FFT-new2.pdf}}%
    \put(0.980302,0.25945524){\color[rgb]{0,0,0}\makebox(0,0)[lt]{\lineheight{1.25}\smash{\begin{tabular}[t]{l}{\SVGunitsize 0.25}\end{tabular}}}}%
    \put(0,0){\includegraphics[width=\unitlength,page=5]{196-k10-long-ORB5-FFT-new2.pdf}}%
    \put(0.980302,0.11022861){\color[rgb]{0,0,0}\makebox(0,0)[lt]{\lineheight{1.25}\smash{\begin{tabular}[t]{l}{\SVGunitsize 0}\end{tabular}}}}%
    \put(0.28996841,0.06710845){\color[rgb]{0,0,0}\makebox(0,0)[t]{\lineheight{1.25}\smash{\begin{tabular}[t]{c}{\SVGunitsize 0.5}\end{tabular}}}}%
    \put(0.43539705,0.06710845){\color[rgb]{0,0,0}\makebox(0,0)[t]{\lineheight{1.25}\smash{\begin{tabular}[t]{c}{\SVGunitsize 1}\end{tabular}}}}%
    \put(0.58089851,0.06710845){\color[rgb]{0,0,0}\makebox(0,0)[t]{\lineheight{1.25}\smash{\begin{tabular}[t]{c}{\SVGunitsize 1.5}\end{tabular}}}}%
    \put(0.72642965,0.06710845){\color[rgb]{0,0,0}\makebox(0,0)[t]{\lineheight{1.25}\smash{\begin{tabular}[t]{c}{\SVGunitsize 2}\end{tabular}}}}%
    \put(0.87197871,0.06710845){\color[rgb]{0,0,0}\makebox(0,0)[t]{\lineheight{1.25}\smash{\begin{tabular}[t]{c}{\SVGunitsize 2.5}\end{tabular}}}}%
    \put(0.08313648,0.24422323){\color[rgb]{0,0,0}\makebox(0,0)[lt]{\lineheight{1.25}\smash{\begin{tabular}[t]{l}\SVGunitsize 10\end{tabular}}}}%
    \put(0.08313648,0.37393272){\color[rgb]{0,0,0}\makebox(0,0)[lt]{\lineheight{1.25}\smash{\begin{tabular}[t]{l}\SVGunitsize 20\end{tabular}}}}%
    \put(0.08313648,0.50368166){\color[rgb]{0,0,0}\makebox(0,0)[lt]{\lineheight{1.25}\smash{\begin{tabular}[t]{l}\SVGunitsize 30\end{tabular}}}}%
    \put(0.08313648,0.63345149){\color[rgb]{0,0,0}\makebox(0,0)[lt]{\lineheight{1.25}\smash{\begin{tabular}[t]{l}\SVGunitsize 40\end{tabular}}}}%
    \put(0,0){\includegraphics[width=\unitlength,page=6]{196-k10-long-ORB5-FFT-new2.pdf}}%
    \put(0.87410754,0.00810592){\color[rgb]{0,0,0}\makebox(0,0)[t]{\lineheight{1.25}\smash{\begin{tabular}[t]{c}\svgunitsize $\times$10$^\text{5}$\end{tabular}}}}%
  \end{picture}%
\endgroup%

%% file: images-simulations/general-simulation-images/Qiu-damping-Lx375.pdf_tex
\begingroup%
  \makeatletter%
  \providecommand\color[2][]{%
    \errmessage{(Inkscape) Color is used for the text in Inkscape, but the package 'color.sty' is not loaded}%
    \renewcommand\color[2][]{}%
  }%
  \providecommand\transparent[1]{%
    \errmessage{(Inkscape) Transparency is used (non-zero) for the text in Inkscape, but the package 'transparent.sty' is not loaded}%
    \renewcommand\transparent[1]{}%
  }%
  \providecommand\rotatebox[2]{#2}%
  \newcommand*\fsize{\dimexpr\f@size pt\relax}%
  \newcommand*\lineheight[1]{\fontsize{\fsize}{#1\fsize}\selectfont}%
  \ifx\svgwidth\undefined%
    \setlength{\unitlength}{395.4667511bp}%
    \ifx\svgscale\undefined%
      \relax%
    \else%
      \setlength{\unitlength}{\unitlength * \real{\svgscale}}%
    \fi%
  \else%
    \setlength{\unitlength}{\svgwidth}%
  \fi%
  \global\let\svgwidth\undefined%
  \global\let\svgscale\undefined%
  \makeatother%
  \begin{picture}(1,0.71550171)%
    \lineheight{1}%
    \setlength\tabcolsep{0pt}%
    \put(0,0){\includegraphics[width=\unitlength,page=1]{Qiu-damping-Lx375.pdf}}%
    \put(0.09187945,0.65102594){\makebox(0,0)[rt]{\lineheight{1.25}\smash{\begin{tabular}[t]{r}\svgunitsize 0\end{tabular}}}}%
    \put(0.13203423,0.00631176){\color[rgb]{0,0,0}\makebox(0,0)[rt]{\lineheight{1.25}\smash{\begin{tabular}[t]{r}\svgunitsize $\times$10$^\text{-5}$\end{tabular}}}}%
    \put(0.09187789,0.5426097){\color[rgb]{0,0,0}\makebox(0,0)[rt]{\lineheight{1.25}\smash{\begin{tabular}[t]{r}\svgunitsize -1\end{tabular}}}}%
    \put(0.09189476,0.43419343){\color[rgb]{0,0,0}\makebox(0,0)[rt]{\lineheight{1.25}\smash{\begin{tabular}[t]{r}\svgunitsize -2\end{tabular}}}}%
    \put(0.0919335,0.32577736){\color[rgb]{0,0,0}\makebox(0,0)[rt]{\lineheight{1.25}\smash{\begin{tabular}[t]{r}\svgunitsize -3\end{tabular}}}}%
    \put(0.09169902,0.21736113){\color[rgb]{0,0,0}\makebox(0,0)[rt]{\lineheight{1.25}\smash{\begin{tabular}[t]{r}\svgunitsize -4\end{tabular}}}}%
    \put(0.09188372,0.10894487){\color[rgb]{0,0,0}\makebox(0,0)[rt]{\lineheight{1.25}\smash{\begin{tabular}[t]{r}\svgunitsize -5\end{tabular}}}}%
    \put(0.02305416,0.3889064){\color[rgb]{0,0,0}\rotatebox{90}{\makebox(0,0)[t]{\lineheight{1.25}\smash{\begin{tabular}[t]{c}\svglabelsize $\gamma_\text{Qiu}$ / $\omega_{ci}$\end{tabular}}}}}%
    \put(1.00109949,0.58228468){\color[rgb]{0,0,0}\makebox(0,0)[rt]{\lineheight{1.25}\smash{\begin{tabular}[t]{r}\svglabelsize $k_r$ / $\left(\frac{2\pi}{a_\text{min}}\right)$\end{tabular}}}}%
    \put(0.45571788,0.07183746){\color[rgb]{0,0,0}\rotatebox{90}{\makebox(0,0)[lt]{\lineheight{1.25}\smash{\begin{tabular}[t]{l}\textcolor{red}{\svglabelsize $\frac{3}{2} k_r^2 \rho_i^2 = 0.3$}\end{tabular}}}}}%
    \put(0.22063809,0.67456865){\color[rgb]{0,0,0}\makebox(0,0)[t]{\lineheight{1.25}\smash{\begin{tabular}[t]{c}\svgunitsize 5\end{tabular}}}}%
    \put(0.33822065,0.67456865){\color[rgb]{0,0,0}\makebox(0,0)[t]{\lineheight{1.25}\smash{\begin{tabular}[t]{c}\svgunitsize 10\end{tabular}}}}%
    \put(0.45580323,0.67456865){\color[rgb]{0,0,0}\makebox(0,0)[t]{\lineheight{1.25}\smash{\begin{tabular}[t]{c}\svgunitsize 15\end{tabular}}}}%
    \put(0.5733481,0.67456869){\color[rgb]{0,0,0}\makebox(0,0)[t]{\lineheight{1.25}\smash{\begin{tabular}[t]{c}\svgunitsize 20\end{tabular}}}}%
    \put(0.69094606,0.67456865){\color[rgb]{0,0,0}\makebox(0,0)[t]{\lineheight{1.25}\smash{\begin{tabular}[t]{c}\svgunitsize 25\end{tabular}}}}%
    \put(0.80837951,0.67456865){\color[rgb]{0,0,0}\makebox(0,0)[t]{\lineheight{1.25}\smash{\begin{tabular}[t]{c}\svgunitsize 30\end{tabular}}}}%
    \put(0.9261112,0.67456865){\color[rgb]{0,0,0}\makebox(0,0)[t]{\lineheight{1.25}\smash{\begin{tabular}[t]{c}\svgunitsize 35\end{tabular}}}}%
  \end{picture}%
\endgroup%

%% file: images-simulations/simulations-k/Growthrate-scheme-k-8-run196-elfield2.pdf_tex
\begingroup%
  \makeatletter%
  \providecommand\color[2][]{%
    \errmessage{(Inkscape) Color is used for the text in Inkscape, but the package 'color.sty' is not loaded}%
    \renewcommand\color[2][]{}%
  }%
  \providecommand\transparent[1]{%
    \errmessage{(Inkscape) Transparency is used (non-zero) for the text in Inkscape, but the package 'transparent.sty' is not loaded}%
    \renewcommand\transparent[1]{}%
  }%
  \providecommand\rotatebox[2]{#2}%
  \newcommand*\fsize{\dimexpr\f@size pt\relax}%
  \newcommand*\lineheight[1]{\fontsize{\fsize}{#1\fsize}\selectfont}%
  \ifx\svgwidth\undefined%
    \setlength{\unitlength}{654.37339783bp}%
    \ifx\svgscale\undefined%
      \relax%
    \else%
      \setlength{\unitlength}{\unitlength * \real{\svgscale}}%
    \fi%
  \else%
    \setlength{\unitlength}{\svgwidth}%
  \fi%
  \global\let\svgwidth\undefined%
  \global\let\svgscale\undefined%
  \makeatother%
  \begin{picture}(1,0.36209459)%
    \lineheight{1}%
    \setlength\tabcolsep{0pt}%
    \put(0,0){\includegraphics[width=\unitlength,page=1]{Growthrate-scheme-k-8-run196-elfield2.pdf}}%
    \put(0.06903624,0.3344763){\color[rgb]{0,0,0}\makebox(0,0)[rt]{\lineheight{1.25}\smash{\begin{tabular}[t]{r}\svgunitsize 1\end{tabular}}}}%
    \put(0.06889026,0.26341598){\color[rgb]{0,0,0}\makebox(0,0)[rt]{\lineheight{1.25}\smash{\begin{tabular}[t]{r}\svgunitsize 0.8\end{tabular}}}}%
    \put(0.06885748,0.1923556){\color[rgb]{0,0,0}\makebox(0,0)[rt]{\lineheight{1.25}\smash{\begin{tabular}[t]{r}\svgunitsize 0.6\end{tabular}}}}%
    \put(0.06881576,0.12129519){\color[rgb]{0,0,0}\makebox(0,0)[rt]{\lineheight{1.25}\smash{\begin{tabular}[t]{r}\svgunitsize 0.4\end{tabular}}}}%
    \put(0.06908391,0.05023498){\color[rgb]{0,0,0}\makebox(0,0)[rt]{\lineheight{1.25}\smash{\begin{tabular}[t]{r}\svgunitsize 0.2\end{tabular}}}}%
    \put(0.07878247,0.03078309){\color[rgb]{0,0,0}\makebox(0,0)[t]{\lineheight{1.25}\smash{\begin{tabular}[t]{c}\svgunitsize 0\end{tabular}}}}%
    \put(0.14534044,0.30212893){\color[rgb]{0,0,0}\makebox(0,0)[lt]{\lineheight{1.25}\smash{\begin{tabular}[t]{l}\svglabelsize Simulation\end{tabular}}}}%
    \put(0.14534044,0.27583858){\color[rgb]{0,0,0}\makebox(0,0)[lt]{\lineheight{1.25}\smash{\begin{tabular}[t]{l}\svglabelsize Fit Region\end{tabular}}}}%
    \put(0.14534044,0.2495482){\color[rgb]{0,0,0}\makebox(0,0)[lt]{\lineheight{1.25}\smash{\begin{tabular}[t]{l}\svglabelsize Radial Mask\end{tabular}}}}%
    \put(0.49724472,0.00432486){\color[rgb]{0,0,0}\makebox(0,0)[t]{\lineheight{1.25}\smash{\begin{tabular}[t]{c}\svglabelsize $t$ / $\omega_{ci}^{-1}$\end{tabular}}}}%
    \put(0.90646133,0.34816191){\color[rgb]{0,0,0}\makebox(0,0)[lt]{\lineheight{1.25}\smash{\begin{tabular}[t]{l}\svglabelsize $E_r$ / $a_0$\end{tabular}}}}%
    \put(0.02361441,0.20719941){\color[rgb]{0,0,0}\rotatebox{90}{\makebox(0,0)[t]{\lineheight{1.25}\smash{\begin{tabular}[t]{c}\svglabelsize $r$ / $a_\text{min}$\end{tabular}}}}}%
    \put(0.17746505,0.03078309){\color[rgb]{0,0,0}\makebox(0,0)[t]{\lineheight{1.25}\smash{\begin{tabular}[t]{c}\svgunitsize 2\end{tabular}}}}%
    \put(0.27557187,0.03078309){\color[rgb]{0,0,0}\makebox(0,0)[t]{\lineheight{1.25}\smash{\begin{tabular}[t]{c}\svgunitsize 4\end{tabular}}}}%
    \put(0.3740553,0.03078309){\color[rgb]{0,0,0}\makebox(0,0)[t]{\lineheight{1.25}\smash{\begin{tabular}[t]{c}\svgunitsize 6\end{tabular}}}}%
    \put(0.47252784,0.03078309){\color[rgb]{0,0,0}\makebox(0,0)[t]{\lineheight{1.25}\smash{\begin{tabular}[t]{c}\svgunitsize 8\end{tabular}}}}%
    \put(0.57094609,0.03078309){\color[rgb]{0,0,0}\makebox(0,0)[t]{\lineheight{1.25}\smash{\begin{tabular}[t]{c}\svgunitsize 10\end{tabular}}}}%
    \put(0.66962864,0.03078309){\color[rgb]{0,0,0}\makebox(0,0)[t]{\lineheight{1.25}\smash{\begin{tabular}[t]{c}\svgunitsize 12\end{tabular}}}}%
    \put(0.76773546,0.03078309){\color[rgb]{0,0,0}\makebox(0,0)[t]{\lineheight{1.25}\smash{\begin{tabular}[t]{c}\svgunitsize 14\end{tabular}}}}%
    \put(0.86621889,0.03078309){\color[rgb]{0,0,0}\makebox(0,0)[t]{\lineheight{1.25}\smash{\begin{tabular}[t]{c}\svgunitsize 16\end{tabular}}}}%
    \put(0.95949264,0.28746925){\makebox(0,0)[lt]{\lineheight{1.25}\smash{\begin{tabular}[t]{l}\svgunitsize 1\end{tabular}}}}%
    \put(0.95949264,0.23960894){\color[rgb]{0,0,0}\makebox(0,0)[lt]{\lineheight{1.25}\smash{\begin{tabular}[t]{l}\svgunitsize 0.5\end{tabular}}}}%
    \put(0.95949264,0.1917486){\color[rgb]{0,0,0}\makebox(0,0)[lt]{\lineheight{1.25}\smash{\begin{tabular}[t]{l}\svgunitsize 0\end{tabular}}}}%
    \put(0.95949264,0.14388827){\color[rgb]{0,0,0}\makebox(0,0)[lt]{\lineheight{1.25}\smash{\begin{tabular}[t]{l}\svgunitsize -0.5\end{tabular}}}}%
    \put(0.95949264,0.09602798){\color[rgb]{0,0,0}\makebox(0,0)[lt]{\lineheight{1.25}\smash{\begin{tabular}[t]{l}\svgunitsize -1\end{tabular}}}}%
    \put(0.89344828,0.00511457){\color[rgb]{0,0,0}\makebox(0,0)[t]{\lineheight{1.25}\smash{\begin{tabular}[t]{c}\svgunitsize $\times$10$^\text{4}$\end{tabular}}}}%
  \end{picture}%
\endgroup%

%% file: images-simulations/simulations-k/Growthrate-scheme-k-8-run196-fit2.pdf_tex
\begingroup%
  \makeatletter%
  \providecommand\color[2][]{%
    \errmessage{(Inkscape) Color is used for the text in Inkscape, but the package 'color.sty' is not loaded}%
    \renewcommand\color[2][]{}%
  }%
  \providecommand\transparent[1]{%
    \errmessage{(Inkscape) Transparency is used (non-zero) for the text in Inkscape, but the package 'transparent.sty' is not loaded}%
    \renewcommand\transparent[1]{}%
  }%
  \providecommand\rotatebox[2]{#2}%
  \newcommand*\fsize{\dimexpr\f@size pt\relax}%
  \newcommand*\lineheight[1]{\fontsize{\fsize}{#1\fsize}\selectfont}%
  \ifx\svgwidth\undefined%
    \setlength{\unitlength}{654.37339783bp}%
    \ifx\svgscale\undefined%
      \relax%
    \else%
      \setlength{\unitlength}{\unitlength * \real{\svgscale}}%
    \fi%
  \else%
    \setlength{\unitlength}{\svgwidth}%
  \fi%
  \global\let\svgwidth\undefined%
  \global\let\svgscale\undefined%
  \makeatother%
  \begin{picture}(1,0.30489468)%
    \lineheight{1}%
    \setlength\tabcolsep{0pt}%
    \put(0,0){\includegraphics[width=\unitlength,page=1]{Growthrate-scheme-k-8-run196-fit2.pdf}}%
    \put(0.06880204,0.26570916){\color[rgb]{0,0,0}\makebox(0,0)[rt]{\lineheight{1.25}\smash{\begin{tabular}[t]{r}\svgunitsize 10$^\text{0}$\end{tabular}}}}%
    \put(0.07878256,0.03078302){\color[rgb]{0,0,0}\makebox(0,0)[t]{\lineheight{1.25}\smash{\begin{tabular}[t]{c}\svgunitsize 0\end{tabular}}}}%
    \put(0.49724532,0.00432485){\color[rgb]{0,0,0}\makebox(0,0)[t]{\lineheight{1.25}\smash{\begin{tabular}[t]{c}\svglabelsize $t$ / $\omega_{ci}^{-1}$\end{tabular}}}}%
    \put(0.02361446,0.17546129){\color[rgb]{0,0,0}\rotatebox{90}{\makebox(0,0)[t]{\lineheight{1.25}\smash{\begin{tabular}[t]{c}\svglabelsize $|\mathfrak{F}[E_r](k=k_\text{pert})|$ (a.u.)\end{tabular}}}}}%
    \put(0.17746511,0.03078302){\color[rgb]{0,0,0}\makebox(0,0)[t]{\lineheight{1.25}\smash{\begin{tabular}[t]{c}\svgunitsize 2\end{tabular}}}}%
    \put(0.27557143,0.03078302){\color[rgb]{0,0,0}\makebox(0,0)[t]{\lineheight{1.25}\smash{\begin{tabular}[t]{c}\svgunitsize 4\end{tabular}}}}%
    \put(0.37405486,0.03078302){\color[rgb]{0,0,0}\makebox(0,0)[t]{\lineheight{1.25}\smash{\begin{tabular}[t]{c}\svgunitsize 6\end{tabular}}}}%
    \put(0.47252844,0.03078302){\color[rgb]{0,0,0}\makebox(0,0)[t]{\lineheight{1.25}\smash{\begin{tabular}[t]{c}\svgunitsize 8\end{tabular}}}}%
    \put(0.57094666,0.03078302){\color[rgb]{0,0,0}\makebox(0,0)[t]{\lineheight{1.25}\smash{\begin{tabular}[t]{c}\svgunitsize 10\end{tabular}}}}%
    \put(0.66962921,0.03078302){\color[rgb]{0,0,0}\makebox(0,0)[t]{\lineheight{1.25}\smash{\begin{tabular}[t]{c}\svgunitsize 12\end{tabular}}}}%
    \put(0.76773603,0.03078302){\color[rgb]{0,0,0}\makebox(0,0)[t]{\lineheight{1.25}\smash{\begin{tabular}[t]{c}\svgunitsize 14\end{tabular}}}}%
    \put(0.86621946,0.03078302){\color[rgb]{0,0,0}\makebox(0,0)[t]{\lineheight{1.25}\smash{\begin{tabular}[t]{c}\svgunitsize 16\end{tabular}}}}%
    \put(0.8934586,0.00510019){\color[rgb]{0,0,0}\makebox(0,0)[t]{\lineheight{1.25}\smash{\begin{tabular}[t]{c}\svgunitsize $\times$10$^\text{4}$\end{tabular}}}}%
    \put(0,0){\includegraphics[width=\unitlength,page=2]{Growthrate-scheme-k-8-run196-fit2.pdf}}%
    \put(0.14534051,0.2843567){\color[rgb]{0,0,0}\makebox(0,0)[lt]{\lineheight{1.25}\smash{\begin{tabular}[t]{l}\svglabelsize FFT coeff.\end{tabular}}}}%
    \put(0.14534051,0.2580666){\color[rgb]{0,0,0}\makebox(0,0)[lt]{\lineheight{1.25}\smash{\begin{tabular}[t]{l}\svglabelsize Envelope\end{tabular}}}}%
    \put(0.14534051,0.23177618){\color[rgb]{0,0,0}\makebox(0,0)[lt]{\lineheight{1.25}\smash{\begin{tabular}[t]{l}\svglabelsize Exp. Fit $\gamma_\text{Fit}\!=\!2.27\,\omega_{ci}$ \end{tabular}}}}%
    \put(0.14534051,0.20548575){\color[rgb]{0,0,0}\makebox(0,0)[lt]{\lineheight{1.25}\smash{\begin{tabular}[t]{l}\svglabelsize Analytic $\gamma_\text{MI}\!=\!2.86\,\omega_{ci}$\end{tabular}}}}%
  \end{picture}%
\endgroup%

%% file: images-simulations/simulations-k/196-Lx375-qiufac-1.pdf_tex
\begingroup%
  \makeatletter%
  \providecommand\color[2][]{%
    \errmessage{(Inkscape) Color is used for the text in Inkscape, but the package 'color.sty' is not loaded}%
    \renewcommand\color[2][]{}%
  }%
  \providecommand\transparent[1]{%
    \errmessage{(Inkscape) Transparency is used (non-zero) for the text in Inkscape, but the package 'transparent.sty' is not loaded}%
    \renewcommand\transparent[1]{}%
  }%
  \providecommand\rotatebox[2]{#2}%
  \newcommand*\fsize{\dimexpr\f@size pt\relax}%
  \newcommand*\lineheight[1]{\fontsize{\fsize}{#1\fsize}\selectfont}%
  \ifx\svgwidth\undefined%
    \setlength{\unitlength}{383.12587738bp}%
    \ifx\svgscale\undefined%
      \relax%
    \else%
      \setlength{\unitlength}{\unitlength * \real{\svgscale}}%
    \fi%
  \else%
    \setlength{\unitlength}{\svgwidth}%
  \fi%
  \global\let\svgwidth\undefined%
  \global\let\svgscale\undefined%
  \makeatother%
  \begin{picture}(1,0.94817721)%
    \lineheight{1}%
    \setlength\tabcolsep{0pt}%
    \put(0,0){\includegraphics[width=\unitlength,page=1]{196-Lx375-qiufac-1.pdf}}%
    \put(0.22147162,0.18719188){\color[rgb]{0,0,0}\makebox(0,0)[lt]{\lineheight{1.25}\smash{\begin{tabular}[t]{l}\svglegendsize $\gamma_\text{MI}$\end{tabular}}}}%
    \put(0.22147162,0.13948227){\color[rgb]{0,0,0}\makebox(0,0)[lt]{\lineheight{1.25}\smash{\begin{tabular}[t]{l}\svglegendsize $\gamma_\text{Qiu}$\end{tabular}}}}%
    \put(0.22147162,0.09177249){\color[rgb]{0,0,0}\makebox(0,0)[lt]{\lineheight{1.25}\smash{\begin{tabular}[t]{l}\svglegendsize $\gamma_\text{MI} + \gamma_\text{Qiu}$\end{tabular}}}}%
    \put(0.22147162,0.04383457){\color[rgb]{0,0,0}\makebox(0,0)[lt]{\lineheight{1.25}\smash{\begin{tabular}[t]{l}\svglegendsize Simulation\end{tabular}}}}%
    \put(0.09483882,0.50765588){\color[rgb]{0,0,0}\makebox(0,0)[rt]{\lineheight{1.25}\smash{\begin{tabular}[t]{r}\svgunitsize 1\end{tabular}}}}%
    \put(0.09483882,0.63294106){\color[rgb]{0,0,0}\makebox(0,0)[rt]{\lineheight{1.25}\smash{\begin{tabular}[t]{r}\svgunitsize 2\end{tabular}}}}%
    \put(0.09483882,0.75822623){\color[rgb]{0,0,0}\makebox(0,0)[rt]{\lineheight{1.25}\smash{\begin{tabular}[t]{r}\svgunitsize 3\end{tabular}}}}%
    \put(0.09483882,0.88351143){\color[rgb]{0,0,0}\makebox(0,0)[rt]{\lineheight{1.25}\smash{\begin{tabular}[t]{r}\svgunitsize 4\end{tabular}}}}%
    \put(0.13628712,0.92438039){\color[rgb]{0,0,0}\makebox(0,0)[rt]{\lineheight{1.25}\smash{\begin{tabular}[t]{r}\svgunitsize $\times$10$^\text{-5}$\end{tabular}}}}%
    \put(0.09485458,0.13180026){\color[rgb]{0,0,0}\makebox(0,0)[rt]{\lineheight{1.25}\smash{\begin{tabular}[t]{r}\svgunitsize -2\end{tabular}}}}%
    \put(0.09489475,0.38237057){\color[rgb]{0,0,0}\makebox(0,0)[rt]{\lineheight{1.25}\smash{\begin{tabular}[t]{r}\svgunitsize 0\end{tabular}}}}%
    \put(0.09465257,0.00651506){\color[rgb]{0,0,0}\makebox(0,0)[rt]{\lineheight{1.25}\smash{\begin{tabular}[t]{r}\svgunitsize -3\end{tabular}}}}%
    \put(0.09484311,0.25708545){\color[rgb]{0,0,0}\makebox(0,0)[rt]{\lineheight{1.25}\smash{\begin{tabular}[t]{r}\svgunitsize -1\end{tabular}}}}%
    \put(0.02379676,0.45422159){\color[rgb]{0,0,0}\rotatebox{90}{\makebox(0,0)[t]{\lineheight{1.25}\smash{\begin{tabular}[t]{c}\svglabelsize $\gamma$ / $\omega_{ci}$\end{tabular}}}}}%
    \put(0,0){\includegraphics[width=\unitlength,page=2]{196-Lx375-qiufac-1.pdf}}%
    \put(0.99931791,0.43102607){\color[rgb]{0,0,0}\makebox(0,0)[rt]{\lineheight{1.25}\smash{\begin{tabular}[t]{r}\svglabelsize $k_\text{pert}$ / $\left(\frac{2\pi}{a_\text{min}}\right)$\end{tabular}}}}%
    \put(0.26615589,0.35354247){\color[rgb]{0,0,0}\makebox(0,0)[t]{\lineheight{1.25}\smash{\begin{tabular}[t]{c}\svgunitsize 5\end{tabular}}}}%
    \put(0.58523268,0.35354247){\color[rgb]{0,0,0}\makebox(0,0)[t]{\lineheight{1.25}\smash{\begin{tabular}[t]{c}\svgunitsize 10\end{tabular}}}}%
    \put(0.90462845,0.35354247){\color[rgb]{0,0,0}\makebox(0,0)[t]{\lineheight{1.25}\smash{\begin{tabular}[t]{c}\svgunitsize 15\end{tabular}}}}%
  \end{picture}%
\endgroup%

%% file: images-simulations/simulations-k/196-Lx375-qiufac-25.pdf_tex
\begingroup%
  \makeatletter%
  \providecommand\color[2][]{%
    \errmessage{(Inkscape) Color is used for the text in Inkscape, but the package 'color.sty' is not loaded}%
    \renewcommand\color[2][]{}%
  }%
  \providecommand\transparent[1]{%
    \errmessage{(Inkscape) Transparency is used (non-zero) for the text in Inkscape, but the package 'transparent.sty' is not loaded}%
    \renewcommand\transparent[1]{}%
  }%
  \providecommand\rotatebox[2]{#2}%
  \newcommand*\fsize{\dimexpr\f@size pt\relax}%
  \newcommand*\lineheight[1]{\fontsize{\fsize}{#1\fsize}\selectfont}%
  \ifx\svgwidth\undefined%
    \setlength{\unitlength}{382.60528564bp}%
    \ifx\svgscale\undefined%
      \relax%
    \else%
      \setlength{\unitlength}{\unitlength * \real{\svgscale}}%
    \fi%
  \else%
    \setlength{\unitlength}{\svgwidth}%
  \fi%
  \global\let\svgwidth\undefined%
  \global\let\svgscale\undefined%
  \makeatother%
  \begin{picture}(1,0.94946741)%
    \lineheight{1}%
    \setlength\tabcolsep{0pt}%
    \put(0,0){\includegraphics[width=\unitlength,page=1]{196-Lx375-qiufac-25.pdf}}%
    \put(0.22177298,0.18744664){\color[rgb]{0,0,0}\makebox(0,0)[lt]{\lineheight{1.25}\smash{\begin{tabular}[t]{l}\svglegendsize $\gamma_\text{MI}$\end{tabular}}}}%
    \put(0.22177298,0.13967212){\color[rgb]{0,0,0}\makebox(0,0)[lt]{\lineheight{1.25}\smash{\begin{tabular}[t]{l}\svglegendsize $2.5\!\times\!\gamma_\text{Qiu}$\end{tabular}}}}%
    \put(0.22177298,0.09189742){\color[rgb]{0,0,0}\makebox(0,0)[lt]{\lineheight{1.25}\smash{\begin{tabular}[t]{l}\svglegendsize $\gamma_\text{MI} + 2.5\!\times\!\gamma_\text{Qiu}$\end{tabular}}}}%
    \put(0.22177298,0.04389427){\color[rgb]{0,0,0}\makebox(0,0)[lt]{\lineheight{1.25}\smash{\begin{tabular}[t]{l}\svglegendsize Simulation\end{tabular}}}}%
    \put(0.09496788,0.50834668){\color[rgb]{0,0,0}\makebox(0,0)[rt]{\lineheight{1.25}\smash{\begin{tabular}[t]{r}\svgunitsize 1\end{tabular}}}}%
    \put(0.09496788,0.63380233){\color[rgb]{0,0,0}\makebox(0,0)[rt]{\lineheight{1.25}\smash{\begin{tabular}[t]{r}\svgunitsize 2\end{tabular}}}}%
    \put(0.09496788,0.75925797){\color[rgb]{0,0,0}\makebox(0,0)[rt]{\lineheight{1.25}\smash{\begin{tabular}[t]{r}\svgunitsize 3\end{tabular}}}}%
    \put(0.09496788,0.88471363){\color[rgb]{0,0,0}\makebox(0,0)[rt]{\lineheight{1.25}\smash{\begin{tabular}[t]{r}\svgunitsize 4\end{tabular}}}}%
    \put(0.13647258,0.92563821){\color[rgb]{0,0,0}\makebox(0,0)[rt]{\lineheight{1.25}\smash{\begin{tabular}[t]{r}\svgunitsize $\times$10$^\text{-5}$\end{tabular}}}}%
    \put(0.09498371,0.13197972){\color[rgb]{0,0,0}\makebox(0,0)[rt]{\lineheight{1.25}\smash{\begin{tabular}[t]{r}\svgunitsize -2\end{tabular}}}}%
    \put(0.0950239,0.3828909){\color[rgb]{0,0,0}\makebox(0,0)[rt]{\lineheight{1.25}\smash{\begin{tabular}[t]{r}\svgunitsize 0\end{tabular}}}}%
    \put(0.09478134,0.00652399){\color[rgb]{0,0,0}\makebox(0,0)[rt]{\lineheight{1.25}\smash{\begin{tabular}[t]{r}\svgunitsize -3\end{tabular}}}}%
    \put(0.09497217,0.25743531){\color[rgb]{0,0,0}\makebox(0,0)[rt]{\lineheight{1.25}\smash{\begin{tabular}[t]{r}\svgunitsize -1\end{tabular}}}}%
    \put(0.02382914,0.45483968){\color[rgb]{0,0,0}\rotatebox{90}{\makebox(0,0)[t]{\lineheight{1.25}\smash{\begin{tabular}[t]{c}\svglabelsize $\gamma$ / $\omega_{ci}$\end{tabular}}}}}%
    \put(0,0){\includegraphics[width=\unitlength,page=2]{196-Lx375-qiufac-25.pdf}}%
    \put(1.00067763,0.43161255){\color[rgb]{0,0,0}\makebox(0,0)[rt]{\lineheight{1.25}\smash{\begin{tabular}[t]{r}\svglabelsize $k_\text{pert}$ / $\left(\frac{2\pi}{a_\text{min}}\right)$\end{tabular}}}}%
    \put(0.26651807,0.35402357){\color[rgb]{0,0,0}\makebox(0,0)[t]{\lineheight{1.25}\smash{\begin{tabular}[t]{c}\svgunitsize 5\end{tabular}}}}%
    \put(0.58602902,0.35402357){\color[rgb]{0,0,0}\makebox(0,0)[t]{\lineheight{1.25}\smash{\begin{tabular}[t]{c}\svgunitsize 10\end{tabular}}}}%
    \put(0.90585941,0.35402357){\color[rgb]{0,0,0}\makebox(0,0)[t]{\lineheight{1.25}\smash{\begin{tabular}[t]{c}\svgunitsize 15\end{tabular}}}}%
    \put(0,0){\includegraphics[width=\unitlength,page=3]{196-Lx375-qiufac-25.pdf}}%
  \end{picture}%
\endgroup%

%% file: images-simulations/damping-Lx-comparison.pdf_tex
\begingroup%
  \makeatletter%
  \providecommand\color[2][]{%
    \errmessage{(Inkscape) Color is used for the text in Inkscape, but the package 'color.sty' is not loaded}%
    \renewcommand\color[2][]{}%
  }%
  \providecommand\transparent[1]{%
    \errmessage{(Inkscape) Transparency is used (non-zero) for the text in Inkscape, but the package 'transparent.sty' is not loaded}%
    \renewcommand\transparent[1]{}%
  }%
  \providecommand\rotatebox[2]{#2}%
  \newcommand*\fsize{\dimexpr\f@size pt\relax}%
  \newcommand*\lineheight[1]{\fontsize{\fsize}{#1\fsize}\selectfont}%
  \ifx\svgwidth\undefined%
    \setlength{\unitlength}{442.44561768bp}%
    \ifx\svgscale\undefined%
      \relax%
    \else%
      \setlength{\unitlength}{\unitlength * \real{\svgscale}}%
    \fi%
  \else%
    \setlength{\unitlength}{\svgwidth}%
  \fi%
  \global\let\svgwidth\undefined%
  \global\let\svgscale\undefined%
  \makeatother%
  \begin{picture}(1,0.80323815)%
    \lineheight{1}%
    \setlength\tabcolsep{0pt}%
    \put(0,0){\includegraphics[width=\unitlength,page=1]{damping-Lx-comparison.pdf}}%
    \put(0.08212352,0.30940772){\color[rgb]{0,0,0}\makebox(0,0)[rt]{\lineheight{1.25}\smash{\begin{tabular}[t]{r}\svgunitsize 1\end{tabular}}}}%
    \put(0.08212352,0.41066308){\color[rgb]{0,0,0}\makebox(0,0)[rt]{\lineheight{1.25}\smash{\begin{tabular}[t]{r}\svgunitsize 2\end{tabular}}}}%
    \put(0.08212352,0.51191846){\color[rgb]{0,0,0}\makebox(0,0)[rt]{\lineheight{1.25}\smash{\begin{tabular}[t]{r}\svgunitsize 3\end{tabular}}}}%
    \put(0.08212352,0.61317384){\color[rgb]{0,0,0}\makebox(0,0)[rt]{\lineheight{1.25}\smash{\begin{tabular}[t]{r}\svgunitsize 4\end{tabular}}}}%
    \put(0.08212352,0.7144292){\color[rgb]{0,0,0}\makebox(0,0)[rt]{\lineheight{1.25}\smash{\begin{tabular}[t]{r}\svgunitsize 5\end{tabular}}}}%
    \put(0.11801494,0.78263183){\color[rgb]{0,0,0}\makebox(0,0)[rt]{\lineheight{1.25}\smash{\begin{tabular}[t]{r}\svgunitsize $\times$10$^\text{-5}$\end{tabular}}}}%
    \put(0.08213703,0.00564159){\color[rgb]{0,0,0}\makebox(0,0)[rt]{\lineheight{1.25}\smash{\begin{tabular}[t]{r}\svgunitsize -2\end{tabular}}}}%
    \put(0.08217221,0.20815227){\color[rgb]{0,0,0}\makebox(0,0)[rt]{\lineheight{1.25}\smash{\begin{tabular}[t]{r}\svgunitsize 0\end{tabular}}}}%
    \put(0.08212733,0.10689691){\color[rgb]{0,0,0}\makebox(0,0)[rt]{\lineheight{1.25}\smash{\begin{tabular}[t]{r}\svgunitsize -1\end{tabular}}}}%
    \put(0.02060627,0.39332303){\color[rgb]{0,0,0}\rotatebox{90}{\makebox(0,0)[t]{\lineheight{1.25}\smash{\begin{tabular}[t]{c}\svglabelsize $\gamma$ / $\omega_{ci}$\end{tabular}}}}}%
    \put(0.20177103,0.1828757){\color[rgb]{0,0,0}\makebox(0,0)[t]{\lineheight{1.25}\smash{\begin{tabular}[t]{c}\svgunitsize 5\end{tabular}}}}%
    \put(0.42058643,0.1828757){\color[rgb]{0,0,0}\makebox(0,0)[t]{\lineheight{1.25}\smash{\begin{tabular}[t]{c}\svgunitsize 10\end{tabular}}}}%
    \put(0.63967801,0.1828757){\color[rgb]{0,0,0}\makebox(0,0)[t]{\lineheight{1.25}\smash{\begin{tabular}[t]{c}\svgunitsize 15\end{tabular}}}}%
    \put(0.85849334,0.1828757){\color[rgb]{0,0,0}\makebox(0,0)[t]{\lineheight{1.25}\smash{\begin{tabular}[t]{c}\svgunitsize 20\end{tabular}}}}%
    \put(0,0){\includegraphics[width=\unitlength,page=2]{damping-Lx-comparison.pdf}}%
    \put(0.86533739,0.24997071){\color[rgb]{0,0,0}\makebox(0,0)[rt]{\lineheight{1.25}\smash{\begin{tabular}[t]{r}\svglabelsize $k_\text{pert}$ / $\left(\frac{2\pi}{a_\text{min}}\right)$\end{tabular}}}}%
    \put(0,0){\includegraphics[width=\unitlength,page=3]{damping-Lx-comparison.pdf}}%
    \put(0.70602842,0.75794641){\color[rgb]{0,0,0}\makebox(0,0)[lt]{\lineheight{1.25}\smash{\begin{tabular}[t]{l}\svglegendsize $\rho_{i1}\!: \gamma_\text{MI}$\end{tabular}}}}%
    \put(0,0){\includegraphics[width=\unitlength,page=4]{damping-Lx-comparison.pdf}}%
    \put(0.7060283,0.71612261){\color[rgb]{0,0,0}\makebox(0,0)[lt]{\lineheight{1.25}\smash{\begin{tabular}[t]{l}\svglegendsize $\rho_{i1}\!: \gamma_\text{MI} + 2.5\!\times\!\gamma_\text{Qiu}$\end{tabular}}}}%
    \put(0,0){\includegraphics[width=\unitlength,page=5]{damping-Lx-comparison.pdf}}%
    \put(0.70602842,0.67429877){\color[rgb]{0,0,0}\makebox(0,0)[lt]{\lineheight{1.25}\smash{\begin{tabular}[t]{l}\svglegendsize $\rho_{i1}\!:$ Simulation\end{tabular}}}}%
    \put(0,0){\includegraphics[width=\unitlength,page=6]{damping-Lx-comparison.pdf}}%
    \put(0.70602842,0.63247495){\color[rgb]{0,0,0}\makebox(0,0)[lt]{\lineheight{1.25}\smash{\begin{tabular}[t]{l}\svglegendsize $\rho_{i2}\!: \gamma_\text{MI}$\end{tabular}}}}%
    \put(0,0){\includegraphics[width=\unitlength,page=7]{damping-Lx-comparison.pdf}}%
    \put(0.7060283,0.59065115){\color[rgb]{0,0,0}\makebox(0,0)[lt]{\lineheight{1.25}\smash{\begin{tabular}[t]{l}\svglegendsize $\rho_{i2}\!: \gamma_\text{MI} + 2.5\!\times\!\gamma_\text{Qiu}$\end{tabular}}}}%
    \put(0,0){\includegraphics[width=\unitlength,page=8]{damping-Lx-comparison.pdf}}%
    \put(0.70602842,0.54882731){\color[rgb]{0,0,0}\makebox(0,0)[lt]{\lineheight{1.25}\smash{\begin{tabular}[t]{l}\svglegendsize $\rho_{i2}\!:$ Simulation\end{tabular}}}}%
    \put(0,0){\includegraphics[width=\unitlength,page=9]{damping-Lx-comparison.pdf}}%
    \put(0.70602842,0.50700349){\color[rgb]{0,0,0}\makebox(0,0)[lt]{\lineheight{1.25}\smash{\begin{tabular}[t]{l}\svglegendsize $\rho_{i3}\!: \gamma_\text{MI}$\end{tabular}}}}%
    \put(0,0){\includegraphics[width=\unitlength,page=10]{damping-Lx-comparison.pdf}}%
    \put(0.7060283,0.46517973){\color[rgb]{0,0,0}\makebox(0,0)[lt]{\lineheight{1.25}\smash{\begin{tabular}[t]{l}\svglegendsize $\rho_{i3}\!: \gamma_\text{MI} + 2.5\!\times\!\gamma_\text{Qiu}$\end{tabular}}}}%
    \put(0,0){\includegraphics[width=\unitlength,page=11]{damping-Lx-comparison.pdf}}%
    \put(0.70602842,0.42335585){\color[rgb]{0,0,0}\makebox(0,0)[lt]{\lineheight{1.25}\smash{\begin{tabular}[t]{l}\svglegendsize $\rho_{i3}\!:$ Simulation\end{tabular}}}}%
  \end{picture}%
\endgroup%

%% file: images-damping/196-k10-long-NLSE-damped-combined.pdf_tex
\begingroup%
  \makeatletter%
  \providecommand\color[2][]{%
    \errmessage{(Inkscape) Color is used for the text in Inkscape, but the package 'color.sty' is not loaded}%
    \renewcommand\color[2][]{}%
  }%
  \providecommand\transparent[1]{%
    \errmessage{(Inkscape) Transparency is used (non-zero) for the text in Inkscape, but the package 'transparent.sty' is not loaded}%
    \renewcommand\transparent[1]{}%
  }%
  \providecommand\rotatebox[2]{#2}%
  \newcommand*\fsize{\dimexpr\f@size pt\relax}%
  \newcommand*\lineheight[1]{\fontsize{\fsize}{#1\fsize}\selectfont}%
  \ifx\svgwidth\undefined%
    \setlength{\unitlength}{808.63806152bp}%
    \ifx\svgscale\undefined%
      \relax%
    \else%
      \setlength{\unitlength}{\unitlength * \real{\svgscale}}%
    \fi%
  \else%
    \setlength{\unitlength}{\svgwidth}%
  \fi%
  \global\let\svgwidth\undefined%
  \global\let\svgscale\undefined%
  \makeatother%
  \begin{picture}(1,0.25962804)%
    \lineheight{1}%
    \setlength\tabcolsep{0pt}%
    \put(0,0){\includegraphics[width=\unitlength,page=1]{196-k10-long-NLSE-damped-combined.pdf}}%
    \put(0.68489465,0.04800932){\color[rgb]{0,0,0}\makebox(0,0)[lt]{\lineheight{1.25}\smash{\begin{tabular}[t]{l}\SVGunitsize \phantom{0}0\end{tabular}}}}%
    \put(0.70810472,0.02790385){\color[rgb]{0,0,0}\makebox(0,0)[t]{\lineheight{1.25}\smash{\begin{tabular}[t]{c}{\SVGunitsize 0}\end{tabular}}}}%
    \put(0.67477139,0.14394074){\color[rgb]{0,0,0}\rotatebox{90}{\makebox(0,0)[t]{\lineheight{1.25}\smash{\begin{tabular}[t]{c}\SVGlabelsize $k_r / (2\pi/a_\text{min})$ \end{tabular}}}}}%
    \put(0.82510574,0.00342742){\color[rgb]{0,0,0}\makebox(0,0)[t]{\lineheight{1.25}\smash{\begin{tabular}[t]{c}\SVGlabelsize $t/\omega_{ci}^{-1}$ \end{tabular}}}}%
    \put(0.88333221,0.2582315){\color[rgb]{0,0,0}\makebox(0,0)[lt]{\lineheight{1.25}\smash{\begin{tabular}[t]{l}{\SVGlabelsize $\mathfrak{F}[E_r]$ (a.u.)}\end{tabular}}}}%
    \put(0.75432714,0.02790385){\color[rgb]{0,0,0}\makebox(0,0)[t]{\lineheight{1.25}\smash{\begin{tabular}[t]{c}{\SVGunitsize 0.5}\end{tabular}}}}%
    \put(0.80053829,0.02790385){\color[rgb]{0,0,0}\makebox(0,0)[t]{\lineheight{1.25}\smash{\begin{tabular}[t]{c}{\SVGunitsize 1}\end{tabular}}}}%
    \put(0.84677255,0.02790385){\color[rgb]{0,0,0}\makebox(0,0)[t]{\lineheight{1.25}\smash{\begin{tabular}[t]{c}{\SVGunitsize 1.5}\end{tabular}}}}%
    \put(0.89301626,0.02790385){\color[rgb]{0,0,0}\makebox(0,0)[t]{\lineheight{1.25}\smash{\begin{tabular}[t]{c}{\SVGunitsize 2}\end{tabular}}}}%
    \put(0.93926574,0.02790385){\color[rgb]{0,0,0}\makebox(0,0)[t]{\lineheight{1.25}\smash{\begin{tabular}[t]{c}{\SVGunitsize 2.5}\end{tabular}}}}%
    \put(0.68489465,0.08969557){\color[rgb]{0,0,0}\makebox(0,0)[lt]{\lineheight{1.25}\smash{\begin{tabular}[t]{l}\SVGunitsize 10\end{tabular}}}}%
    \put(0.68489465,0.13137557){\color[rgb]{0,0,0}\makebox(0,0)[lt]{\lineheight{1.25}\smash{\begin{tabular}[t]{l}\SVGunitsize 20\end{tabular}}}}%
    \put(0.68489465,0.17306815){\color[rgb]{0,0,0}\makebox(0,0)[lt]{\lineheight{1.25}\smash{\begin{tabular}[t]{l}\SVGunitsize 30\end{tabular}}}}%
    \put(0.68489465,0.21476733){\color[rgb]{0,0,0}\makebox(0,0)[lt]{\lineheight{1.25}\smash{\begin{tabular}[t]{l}\SVGunitsize 40\end{tabular}}}}%
    \put(0,0){\includegraphics[width=\unitlength,page=2]{196-k10-long-NLSE-damped-combined.pdf}}%
    \put(0.05297772,0.23530555){\color[rgb]{0,0,0}\makebox(0,0)[rt]{\lineheight{1.25}\smash{\begin{tabular}[t]{r}\svgunitsize 1\end{tabular}}}}%
    \put(0.05285958,0.18751831){\color[rgb]{0,0,0}\makebox(0,0)[rt]{\lineheight{1.25}\smash{\begin{tabular}[t]{r}\svgunitsize 0.8\end{tabular}}}}%
    \put(0.05283306,0.13973104){\color[rgb]{0,0,0}\makebox(0,0)[rt]{\lineheight{1.25}\smash{\begin{tabular}[t]{r}\svgunitsize 0.6\end{tabular}}}}%
    \put(0.0527993,0.09194375){\color[rgb]{0,0,0}\makebox(0,0)[rt]{\lineheight{1.25}\smash{\begin{tabular}[t]{r}\svgunitsize 0.4\end{tabular}}}}%
    \put(0.05301631,0.04421096){\color[rgb]{0,0,0}\makebox(0,0)[rt]{\lineheight{1.25}\smash{\begin{tabular}[t]{r}\svgunitsize 0.2\end{tabular}}}}%
    \put(0.06086464,0.02862056){\color[rgb]{0,0,0}\makebox(0,0)[t]{\lineheight{1.25}\smash{\begin{tabular}[t]{c}\svgunitsize 0\end{tabular}}}}%
    \put(0.32605592,0.0034998){\color[rgb]{0,0,0}\makebox(0,0)[t]{\lineheight{1.25}\smash{\begin{tabular}[t]{c}\svglabelsize $t$ / $\omega_{ci}^{-1}$\end{tabular}}}}%
    \put(0.57668463,0.25835331){\color[rgb]{0,0,0}\makebox(0,0)[lt]{\lineheight{1.25}\smash{\begin{tabular}[t]{l}\svglabelsize $E_r$ / $a_0$\end{tabular}}}}%
    \put(0.01993104,0.14501879){\color[rgb]{0,0,0}\rotatebox{90}{\makebox(0,0)[t]{\lineheight{1.25}\smash{\begin{tabular}[t]{c}\svglabelsize $r$ / $a_\text{min}$\end{tabular}}}}}%
    \put(0.16734948,0.02862056){\color[rgb]{0,0,0}\makebox(0,0)[t]{\lineheight{1.25}\smash{\begin{tabular}[t]{c}\svgunitsize 0.5\end{tabular}}}}%
    \put(0.27376683,0.02862056){\color[rgb]{0,0,0}\makebox(0,0)[t]{\lineheight{1.25}\smash{\begin{tabular}[t]{c}\svgunitsize 1\end{tabular}}}}%
    \put(0.38012561,0.02862056){\color[rgb]{0,0,0}\makebox(0,0)[t]{\lineheight{1.25}\smash{\begin{tabular}[t]{c}\svgunitsize 1.5\end{tabular}}}}%
    \put(0.48658988,0.02862056){\color[rgb]{0,0,0}\makebox(0,0)[t]{\lineheight{1.25}\smash{\begin{tabular}[t]{c}\svgunitsize 2\end{tabular}}}}%
    \put(0.59290173,0.02862056){\color[rgb]{0,0,0}\makebox(0,0)[t]{\lineheight{1.25}\smash{\begin{tabular}[t]{c}\svgunitsize 2.5\end{tabular}}}}%
    \put(0.62330861,0.2291442){\color[rgb]{0,0,0}\makebox(0,0)[lt]{\lineheight{1.25}\smash{\begin{tabular}[t]{l}\svgunitsize 1.5\end{tabular}}}}%
    \put(0.62330861,0.04485204){\color[rgb]{0,0,0}\makebox(0,0)[lt]{\lineheight{1.25}\smash{\begin{tabular}[t]{l}\svgunitsize 0\end{tabular}}}}%
    \put(0.57357283,0.0078488){\color[rgb]{0,0,0}\makebox(0,0)[t]{\lineheight{1.25}\smash{\begin{tabular}[t]{c}\svgunitsize $\times$10$^\text{5}$\end{tabular}}}}%
    \put(0,0){\includegraphics[width=\unitlength,page=3]{196-k10-long-NLSE-damped-combined.pdf}}%
    \put(0.62330861,0.16771348){\color[rgb]{0,0,0}\makebox(0,0)[lt]{\lineheight{1.25}\smash{\begin{tabular}[t]{l}\svgunitsize 1\end{tabular}}}}%
    \put(0,0){\includegraphics[width=\unitlength,page=4]{196-k10-long-NLSE-damped-combined.pdf}}%
    \put(0.62330861,0.10628275){\color[rgb]{0,0,0}\makebox(0,0)[lt]{\lineheight{1.25}\smash{\begin{tabular}[t]{l}\svgunitsize 0.5\end{tabular}}}}%
    \put(0.95036368,0.0078488){\color[rgb]{0,0,0}\makebox(0,0)[t]{\lineheight{1.25}\smash{\begin{tabular}[t]{c}\svgunitsize $\times$10$^\text{5}$\end{tabular}}}}%
    \put(0.97508095,0.04493411){\color[rgb]{0,0,0}\makebox(0,0)[lt]{\lineheight{1.25}\smash{\begin{tabular}[t]{l}\svgunitsize 0\end{tabular}}}}%
    \put(0,0){\includegraphics[width=\unitlength,page=5]{196-k10-long-NLSE-damped-combined.pdf}}%
    \put(0.97508095,0.09232808){\color[rgb]{0,0,0}\makebox(0,0)[lt]{\lineheight{1.25}\smash{\begin{tabular}[t]{l}\svgunitsize 0.25\end{tabular}}}}%
    \put(0,0){\includegraphics[width=\unitlength,page=6]{196-k10-long-NLSE-damped-combined.pdf}}%
    \put(0.97508095,0.13974605){\color[rgb]{0,0,0}\makebox(0,0)[lt]{\lineheight{1.25}\smash{\begin{tabular}[t]{l}\svgunitsize 0.5\end{tabular}}}}%
    \put(0,0){\includegraphics[width=\unitlength,page=7]{196-k10-long-NLSE-damped-combined.pdf}}%
    \put(0.97508095,0.18716401){\color[rgb]{0,0,0}\makebox(0,0)[lt]{\lineheight{1.25}\smash{\begin{tabular}[t]{l}\svgunitsize 0.75\end{tabular}}}}%
    \put(0,0){\includegraphics[width=\unitlength,page=8]{196-k10-long-NLSE-damped-combined.pdf}}%
    \put(0.97508095,0.23600122){\color[rgb]{0,0,0}\makebox(0,0)[lt]{\lineheight{1.25}\smash{\begin{tabular}[t]{l}\svgunitsize 1\end{tabular}}}}%
  \end{picture}%
\endgroup%

%% file: images-simulations/Self-focusing-Gauss-NLSE.pdf_tex
\begingroup%
  \makeatletter%
  \providecommand\color[2][]{%
    \errmessage{(Inkscape) Color is used for the text in Inkscape, but the package 'color.sty' is not loaded}%
    \renewcommand\color[2][]{}%
  }%
  \providecommand\transparent[1]{%
    \errmessage{(Inkscape) Transparency is used (non-zero) for the text in Inkscape, but the package 'transparent.sty' is not loaded}%
    \renewcommand\transparent[1]{}%
  }%
  \providecommand\rotatebox[2]{#2}%
  \newcommand*\fsize{\dimexpr\f@size pt\relax}%
  \newcommand*\lineheight[1]{\fontsize{\fsize}{#1\fsize}\selectfont}%
  \ifx\svgwidth\undefined%
    \setlength{\unitlength}{654.37339783bp}%
    \ifx\svgscale\undefined%
      \relax%
    \else%
      \setlength{\unitlength}{\unitlength * \real{\svgscale}}%
    \fi%
  \else%
    \setlength{\unitlength}{\svgwidth}%
  \fi%
  \global\let\svgwidth\undefined%
  \global\let\svgscale\undefined%
  \makeatother%
  \begin{picture}(1,0.31395694)%
    \lineheight{1}%
    \setlength\tabcolsep{0pt}%
    \put(0,0){\includegraphics[width=\unitlength,page=1]{Self-focusing-Gauss-NLSE.pdf}}%
    \put(0.0688585,0.27542728){\color[rgb]{0,0,0}\makebox(0,0)[rt]{\lineheight{1.25}\smash{\begin{tabular}[t]{r}\svgunitsize 0.7\end{tabular}}}}%
    \put(0.06882573,0.22174982){\color[rgb]{0,0,0}\makebox(0,0)[rt]{\lineheight{1.25}\smash{\begin{tabular}[t]{r}\svgunitsize 0.6\end{tabular}}}}%
    \put(0.06878401,0.16803283){\color[rgb]{0,0,0}\makebox(0,0)[rt]{\lineheight{1.25}\smash{\begin{tabular}[t]{r}\svgunitsize 0.5\end{tabular}}}}%
    \put(0,0){\includegraphics[width=\unitlength,page=2]{Self-focusing-Gauss-NLSE.pdf}}%
    \put(0.06878401,0.11444207){\color[rgb]{0,0,0}\makebox(0,0)[rt]{\lineheight{1.25}\smash{\begin{tabular}[t]{r}\svgunitsize 0.4\end{tabular}}}}%
    \put(0,0){\includegraphics[width=\unitlength,page=3]{Self-focusing-Gauss-NLSE.pdf}}%
    \put(0.06878405,0.06062955){\color[rgb]{0,0,0}\makebox(0,0)[rt]{\lineheight{1.25}\smash{\begin{tabular}[t]{r}\svgunitsize 0.3\end{tabular}}}}%
    \put(0,0){\includegraphics[width=\unitlength,page=4]{Self-focusing-Gauss-NLSE.pdf}}%
    \put(0.07875073,0.03078312){\color[rgb]{0,0,0}\makebox(0,0)[t]{\lineheight{1.25}\smash{\begin{tabular}[t]{c}\svgunitsize 0\end{tabular}}}}%
    \put(0.49721298,0.00432485){\color[rgb]{0,0,0}\makebox(0,0)[t]{\lineheight{1.25}\smash{\begin{tabular}[t]{c}\svglabelsize $t$ / $\omega_{ci}^{-1}$\end{tabular}}}}%
    \put(0.90642954,0.30002426){\color[rgb]{0,0,0}\makebox(0,0)[lt]{\lineheight{1.25}\smash{\begin{tabular}[t]{l}\svglabelsize $E_r$ / $a_0$\end{tabular}}}}%
    \put(0.02358264,0.1746215){\color[rgb]{0,0,0}\rotatebox{90}{\makebox(0,0)[t]{\lineheight{1.25}\smash{\begin{tabular}[t]{c}\svglabelsize $r$ / $a_\text{min}$\end{tabular}}}}}%
    \put(0.95946091,0.25979015){\color[rgb]{0,0,0}\makebox(0,0)[lt]{\lineheight{1.25}\smash{\begin{tabular}[t]{l}\svgunitsize 1.5\end{tabular}}}}%
    \put(0.95946091,0.05084103){\color[rgb]{0,0,0}\makebox(0,0)[lt]{\lineheight{1.25}\smash{\begin{tabular}[t]{l}\svgunitsize 0\end{tabular}}}}%
    \put(0.89341655,0.00511451){\color[rgb]{0,0,0}\makebox(0,0)[t]{\lineheight{1.25}\smash{\begin{tabular}[t]{c}\svgunitsize $\times$10$^\text{5}$\end{tabular}}}}%
    \put(0,0){\includegraphics[width=\unitlength,page=5]{Self-focusing-Gauss-NLSE.pdf}}%
    \put(0.95946091,0.18996414){\color[rgb]{0,0,0}\makebox(0,0)[lt]{\lineheight{1.25}\smash{\begin{tabular}[t]{l}\svgunitsize 1\end{tabular}}}}%
    \put(0,0){\includegraphics[width=\unitlength,page=6]{Self-focusing-Gauss-NLSE.pdf}}%
    \put(0.95946091,0.12032749){\color[rgb]{0,0,0}\makebox(0,0)[lt]{\lineheight{1.25}\smash{\begin{tabular}[t]{l}\svgunitsize 0.5\end{tabular}}}}%
    \put(0.43929167,0.03078312){\color[rgb]{0,0,0}\makebox(0,0)[t]{\lineheight{1.25}\smash{\begin{tabular}[t]{c}\svgunitsize 0.6\end{tabular}}}}%
    \put(0,0){\includegraphics[width=\unitlength,page=7]{Self-focusing-Gauss-NLSE.pdf}}%
    \put(0.68044621,0.03078312){\color[rgb]{0,0,0}\makebox(0,0)[t]{\lineheight{1.25}\smash{\begin{tabular}[t]{c}\svgunitsize 1\end{tabular}}}}%
    \put(0,0){\includegraphics[width=\unitlength,page=8]{Self-focusing-Gauss-NLSE.pdf}}%
    \put(0.19841271,0.03078312){\color[rgb]{0,0,0}\makebox(0,0)[t]{\lineheight{1.25}\smash{\begin{tabular}[t]{c}\svgunitsize 0.2\end{tabular}}}}%
    \put(0,0){\includegraphics[width=\unitlength,page=9]{Self-focusing-Gauss-NLSE.pdf}}%
    \put(0.31880831,0.03078312){\color[rgb]{0,0,0}\makebox(0,0)[t]{\lineheight{1.25}\smash{\begin{tabular}[t]{c}\svgunitsize 0.4\end{tabular}}}}%
    \put(0,0){\includegraphics[width=\unitlength,page=10]{Self-focusing-Gauss-NLSE.pdf}}%
    \put(0.55851703,0.03078312){\color[rgb]{0,0,0}\makebox(0,0)[t]{\lineheight{1.25}\smash{\begin{tabular}[t]{c}\svgunitsize 0.8\end{tabular}}}}%
    \put(0,0){\includegraphics[width=\unitlength,page=11]{Self-focusing-Gauss-NLSE.pdf}}%
    \put(0.79890381,0.03078312){\color[rgb]{0,0,0}\makebox(0,0)[t]{\lineheight{1.25}\smash{\begin{tabular}[t]{c}\svgunitsize 1.2\end{tabular}}}}%
    \put(0,0){\includegraphics[width=\unitlength,page=12]{Self-focusing-Gauss-NLSE.pdf}}%
    \put(0.91847252,0.03078312){\color[rgb]{0,0,0}\makebox(0,0)[t]{\lineheight{1.25}\smash{\begin{tabular}[t]{c}\svgunitsize 1.4\end{tabular}}}}%
    \put(0,0){\includegraphics[width=\unitlength,page=13]{Self-focusing-Gauss-NLSE.pdf}}%
  \end{picture}%
\endgroup%

%% file: images-simulations/Self-focusing-Gauss-ORB5.pdf_tex
\begingroup%
  \makeatletter%
  \providecommand\color[2][]{%
    \errmessage{(Inkscape) Color is used for the text in Inkscape, but the package 'color.sty' is not loaded}%
    \renewcommand\color[2][]{}%
  }%
  \providecommand\transparent[1]{%
    \errmessage{(Inkscape) Transparency is used (non-zero) for the text in Inkscape, but the package 'transparent.sty' is not loaded}%
    \renewcommand\transparent[1]{}%
  }%
  \providecommand\rotatebox[2]{#2}%
  \newcommand*\fsize{\dimexpr\f@size pt\relax}%
  \newcommand*\lineheight[1]{\fontsize{\fsize}{#1\fsize}\selectfont}%
  \ifx\svgwidth\undefined%
    \setlength{\unitlength}{654.37339783bp}%
    \ifx\svgscale\undefined%
      \relax%
    \else%
      \setlength{\unitlength}{\unitlength * \real{\svgscale}}%
    \fi%
  \else%
    \setlength{\unitlength}{\svgwidth}%
  \fi%
  \global\let\svgwidth\undefined%
  \global\let\svgscale\undefined%
  \makeatother%
  \begin{picture}(1,0.31395697)%
    \lineheight{1}%
    \setlength\tabcolsep{0pt}%
    \put(0,0){\includegraphics[width=\unitlength,page=1]{Self-focusing-Gauss-ORB5.pdf}}%
    \put(0.0688585,0.27542731){\color[rgb]{0,0,0}\makebox(0,0)[rt]{\lineheight{1.25}\smash{\begin{tabular}[t]{r}\svgunitsize 0.7\end{tabular}}}}%
    \put(0.06882573,0.22174986){\color[rgb]{0,0,0}\makebox(0,0)[rt]{\lineheight{1.25}\smash{\begin{tabular}[t]{r}\svgunitsize 0.6\end{tabular}}}}%
    \put(0.06878401,0.16803287){\color[rgb]{0,0,0}\makebox(0,0)[rt]{\lineheight{1.25}\smash{\begin{tabular}[t]{r}\svgunitsize 0.5\end{tabular}}}}%
    \put(0,0){\includegraphics[width=\unitlength,page=2]{Self-focusing-Gauss-ORB5.pdf}}%
    \put(0.06878401,0.11444208){\color[rgb]{0,0,0}\makebox(0,0)[rt]{\lineheight{1.25}\smash{\begin{tabular}[t]{r}\svgunitsize 0.4\end{tabular}}}}%
    \put(0,0){\includegraphics[width=\unitlength,page=3]{Self-focusing-Gauss-ORB5.pdf}}%
    \put(0.06878405,0.06062958){\color[rgb]{0,0,0}\makebox(0,0)[rt]{\lineheight{1.25}\smash{\begin{tabular}[t]{r}\svgunitsize 0.3\end{tabular}}}}%
    \put(0,0){\includegraphics[width=\unitlength,page=4]{Self-focusing-Gauss-ORB5.pdf}}%
    \put(0.07875073,0.03078314){\color[rgb]{0,0,0}\makebox(0,0)[t]{\lineheight{1.25}\smash{\begin{tabular}[t]{c}\svgunitsize 0\end{tabular}}}}%
    \put(0.49721298,0.00432487){\color[rgb]{0,0,0}\makebox(0,0)[t]{\lineheight{1.25}\smash{\begin{tabular}[t]{c}\svglabelsize $t$ / $\omega_{ci}^{-1}$\end{tabular}}}}%
    \put(0.90642954,0.30002429){\color[rgb]{0,0,0}\makebox(0,0)[lt]{\lineheight{1.25}\smash{\begin{tabular}[t]{l}\svglabelsize $E_r$ / $a_0$\end{tabular}}}}%
    \put(0.02358264,0.17462154){\color[rgb]{0,0,0}\rotatebox{90}{\makebox(0,0)[t]{\lineheight{1.25}\smash{\begin{tabular}[t]{c}\svglabelsize $r$ / $a_\text{min}$\end{tabular}}}}}%
    \put(0.95946091,0.05084106){\color[rgb]{0,0,0}\makebox(0,0)[lt]{\lineheight{1.25}\smash{\begin{tabular}[t]{l}\svgunitsize 0\end{tabular}}}}%
    \put(0.89341655,0.00511455){\color[rgb]{0,0,0}\makebox(0,0)[t]{\lineheight{1.25}\smash{\begin{tabular}[t]{c}\svgunitsize $\times$10$^\text{5}$\end{tabular}}}}%
    \put(0,0){\includegraphics[width=\unitlength,page=5]{Self-focusing-Gauss-ORB5.pdf}}%
    \put(0.95946091,0.09210609){\color[rgb]{0,0,0}\makebox(0,0)[lt]{\lineheight{1.25}\smash{\begin{tabular}[t]{l}\svgunitsize 0.2\end{tabular}}}}%
    \put(0,0){\includegraphics[width=\unitlength,page=6]{Self-focusing-Gauss-ORB5.pdf}}%
    \put(0.95946091,0.13335181){\color[rgb]{0,0,0}\makebox(0,0)[lt]{\lineheight{1.25}\smash{\begin{tabular}[t]{l}\svgunitsize 0.4\end{tabular}}}}%
    \put(0,0){\includegraphics[width=\unitlength,page=7]{Self-focusing-Gauss-ORB5.pdf}}%
    \put(0.95946091,0.17461481){\color[rgb]{0,0,0}\makebox(0,0)[lt]{\lineheight{1.25}\smash{\begin{tabular}[t]{l}\svgunitsize 0.6\end{tabular}}}}%
    \put(0,0){\includegraphics[width=\unitlength,page=8]{Self-focusing-Gauss-ORB5.pdf}}%
    \put(0.95946091,0.215946){\color[rgb]{0,0,0}\makebox(0,0)[lt]{\lineheight{1.25}\smash{\begin{tabular}[t]{l}\svgunitsize 0.8\end{tabular}}}}%
    \put(0,0){\includegraphics[width=\unitlength,page=9]{Self-focusing-Gauss-ORB5.pdf}}%
    \put(0.95946091,0.25722178){\color[rgb]{0,0,0}\makebox(0,0)[lt]{\lineheight{1.25}\smash{\begin{tabular}[t]{l}\svgunitsize 1\end{tabular}}}}%
    \put(0.43929167,0.03078314){\color[rgb]{0,0,0}\makebox(0,0)[t]{\lineheight{1.25}\smash{\begin{tabular}[t]{c}\svgunitsize 0.6\end{tabular}}}}%
    \put(0,0){\includegraphics[width=\unitlength,page=10]{Self-focusing-Gauss-ORB5.pdf}}%
    \put(0.68044621,0.03078319){\color[rgb]{0,0,0}\makebox(0,0)[t]{\lineheight{1.25}\smash{\begin{tabular}[t]{c}\svgunitsize 1\end{tabular}}}}%
    \put(0,0){\includegraphics[width=\unitlength,page=11]{Self-focusing-Gauss-ORB5.pdf}}%
    \put(0.19841271,0.03078314){\color[rgb]{0,0,0}\makebox(0,0)[t]{\lineheight{1.25}\smash{\begin{tabular}[t]{c}\svgunitsize 0.2\end{tabular}}}}%
    \put(0,0){\includegraphics[width=\unitlength,page=12]{Self-focusing-Gauss-ORB5.pdf}}%
    \put(0.31880831,0.03078314){\color[rgb]{0,0,0}\makebox(0,0)[t]{\lineheight{1.25}\smash{\begin{tabular}[t]{c}\svgunitsize 0.4\end{tabular}}}}%
    \put(0,0){\includegraphics[width=\unitlength,page=13]{Self-focusing-Gauss-ORB5.pdf}}%
    \put(0.55851703,0.03078314){\color[rgb]{0,0,0}\makebox(0,0)[t]{\lineheight{1.25}\smash{\begin{tabular}[t]{c}\svgunitsize 0.8\end{tabular}}}}%
    \put(0,0){\includegraphics[width=\unitlength,page=14]{Self-focusing-Gauss-ORB5.pdf}}%
    \put(0.79890381,0.03078314){\color[rgb]{0,0,0}\makebox(0,0)[t]{\lineheight{1.25}\smash{\begin{tabular}[t]{c}\svgunitsize 1.2\end{tabular}}}}%
    \put(0,0){\includegraphics[width=\unitlength,page=15]{Self-focusing-Gauss-ORB5.pdf}}%
    \put(0.91847252,0.03078314){\color[rgb]{0,0,0}\makebox(0,0)[t]{\lineheight{1.25}\smash{\begin{tabular}[t]{c}\svgunitsize 1.4\end{tabular}}}}%
    \put(0,0){\includegraphics[width=\unitlength,page=16]{Self-focusing-Gauss-ORB5.pdf}}%
  \end{picture}%
\endgroup%

%% file: images-simulations/Self-focusing-Gauss-NLSE-damped.pdf_tex
\begingroup%
  \makeatletter%
  \providecommand\color[2][]{%
    \errmessage{(Inkscape) Color is used for the text in Inkscape, but the package 'color.sty' is not loaded}%
    \renewcommand\color[2][]{}%
  }%
  \providecommand\transparent[1]{%
    \errmessage{(Inkscape) Transparency is used (non-zero) for the text in Inkscape, but the package 'transparent.sty' is not loaded}%
    \renewcommand\transparent[1]{}%
  }%
  \providecommand\rotatebox[2]{#2}%
  \newcommand*\fsize{\dimexpr\f@size pt\relax}%
  \newcommand*\lineheight[1]{\fontsize{\fsize}{#1\fsize}\selectfont}%
  \ifx\svgwidth\undefined%
    \setlength{\unitlength}{654.37339783bp}%
    \ifx\svgscale\undefined%
      \relax%
    \else%
      \setlength{\unitlength}{\unitlength * \real{\svgscale}}%
    \fi%
  \else%
    \setlength{\unitlength}{\svgwidth}%
  \fi%
  \global\let\svgwidth\undefined%
  \global\let\svgscale\undefined%
  \makeatother%
  \begin{picture}(1,0.31395694)%
    \lineheight{1}%
    \setlength\tabcolsep{0pt}%
    \put(0,0){\includegraphics[width=\unitlength,page=1]{Self-focusing-Gauss-NLSE-damped.pdf}}%
    \put(0.0688585,0.27542728){\color[rgb]{0,0,0}\makebox(0,0)[rt]{\lineheight{1.25}\smash{\begin{tabular}[t]{r}\svgunitsize 0.7\end{tabular}}}}%
    \put(0.06882573,0.22174982){\color[rgb]{0,0,0}\makebox(0,0)[rt]{\lineheight{1.25}\smash{\begin{tabular}[t]{r}\svgunitsize 0.6\end{tabular}}}}%
    \put(0.06878401,0.16803283){\color[rgb]{0,0,0}\makebox(0,0)[rt]{\lineheight{1.25}\smash{\begin{tabular}[t]{r}\svgunitsize 0.5\end{tabular}}}}%
    \put(0,0){\includegraphics[width=\unitlength,page=2]{Self-focusing-Gauss-NLSE-damped.pdf}}%
    \put(0.06878401,0.11444207){\color[rgb]{0,0,0}\makebox(0,0)[rt]{\lineheight{1.25}\smash{\begin{tabular}[t]{r}\svgunitsize 0.4\end{tabular}}}}%
    \put(0,0){\includegraphics[width=\unitlength,page=3]{Self-focusing-Gauss-NLSE-damped.pdf}}%
    \put(0.06878405,0.06062955){\color[rgb]{0,0,0}\makebox(0,0)[rt]{\lineheight{1.25}\smash{\begin{tabular}[t]{r}\svgunitsize 0.3\end{tabular}}}}%
    \put(0,0){\includegraphics[width=\unitlength,page=4]{Self-focusing-Gauss-NLSE-damped.pdf}}%
    \put(0.07875073,0.03078312){\color[rgb]{0,0,0}\makebox(0,0)[t]{\lineheight{1.25}\smash{\begin{tabular}[t]{c}\svgunitsize 0\end{tabular}}}}%
    \put(0.49721298,0.00432485){\color[rgb]{0,0,0}\makebox(0,0)[t]{\lineheight{1.25}\smash{\begin{tabular}[t]{c}\svglabelsize $t$ / $\omega_{ci}$\end{tabular}}}}%
    \put(0.90642954,0.30002426){\color[rgb]{0,0,0}\makebox(0,0)[lt]{\lineheight{1.25}\smash{\begin{tabular}[t]{l}\svglabelsize $E_r$ / $a_0$\end{tabular}}}}%
    \put(0.02358264,0.17462151){\color[rgb]{0,0,0}\rotatebox{90}{\makebox(0,0)[t]{\lineheight{1.25}\smash{\begin{tabular}[t]{c}\svglabelsize $r$ / $a_\text{min}$\end{tabular}}}}}%
    \put(0.95946091,0.05084103){\color[rgb]{0,0,0}\makebox(0,0)[lt]{\lineheight{1.25}\smash{\begin{tabular}[t]{l}\svgunitsize 0\end{tabular}}}}%
    \put(0.89341655,0.00511451){\color[rgb]{0,0,0}\makebox(0,0)[t]{\lineheight{1.25}\smash{\begin{tabular}[t]{c}\svgunitsize $\times$10$^\text{5}$\end{tabular}}}}%
    \put(0,0){\includegraphics[width=\unitlength,page=5]{Self-focusing-Gauss-NLSE-damped.pdf}}%
    \put(0.95946091,0.24478122){\color[rgb]{0,0,0}\makebox(0,0)[lt]{\lineheight{1.25}\smash{\begin{tabular}[t]{l}\svgunitsize 1\end{tabular}}}}%
    \put(0.43929167,0.03078312){\color[rgb]{0,0,0}\makebox(0,0)[t]{\lineheight{1.25}\smash{\begin{tabular}[t]{c}\svgunitsize 0.6\end{tabular}}}}%
    \put(0,0){\includegraphics[width=\unitlength,page=6]{Self-focusing-Gauss-NLSE-damped.pdf}}%
    \put(0.68044621,0.03078312){\color[rgb]{0,0,0}\makebox(0,0)[t]{\lineheight{1.25}\smash{\begin{tabular}[t]{c}\svgunitsize 1\end{tabular}}}}%
    \put(0,0){\includegraphics[width=\unitlength,page=7]{Self-focusing-Gauss-NLSE-damped.pdf}}%
    \put(0.19841271,0.03078312){\color[rgb]{0,0,0}\makebox(0,0)[t]{\lineheight{1.25}\smash{\begin{tabular}[t]{c}\svgunitsize 0.2\end{tabular}}}}%
    \put(0,0){\includegraphics[width=\unitlength,page=8]{Self-focusing-Gauss-NLSE-damped.pdf}}%
    \put(0.31880831,0.03078312){\color[rgb]{0,0,0}\makebox(0,0)[t]{\lineheight{1.25}\smash{\begin{tabular}[t]{c}\svgunitsize 0.4\end{tabular}}}}%
    \put(0,0){\includegraphics[width=\unitlength,page=9]{Self-focusing-Gauss-NLSE-damped.pdf}}%
    \put(0.55851703,0.03078312){\color[rgb]{0,0,0}\makebox(0,0)[t]{\lineheight{1.25}\smash{\begin{tabular}[t]{c}\svgunitsize 0.8\end{tabular}}}}%
    \put(0,0){\includegraphics[width=\unitlength,page=10]{Self-focusing-Gauss-NLSE-damped.pdf}}%
    \put(0.79890381,0.03078312){\color[rgb]{0,0,0}\makebox(0,0)[t]{\lineheight{1.25}\smash{\begin{tabular}[t]{c}\svgunitsize 1.2\end{tabular}}}}%
    \put(0,0){\includegraphics[width=\unitlength,page=11]{Self-focusing-Gauss-NLSE-damped.pdf}}%
    \put(0.91847252,0.03078312){\color[rgb]{0,0,0}\makebox(0,0)[t]{\lineheight{1.25}\smash{\begin{tabular}[t]{c}\svgunitsize 1.4\end{tabular}}}}%
    \put(0,0){\includegraphics[width=\unitlength,page=12]{Self-focusing-Gauss-NLSE-damped.pdf}}%
    \put(0.95946091,0.20599317){\color[rgb]{0,0,0}\makebox(0,0)[lt]{\lineheight{1.25}\smash{\begin{tabular}[t]{l}\svgunitsize 0.8\end{tabular}}}}%
    \put(0,0){\includegraphics[width=\unitlength,page=13]{Self-focusing-Gauss-NLSE-damped.pdf}}%
    \put(0.95946091,0.16720513){\color[rgb]{0,0,0}\makebox(0,0)[lt]{\lineheight{1.25}\smash{\begin{tabular}[t]{l}\svgunitsize 0.6\end{tabular}}}}%
    \put(0,0){\includegraphics[width=\unitlength,page=14]{Self-focusing-Gauss-NLSE-damped.pdf}}%
    \put(0.95946091,0.1284171){\color[rgb]{0,0,0}\makebox(0,0)[lt]{\lineheight{1.25}\smash{\begin{tabular}[t]{l}\svgunitsize 0.4\end{tabular}}}}%
    \put(0,0){\includegraphics[width=\unitlength,page=15]{Self-focusing-Gauss-NLSE-damped.pdf}}%
    \put(0.95946091,0.08962907){\color[rgb]{0,0,0}\makebox(0,0)[lt]{\lineheight{1.25}\smash{\begin{tabular}[t]{l}\svgunitsize 0.2\end{tabular}}}}%
    \put(0,0){\includegraphics[width=\unitlength,page=16]{Self-focusing-Gauss-NLSE-damped.pdf}}%
    \put(0.95946091,0.28356923){\color[rgb]{0,0,0}\makebox(0,0)[lt]{\lineheight{1.25}\smash{\begin{tabular}[t]{l}\svgunitsize 1.2\end{tabular}}}}%
    \put(0,0){\includegraphics[width=\unitlength,page=17]{Self-focusing-Gauss-NLSE-damped.pdf}}%
  \end{picture}%
\endgroup%

%% file: images-simulations/breather2.pdf_tex
\begingroup%
  \makeatletter%
  \providecommand\color[2][]{%
    \errmessage{(Inkscape) Color is used for the text in Inkscape, but the package 'color.sty' is not loaded}%
    \renewcommand\color[2][]{}%
  }%
  \providecommand\transparent[1]{%
    \errmessage{(Inkscape) Transparency is used (non-zero) for the text in Inkscape, but the package 'transparent.sty' is not loaded}%
    \renewcommand\transparent[1]{}%
  }%
  \providecommand\rotatebox[2]{#2}%
  \newcommand*\fsize{\dimexpr\f@size pt\relax}%
  \newcommand*\lineheight[1]{\fontsize{\fsize}{#1\fsize}\selectfont}%
  \ifx\svgwidth\undefined%
    \setlength{\unitlength}{654.37338521bp}%
    \ifx\svgscale\undefined%
      \relax%
    \else%
      \setlength{\unitlength}{\unitlength * \real{\svgscale}}%
    \fi%
  \else%
    \setlength{\unitlength}{\svgwidth}%
  \fi%
  \global\let\svgwidth\undefined%
  \global\let\svgscale\undefined%
  \makeatother%
  \begin{picture}(1,0.21455683)%
    \lineheight{1}%
    \setlength\tabcolsep{0pt}%
    \put(0,0){\includegraphics[width=\unitlength,page=1]{breather2.pdf}}%
    \put(0.90439575,0.20062413){\color[rgb]{0,0,0}\makebox(0,0)[lt]{\lineheight{1.25}\smash{\begin{tabular}[t]{l}\svglabelsize $E_r$ / $a_0$\end{tabular}}}}%
    \put(0,0){\includegraphics[width=\unitlength,page=2]{breather2.pdf}}%
    \put(0.06905297,0.05004869){\color[rgb]{0,0,0}\makebox(0,0)[rt]{\lineheight{1.25}\smash{\begin{tabular}[t]{r}\svgunitsize 0.6\end{tabular}}}}%
    \put(0.06905281,0.0843934){\color[rgb]{0,0,0}\makebox(0,0)[rt]{\lineheight{1.25}\smash{\begin{tabular}[t]{r}\svgunitsize 0.7\end{tabular}}}}%
    \put(0.06905281,0.11837082){\color[rgb]{0,0,0}\makebox(0,0)[rt]{\lineheight{1.25}\smash{\begin{tabular}[t]{r}\svgunitsize 0.8\end{tabular}}}}%
    \put(0.06905281,0.15258223){\color[rgb]{0,0,0}\makebox(0,0)[rt]{\lineheight{1.25}\smash{\begin{tabular}[t]{r}\svgunitsize 0.9\end{tabular}}}}%
    \put(0.06905282,0.18666969){\color[rgb]{0,0,0}\makebox(0,0)[rt]{\lineheight{1.25}\smash{\begin{tabular}[t]{r}\svgunitsize 1\end{tabular}}}}%
    \put(0.07875028,0.03078309){\color[rgb]{0,0,0}\makebox(0,0)[t]{\lineheight{1.25}\smash{\begin{tabular}[t]{c}\svgunitsize 0\end{tabular}}}}%
    \put(0.49721781,0.00432486){\color[rgb]{0,0,0}\makebox(0,0)[t]{\lineheight{1.25}\smash{\begin{tabular}[t]{c}\svglabelsize $t$ / $\omega_{ci}^{-1}$\end{tabular}}}}%
    \put(0.02358771,0.12458425){\color[rgb]{0,0,0}\rotatebox{90}{\makebox(0,0)[t]{\lineheight{1.25}\smash{\begin{tabular}[t]{c}\svglabelsize $r$ / $a_\text{min}$\end{tabular}}}}}%
    \put(0.19642493,0.03078309){\color[rgb]{0,0,0}\makebox(0,0)[t]{\lineheight{1.25}\smash{\begin{tabular}[t]{c}\svgunitsize 0.5\end{tabular}}}}%
    \put(0,0){\includegraphics[width=\unitlength,page=3]{breather2.pdf}}%
    \put(0.1188447,0.08597278){\color[rgb]{0,0,0}\makebox(0,0)[lt]{\lineheight{1.25}\smash{\begin{tabular}[t]{l}\svgunitsize $r_1\!\approx\!0.71\,a_\text{min}$\end{tabular}}}}%
    \put(0.1188447,0.06813126){\color[rgb]{0,0,0}\makebox(0,0)[lt]{\lineheight{1.25}\smash{\begin{tabular}[t]{l}\svgunitsize $r_2\!\approx\!0.87\,a_\text{min}$\end{tabular}}}}%
    \put(0.1188447,0.10330391){\color[rgb]{0,0,0}\makebox(0,0)[lt]{\lineheight{1.25}\smash{\begin{tabular}[t]{l}\svgunitsize GK sim.\end{tabular}}}}%
    \put(0,0){\includegraphics[width=\unitlength,page=4]{breather2.pdf}}%
    \put(0.31446877,0.03078307){\color[rgb]{0,0,0}\makebox(0,0)[t]{\lineheight{1.25}\smash{\begin{tabular}[t]{c}\svgunitsize 1\end{tabular}}}}%
    \put(0,0){\includegraphics[width=\unitlength,page=5]{breather2.pdf}}%
    \put(0.43263076,0.03078307){\color[rgb]{0,0,0}\makebox(0,0)[t]{\lineheight{1.25}\smash{\begin{tabular}[t]{c}\svgunitsize 1.5\end{tabular}}}}%
    \put(0,0){\includegraphics[width=\unitlength,page=6]{breather2.pdf}}%
    \put(0.55068871,0.03078307){\color[rgb]{0,0,0}\makebox(0,0)[t]{\lineheight{1.25}\smash{\begin{tabular}[t]{c}\svgunitsize 2\end{tabular}}}}%
    \put(0,0){\includegraphics[width=\unitlength,page=7]{breather2.pdf}}%
    \put(0.66877625,0.03078307){\color[rgb]{0,0,0}\makebox(0,0)[t]{\lineheight{1.25}\smash{\begin{tabular}[t]{c}\svgunitsize 2.5\end{tabular}}}}%
    \put(0,0){\includegraphics[width=\unitlength,page=8]{breather2.pdf}}%
    \put(0.78696984,0.03078307){\color[rgb]{0,0,0}\makebox(0,0)[t]{\lineheight{1.25}\smash{\begin{tabular}[t]{c}\svgunitsize 3\end{tabular}}}}%
    \put(0,0){\includegraphics[width=\unitlength,page=9]{breather2.pdf}}%
    \put(0.9050515,0.03078307){\color[rgb]{0,0,0}\makebox(0,0)[t]{\lineheight{1.25}\smash{\begin{tabular}[t]{c}\svgunitsize 3.5\end{tabular}}}}%
    \put(0.8934209,0.00511424){\color[rgb]{0,0,0}\makebox(0,0)[t]{\lineheight{1.25}\smash{\begin{tabular}[t]{c}\svgunitsize $\times$10$^\text{5}$\end{tabular}}}}%
    \put(0,0){\includegraphics[width=\unitlength,page=10]{breather2.pdf}}%
    \put(0.95746373,0.17081944){\color[rgb]{0,0,0}\makebox(0,0)[lt]{\lineheight{1.25}\smash{\begin{tabular}[t]{l}\svgunitsize 1.5\end{tabular}}}}%
    \put(0,0){\includegraphics[width=\unitlength,page=11]{breather2.pdf}}%
    \put(0.95746373,0.05081823){\color[rgb]{0,0,0}\makebox(0,0)[lt]{\lineheight{1.25}\smash{\begin{tabular}[t]{l}\svgunitsize 0\end{tabular}}}}%
    \put(0,0){\includegraphics[width=\unitlength,page=12]{breather2.pdf}}%
    \put(0.95746373,0.13081904){\color[rgb]{0,0,0}\makebox(0,0)[lt]{\lineheight{1.25}\smash{\begin{tabular}[t]{l}\svgunitsize 1\end{tabular}}}}%
    \put(0,0){\includegraphics[width=\unitlength,page=13]{breather2.pdf}}%
    \put(0.95746373,0.09081864){\color[rgb]{0,0,0}\makebox(0,0)[lt]{\lineheight{1.25}\smash{\begin{tabular}[t]{l}\svgunitsize 0.5\end{tabular}}}}%
    \put(0,0){\includegraphics[width=\unitlength,page=14]{breather2.pdf}}%
  \end{picture}%
\endgroup%

%% file: images-simulations/breather3.pdf_tex
\begingroup%
  \makeatletter%
  \providecommand\color[2][]{%
    \errmessage{(Inkscape) Color is used for the text in Inkscape, but the package 'color.sty' is not loaded}%
    \renewcommand\color[2][]{}%
  }%
  \providecommand\transparent[1]{%
    \errmessage{(Inkscape) Transparency is used (non-zero) for the text in Inkscape, but the package 'transparent.sty' is not loaded}%
    \renewcommand\transparent[1]{}%
  }%
  \providecommand\rotatebox[2]{#2}%
  \newcommand*\fsize{\dimexpr\f@size pt\relax}%
  \newcommand*\lineheight[1]{\fontsize{\fsize}{#1\fsize}\selectfont}%
  \ifx\svgwidth\undefined%
    \setlength{\unitlength}{654.37338521bp}%
    \ifx\svgscale\undefined%
      \relax%
    \else%
      \setlength{\unitlength}{\unitlength * \real{\svgscale}}%
    \fi%
  \else%
    \setlength{\unitlength}{\svgwidth}%
  \fi%
  \global\let\svgwidth\undefined%
  \global\let\svgscale\undefined%
  \makeatother%
  \begin{picture}(1,0.20060244)%
    \lineheight{1}%
    \setlength\tabcolsep{0pt}%
    \put(0,0){\includegraphics[width=\unitlength,page=1]{breather3.pdf}}%
    \put(0.06905319,0.05004865){\color[rgb]{0,0,0}\makebox(0,0)[rt]{\lineheight{1.25}\smash{\begin{tabular}[t]{r}\svgunitsize 0\end{tabular}}}}%
    \put(0.06905288,0.08439342){\color[rgb]{0,0,0}\makebox(0,0)[rt]{\lineheight{1.25}\smash{\begin{tabular}[t]{r}\svgunitsize 0.5\end{tabular}}}}%
    \put(0.06905288,0.11837087){\color[rgb]{0,0,0}\makebox(0,0)[rt]{\lineheight{1.25}\smash{\begin{tabular}[t]{r}\svgunitsize 1\end{tabular}}}}%
    \put(0.06905288,0.15258226){\color[rgb]{0,0,0}\makebox(0,0)[rt]{\lineheight{1.25}\smash{\begin{tabular}[t]{r}\svgunitsize 1.5\end{tabular}}}}%
    \put(0.06905288,0.18666977){\color[rgb]{0,0,0}\makebox(0,0)[rt]{\lineheight{1.25}\smash{\begin{tabular}[t]{r}\svgunitsize 2\end{tabular}}}}%
    \put(0.07875045,0.0307831){\color[rgb]{0,0,0}\makebox(0,0)[t]{\lineheight{1.25}\smash{\begin{tabular}[t]{c}\svgunitsize 0\end{tabular}}}}%
    \put(0.49721798,0.00432486){\color[rgb]{0,0,0}\makebox(0,0)[t]{\lineheight{1.25}\smash{\begin{tabular}[t]{c}\svglabelsize $t$ / $\omega_{ci}^{-1}$\end{tabular}}}}%
    \put(0.02358778,0.1245843){\color[rgb]{0,0,0}\rotatebox{90}{\makebox(0,0)[t]{\lineheight{1.25}\smash{\begin{tabular}[t]{c}\svglabelsize $E_r$ / $a_0$\end{tabular}}}}}%
    \put(0.19642509,0.0307831){\color[rgb]{0,0,0}\makebox(0,0)[t]{\lineheight{1.25}\smash{\begin{tabular}[t]{c}\svgunitsize 0.5\end{tabular}}}}%
    \put(0.11884488,0.0859728){\color[rgb]{0,0,0}\makebox(0,0)[lt]{\lineheight{1.25}\smash{\begin{tabular}[t]{l}\svgunitsize $r_2\!\approx\!0.87\,a_\text{min}$\end{tabular}}}}%
    \put(0.11884488,0.06813129){\color[rgb]{0,0,0}\makebox(0,0)[lt]{\lineheight{1.25}\smash{\begin{tabular}[t]{l}\svgunitsize Envelope\end{tabular}}}}%
    \put(0.11884488,0.1033889){\color[rgb]{0,0,0}\makebox(0,0)[lt]{\lineheight{1.25}\smash{\begin{tabular}[t]{l}\svgunitsize Envelope\end{tabular}}}}%
    \put(0.11884923,0.12114529){\color[rgb]{0,0,0}\makebox(0,0)[lt]{\lineheight{1.25}\smash{\begin{tabular}[t]{l}\svgunitsize $r_1\!\approx\!0.71\,a_\text{min}$\end{tabular}}}}%
    \put(0,0){\includegraphics[width=\unitlength,page=2]{breather3.pdf}}%
    \put(0.3144689,0.03078309){\color[rgb]{0,0,0}\makebox(0,0)[t]{\lineheight{1.25}\smash{\begin{tabular}[t]{c}\svgunitsize 1\end{tabular}}}}%
    \put(0,0){\includegraphics[width=\unitlength,page=3]{breather3.pdf}}%
    \put(0.43263089,0.03078309){\color[rgb]{0,0,0}\makebox(0,0)[t]{\lineheight{1.25}\smash{\begin{tabular}[t]{c}\svgunitsize 1.5\end{tabular}}}}%
    \put(0,0){\includegraphics[width=\unitlength,page=4]{breather3.pdf}}%
    \put(0.55068884,0.03078309){\color[rgb]{0,0,0}\makebox(0,0)[t]{\lineheight{1.25}\smash{\begin{tabular}[t]{c}\svgunitsize 2\end{tabular}}}}%
    \put(0,0){\includegraphics[width=\unitlength,page=5]{breather3.pdf}}%
    \put(0.66877638,0.03078309){\color[rgb]{0,0,0}\makebox(0,0)[t]{\lineheight{1.25}\smash{\begin{tabular}[t]{c}\svgunitsize 2.5\end{tabular}}}}%
    \put(0,0){\includegraphics[width=\unitlength,page=6]{breather3.pdf}}%
    \put(0.78696997,0.03078309){\color[rgb]{0,0,0}\makebox(0,0)[t]{\lineheight{1.25}\smash{\begin{tabular}[t]{c}\svgunitsize 3\end{tabular}}}}%
    \put(0,0){\includegraphics[width=\unitlength,page=7]{breather3.pdf}}%
    \put(0.90505163,0.03078309){\color[rgb]{0,0,0}\makebox(0,0)[t]{\lineheight{1.25}\smash{\begin{tabular}[t]{c}\svgunitsize 3.5\end{tabular}}}}%
    \put(0.893421,0.00511427){\color[rgb]{0,0,0}\makebox(0,0)[t]{\lineheight{1.25}\smash{\begin{tabular}[t]{c}\svgunitsize $\times$10$^\text{5}$\end{tabular}}}}%
  \end{picture}%
\endgroup%

%% file: images-appendix/alpha-NL-vs-taue-result-15-05-23.pdf_tex
\begingroup%
  \makeatletter%
  \providecommand\color[2][]{%
    \errmessage{(Inkscape) Color is used for the text in Inkscape, but the package 'color.sty' is not loaded}%
    \renewcommand\color[2][]{}%
  }%
  \providecommand\transparent[1]{%
    \errmessage{(Inkscape) Transparency is used (non-zero) for the text in Inkscape, but the package 'transparent.sty' is not loaded}%
    \renewcommand\transparent[1]{}%
  }%
  \providecommand\rotatebox[2]{#2}%
  \newcommand*\fsize{\dimexpr\f@size pt\relax}%
  \newcommand*\lineheight[1]{\fontsize{\fsize}{#1\fsize}\selectfont}%
  \ifx\svgwidth\undefined%
    \setlength{\unitlength}{476.25bp}%
    \ifx\svgscale\undefined%
      \relax%
    \else%
      \setlength{\unitlength}{\unitlength * \real{\svgscale}}%
    \fi%
  \else%
    \setlength{\unitlength}{\svgwidth}%
  \fi%
  \global\let\svgwidth\undefined%
  \global\let\svgscale\undefined%
  \makeatother%
  \begin{picture}(1,0.62796198)%
    \lineheight{1}%
    \setlength\tabcolsep{0pt}%
    \put(0,0){\includegraphics[width=\unitlength,page=1]{alpha-NL-vs-taue-result-15-05-23.pdf}}%
    \put(0.14184178,0.06109036){\makebox(0,0)[lt]{\lineheight{1.25}\smash{\begin{tabular}[t]{l}\SVGunitsize 0\end{tabular}}}}%
    \put(0.31619501,0.06109036){\makebox(0,0)[lt]{\lineheight{1.25}\smash{\begin{tabular}[t]{l}\SVGunitsize 5\end{tabular}}}}%
    \put(0.48424902,0.06109036){\makebox(0,0)[lt]{\lineheight{1.25}\smash{\begin{tabular}[t]{l}\SVGunitsize 10\end{tabular}}}}%
    \put(0.65860225,0.06109036){\makebox(0,0)[lt]{\lineheight{1.25}\smash{\begin{tabular}[t]{l}\SVGunitsize 15\end{tabular}}}}%
    \put(0.83295532,0.06109036){\makebox(0,0)[lt]{\lineheight{1.25}\smash{\begin{tabular}[t]{l}\SVGunitsize 20\end{tabular}}}}%
    \put(0.42215706,0.00544721){\makebox(0,0)[lt]{\lineheight{1.25}\smash{\begin{tabular}[t]{l}\SVGlabelsize $\tau_e = T_e/T_i$\end{tabular}}}}%
    \put(0,0){\includegraphics[width=\unitlength,page=2]{alpha-NL-vs-taue-result-15-05-23.pdf}}%
    \put(0.11400392,0.09862312){\makebox(0,0)[lt]{\lineheight{1.25}\smash{\begin{tabular}[t]{l}\SVGunitsize 0\end{tabular}}}}%
    \put(0.07463384,0.18418737){\makebox(0,0)[lt]{\lineheight{1.25}\smash{\begin{tabular}[t]{l}\SVGunitsize 200\end{tabular}}}}%
    \put(0.07463384,0.26975178){\makebox(0,0)[lt]{\lineheight{1.25}\smash{\begin{tabular}[t]{l}\SVGunitsize 400\end{tabular}}}}%
    \put(0.07463384,0.35531603){\makebox(0,0)[lt]{\lineheight{1.25}\smash{\begin{tabular}[t]{l}\SVGunitsize 600\end{tabular}}}}%
    \put(0.07463384,0.44088028){\makebox(0,0)[lt]{\lineheight{1.25}\smash{\begin{tabular}[t]{l}\SVGunitsize 800\end{tabular}}}}%
    \put(0.05549013,0.52644468){\makebox(0,0)[lt]{\lineheight{1.25}\smash{\begin{tabular}[t]{l}\SVGunitsize 1000\end{tabular}}}}%
    \put(0.05549013,0.61200894){\makebox(0,0)[lt]{\lineheight{1.25}\smash{\begin{tabular}[t]{l}\SVGunitsize 1200\end{tabular}}}}%
    \put(0,0){\includegraphics[width=\unitlength,page=3]{alpha-NL-vs-taue-result-15-05-23.pdf}}%
    \put(0.0208445,0.27500127){\rotatebox{90}{\makebox(0,0)[lt]{\lineheight{1.25}\smash{\begin{tabular}[t]{l}\SVGlabelsize $\alpha_\text{NL} (\omega_\text{ci})$\end{tabular}}}}}%
  \end{picture}%
\endgroup%

%% file: images-appendix/alpha-NL-vs-rhoi-amin-result-15-05-23.pdf_tex
\begingroup%
  \makeatletter%
  \providecommand\color[2][]{%
    \errmessage{(Inkscape) Color is used for the text in Inkscape, but the package 'color.sty' is not loaded}%
    \renewcommand\color[2][]{}%
  }%
  \providecommand\transparent[1]{%
    \errmessage{(Inkscape) Transparency is used (non-zero) for the text in Inkscape, but the package 'transparent.sty' is not loaded}%
    \renewcommand\transparent[1]{}%
  }%
  \providecommand\rotatebox[2]{#2}%
  \newcommand*\fsize{\dimexpr\f@size pt\relax}%
  \newcommand*\lineheight[1]{\fontsize{\fsize}{#1\fsize}\selectfont}%
  \ifx\svgwidth\undefined%
    \setlength{\unitlength}{476.24990845bp}%
    \ifx\svgscale\undefined%
      \relax%
    \else%
      \setlength{\unitlength}{\unitlength * \real{\svgscale}}%
    \fi%
  \else%
    \setlength{\unitlength}{\svgwidth}%
  \fi%
  \global\let\svgwidth\undefined%
  \global\let\svgscale\undefined%
  \makeatother%
  \begin{picture}(1,0.62148738)%
    \lineheight{1}%
    \setlength\tabcolsep{0pt}%
    \put(0,0){\includegraphics[width=\unitlength,page=1]{alpha-NL-vs-rhoi-amin-result-15-05-23.pdf}}%
    \put(0.78511022,0.01943136){\color[rgb]{0.14901961,0.14901961,0.14901961}\makebox(0,0)[lt]{\lineheight{1.25}\smash{\begin{tabular}[t]{l}\SVGunitsize $\cdot$10$^\text{-3}$\end{tabular}}}}%
    \put(0.07136316,0.21049251){\makebox(0,0)[lt]{\lineheight{1.25}\smash{\begin{tabular}[t]{l}\SVGunitsize 200\end{tabular}}}}%
    \put(0.07136316,0.3214949){\makebox(0,0)[lt]{\lineheight{1.25}\smash{\begin{tabular}[t]{l}\SVGunitsize 400\end{tabular}}}}%
    \put(0.07136316,0.43249728){\makebox(0,0)[lt]{\lineheight{1.25}\smash{\begin{tabular}[t]{l}\SVGunitsize 600\end{tabular}}}}%
    \put(0.07136316,0.5434995){\makebox(0,0)[lt]{\lineheight{1.25}\smash{\begin{tabular}[t]{l}\SVGunitsize 800\end{tabular}}}}%
    \put(0.11073325,0.09999219){\makebox(0,0)[lt]{\lineheight{1.25}\smash{\begin{tabular}[t]{l}\SVGunitsize 0\end{tabular}}}}%
    \put(0.13664501,0.06109039){\makebox(0,0)[lt]{\lineheight{1.25}\smash{\begin{tabular}[t]{l}\SVGunitsize 3\end{tabular}}}}%
    \put(0.43335597,0.06109428){\makebox(0,0)[lt]{\lineheight{1.25}\smash{\begin{tabular}[t]{l}\SVGunitsize 5\end{tabular}}}}%
    \put(0.44616297,0.00544723){\makebox(0,0)[lt]{\lineheight{1.25}\smash{\begin{tabular}[t]{l}\SVGlabelsize $\rho_i/a_\text{min}$ \end{tabular}}}}%
    \put(0,0){\includegraphics[width=\unitlength,page=2]{alpha-NL-vs-rhoi-amin-result-15-05-23.pdf}}%
    \put(0.01754844,0.27617392){\color[rgb]{0.14901961,0.14901961,0.14901961}\rotatebox{90}{\makebox(0,0)[lt]{\lineheight{1.25}\smash{\begin{tabular}[t]{l}\SVGlabelsize $\alpha_\text{NL} (\omega_\text{ci})$\end{tabular}}}}}%
    \put(0.285003,0.061095){\makebox(0,0)[lt]{\lineheight{1.25}\smash{\begin{tabular}[t]{l}\SVGunitsize 4\end{tabular}}}}%
    \put(0,0){\includegraphics[width=\unitlength,page=3]{alpha-NL-vs-rhoi-amin-result-15-05-23.pdf}}%
    \put(0.73007144,0.061095){\makebox(0,0)[lt]{\lineheight{1.25}\smash{\begin{tabular}[t]{l}\SVGunitsize 7\end{tabular}}}}%
    \put(0.58171847,0.06109573){\makebox(0,0)[lt]{\lineheight{1.25}\smash{\begin{tabular}[t]{l}\SVGunitsize 6\end{tabular}}}}%
    \put(0,0){\includegraphics[width=\unitlength,page=4]{alpha-NL-vs-rhoi-amin-result-15-05-23.pdf}}%
    \put(0.87842801,0.0610944){\makebox(0,0)[lt]{\lineheight{1.25}\smash{\begin{tabular}[t]{l}\SVGunitsize 8\end{tabular}}}}%
  \end{picture}%
\endgroup%

%% file: images-appendix/alpha-NL-vs-r0-result-31-05-23.pdf_tex
\begingroup%
  \makeatletter%
  \providecommand\color[2][]{%
    \errmessage{(Inkscape) Color is used for the text in Inkscape, but the package 'color.sty' is not loaded}%
    \renewcommand\color[2][]{}%
  }%
  \providecommand\transparent[1]{%
    \errmessage{(Inkscape) Transparency is used (non-zero) for the text in Inkscape, but the package 'transparent.sty' is not loaded}%
    \renewcommand\transparent[1]{}%
  }%
  \providecommand\rotatebox[2]{#2}%
  \newcommand*\fsize{\dimexpr\f@size pt\relax}%
  \newcommand*\lineheight[1]{\fontsize{\fsize}{#1\fsize}\selectfont}%
  \ifx\svgwidth\undefined%
    \setlength{\unitlength}{476.25bp}%
    \ifx\svgscale\undefined%
      \relax%
    \else%
      \setlength{\unitlength}{\unitlength * \real{\svgscale}}%
    \fi%
  \else%
    \setlength{\unitlength}{\svgwidth}%
  \fi%
  \global\let\svgwidth\undefined%
  \global\let\svgscale\undefined%
  \makeatother%
  \begin{picture}(1,0.62148091)%
    \lineheight{1}%
    \setlength\tabcolsep{0pt}%
    \put(0,0){\includegraphics[width=\unitlength,page=1]{alpha-NL-vs-r0-result-31-05-23.pdf}}%
    \put(0.27468952,0.0610904){\makebox(0,0)[lt]{\lineheight{1.25}\smash{\begin{tabular}[t]{l}\SVGunitsize 0.2\end{tabular}}}}%
    \put(0.42114622,0.0610904){\makebox(0,0)[lt]{\lineheight{1.25}\smash{\begin{tabular}[t]{l}\SVGunitsize 0.4\end{tabular}}}}%
    \put(0.56760291,0.0610904){\makebox(0,0)[lt]{\lineheight{1.25}\smash{\begin{tabular}[t]{l}\SVGunitsize 0.6\end{tabular}}}}%
    \put(0.71405969,0.06108723){\color[rgb]{0.14901961,0.14901961,0.14901961}\makebox(0,0)[lt]{\lineheight{1.25}\smash{\begin{tabular}[t]{l}\SVGunitsize 0.8\end{tabular}}}}%
    \put(0.86996518,0.06109015){\color[rgb]{0.14901961,0.14901961,0.14901961}\makebox(0,0)[lt]{\lineheight{1.25}\smash{\begin{tabular}[t]{l}\SVGunitsize 1\end{tabular}}}}%
    \put(0.07109171,0.17851343){\makebox(0,0)[lt]{\lineheight{1.25}\smash{\begin{tabular}[t]{l}\SVGunitsize 500\end{tabular}}}}%
    \put(0.05219407,0.25629909){\makebox(0,0)[lt]{\lineheight{1.25}\smash{\begin{tabular}[t]{l}\SVGunitsize 1000\end{tabular}}}}%
    \put(0.05219407,0.33408474){\makebox(0,0)[lt]{\lineheight{1.25}\smash{\begin{tabular}[t]{l}\SVGunitsize 1500\end{tabular}}}}%
    \put(0.05219407,0.41187058){\makebox(0,0)[lt]{\lineheight{1.25}\smash{\begin{tabular}[t]{l}\SVGunitsize 2000\end{tabular}}}}%
    \put(0.05219407,0.5674419){\makebox(0,0)[lt]{\lineheight{1.25}\smash{\begin{tabular}[t]{l}\SVGunitsize 3000\end{tabular}}}}%
    \put(0.05219407,0.49102148){\makebox(0,0)[lt]{\lineheight{1.25}\smash{\begin{tabular}[t]{l}\SVGunitsize 2500\end{tabular}}}}%
    \put(0.11046178,0.09940838){\makebox(0,0)[lt]{\lineheight{1.25}\smash{\begin{tabular}[t]{l}\SVGunitsize 0\end{tabular}}}}%
    \put(0.13768165,0.0610904){\makebox(0,0)[lt]{\lineheight{1.25}\smash{\begin{tabular}[t]{l}\SVGunitsize 0\end{tabular}}}}%
    \put(0.44368926,0.00544725){\makebox(0,0)[lt]{\lineheight{1.25}\smash{\begin{tabular}[t]{l}\SVGlabelsize $r_0 / a_\text{min}$\end{tabular}}}}%
    \put(0,0){\includegraphics[width=\unitlength,page=2]{alpha-NL-vs-r0-result-31-05-23.pdf}}%
    \put(0.01754843,0.27579391){\rotatebox{90}{\makebox(0,0)[lt]{\lineheight{1.25}\smash{\begin{tabular}[t]{l}\svglabelsize $\alpha_\text{NL} (\omega_\text{ci})$\end{tabular}}}}}%
  \end{picture}%
\endgroup%

%% file: images-theory/MI-understanding-0-altered-edge-mzcolormap-2-edit.pdf_tex
\begingroup%
  \makeatletter%
  \providecommand\color[2][]{%
    \errmessage{(Inkscape) Color is used for the text in Inkscape, but the package 'color.sty' is not loaded}%
    \renewcommand\color[2][]{}%
  }%
  \providecommand\transparent[1]{%
    \errmessage{(Inkscape) Transparency is used (non-zero) for the text in Inkscape, but the package 'transparent.sty' is not loaded}%
    \renewcommand\transparent[1]{}%
  }%
  \providecommand\rotatebox[2]{#2}%
  \newcommand*\fsize{\dimexpr\f@size pt\relax}%
  \newcommand*\lineheight[1]{\fontsize{\fsize}{#1\fsize}\selectfont}%
  \ifx\svgwidth\undefined%
    \setlength{\unitlength}{1344.26600178bp}%
    \ifx\svgscale\undefined%
      \relax%
    \else%
      \setlength{\unitlength}{\unitlength * \real{\svgscale}}%
    \fi%
  \else%
    \setlength{\unitlength}{\svgwidth}%
  \fi%
  \global\let\svgwidth\undefined%
  \global\let\svgscale\undefined%
  \makeatother%
  \begin{picture}(1,0.26175072)%
    \lineheight{1}%
    \setlength\tabcolsep{0pt}%
    \put(0,0){\includegraphics[width=\unitlength,page=1]{MI-understanding-0-altered-edge-mzcolormap-2-edit.pdf}}%
    \put(0.0365816,0.20754746){\color[rgb]{0,0,0}\makebox(0,0)[lt]{\lineheight{1.25}\smash{\begin{tabular}[t]{l}\SVGunitsize 0.6\end{tabular}}}}%
    \put(0.95881568,0.20197012){\color[rgb]{0,0,0}\makebox(0,0)[lt]{\lineheight{1.25}\smash{\begin{tabular}[t]{l}{\SVGunitsize 1}\end{tabular}}}}%
    \put(0.9118166,0.24479505){\color[rgb]{0,0,0}\makebox(0,0)[lt]{\lineheight{1.25}\smash{\begin{tabular}[t]{l}{\SVGlabelsize $\text{Re}[\psi]$/$a_0$}\end{tabular}}}}%
    \put(0.95881566,0.0663863){\color[rgb]{0,0,0}\makebox(0,0)[lt]{\lineheight{1.25}\smash{\begin{tabular}[t]{l}{\SVGunitsize -1}\end{tabular}}}}%
    \put(0.95881566,0.13567364){\color[rgb]{0,0,0}\makebox(0,0)[lt]{\lineheight{1.25}\smash{\begin{tabular}[t]{l}{\SVGunitsize 0}\end{tabular}}}}%
    \put(0,0){\includegraphics[width=\unitlength,page=2]{MI-understanding-0-altered-edge-mzcolormap-2-edit.pdf}}%
    \put(0.0365816,0.05512584){\color[rgb]{0,0,0}\makebox(0,0)[lt]{\lineheight{1.25}\smash{\begin{tabular}[t]{l}\SVGunitsize 0.4\end{tabular}}}}%
    \put(0.07449232,0.03088755){\color[rgb]{0,0,0}\makebox(0,0)[lt]{\lineheight{1.25}\smash{\begin{tabular}[t]{l}{\SVGunitsize 0}\end{tabular}}}}%
    \put(0.17840684,0.03088755){\color[rgb]{0,0,0}\makebox(0,0)[lt]{\lineheight{1.25}\smash{\begin{tabular}[t]{l}{\SVGunitsize 1}\end{tabular}}}}%
    \put(0.28200415,0.03088755){\color[rgb]{0,0,0}\makebox(0,0)[lt]{\lineheight{1.25}\smash{\begin{tabular}[t]{l}{\SVGunitsize 2}\end{tabular}}}}%
    \put(0.38578519,0.03088755){\color[rgb]{0,0,0}\makebox(0,0)[lt]{\lineheight{1.25}\smash{\begin{tabular}[t]{l}{\SVGunitsize 3}\end{tabular}}}}%
    \put(0.4896878,0.03088755){\color[rgb]{0,0,0}\makebox(0,0)[lt]{\lineheight{1.25}\smash{\begin{tabular}[t]{l}{\SVGunitsize 4}\end{tabular}}}}%
    \put(0.59322101,0.03088755){\color[rgb]{0,0,0}\makebox(0,0)[lt]{\lineheight{1.25}\smash{\begin{tabular}[t]{l}{\SVGunitsize 5}\end{tabular}}}}%
    \put(0.69705815,0.03088755){\color[rgb]{0,0,0}\makebox(0,0)[lt]{\lineheight{1.25}\smash{\begin{tabular}[t]{l}{\SVGunitsize 6}\end{tabular}}}}%
    \put(0.80080363,0.03088755){\color[rgb]{0,0,0}\makebox(0,0)[lt]{\lineheight{1.25}\smash{\begin{tabular}[t]{l}{\SVGunitsize 7}\end{tabular}}}}%
    \put(0.90458804,0.03088755){\color[rgb]{0,0,0}\makebox(0,0)[lt]{\lineheight{1.25}\smash{\begin{tabular}[t]{l}{\SVGunitsize 8}\end{tabular}}}}%
    \put(0.0365816,0.13219116){\color[rgb]{0,0,0}\makebox(0,0)[lt]{\lineheight{1.25}\smash{\begin{tabular}[t]{l}\SVGunitsize 0.5\end{tabular}}}}%
    \put(0.02044052,0.09859009){\color[rgb]{0,0,0}\rotatebox{90}{\makebox(0,0)[lt]{\lineheight{1.25}\smash{\begin{tabular}[t]{l}\SVGlabelsize $r$ (a.u.)\end{tabular}}}}}%
    \put(0.45921353,0.00340431){\color[rgb]{0,0,0}\makebox(0,0)[lt]{\lineheight{1.25}\smash{\begin{tabular}[t]{l}\SVGlabelsize $t$ (a.u.)\end{tabular}}}}%
    \put(0,0){\includegraphics[width=\unitlength,page=3]{MI-understanding-0-altered-edge-mzcolormap-2-edit.pdf}}%
    \put(0.70654793,0.19070144){\color[rgb]{0,0,0}\makebox(0,0)[lt]{\lineheight{1.25}\smash{\begin{tabular}[t]{l}{\SVGunitsize anomalous disp.}\end{tabular}}}}%
    \put(0,0){\includegraphics[width=\unitlength,page=4]{MI-understanding-0-altered-edge-mzcolormap-2-edit.pdf}}%
    \put(0.52661518,0.19070144){\color[rgb]{0,0,0}\makebox(0,0)[lt]{\lineheight{1.25}\smash{\begin{tabular}[t]{l}{\SVGmathsize $\calG\!>\!0$}\end{tabular}}}}%
    \put(0,0){\includegraphics[width=\unitlength,page=5]{MI-understanding-0-altered-edge-mzcolormap-2-edit.pdf}}%
    \put(0.60560077,0.19070144){\color[rgb]{0,0,0}\makebox(0,0)[lt]{\lineheight{1.25}\smash{\begin{tabular}[t]{l}{\SVGmathsize $\alpha_\text{NL}\!=\!0$}\end{tabular}}}}%
    \put(0,0){\includegraphics[width=\unitlength,page=6]{MI-understanding-0-altered-edge-mzcolormap-2-edit.pdf}}%
  \end{picture}%
\endgroup%